\documentclass[traditabstract]{aa}

\usepackage{aas_macros}
\usepackage{aecompl}

\usepackage[round]{natbib}
\usepackage{array}
\usepackage{float}
\usepackage{amsmath}
\usepackage{graphicx}
\usepackage{epsfig}
\usepackage{amssymb}
\usepackage[latin1]{inputenc}
\usepackage{color}
\usepackage{subfigure}
\usepackage{sidecap}



\begin{document}

\bibliographystyle{plainnat}

\title{Protoneutron star evolution and the neutrino-driven wind
in general relativistic neutrino radiation hydrodynamics simulations}

\titlerunning{protoneutron star evolution and the neutrino-driven wind}

\author{T. Fischer
        \inst{1},
	S. C. Whitehouse
	\inst{1},
        A. Mezzacappa
        \inst{2},
        F.-K. Thielemann
        \inst{1}$^,$\inst{3}
        and 
        M. Liebend\"orfer
        \inst{1}
}

\authorrunning{Fischer et al.}

\institute{
Department of Physics, University of Basel,
Klingelbergstrasse 82, 4056 Basel, Switzerland
\and
Physics Division, Oak Ridge National Laboratory,
Oak Ridge, Tennessee 37831-1200, United States of America
\and
GSI Helmholtzzentrum f\"ur Schwerionenforschung GmbH,
Planckstrasse1, 64291 Darmstadt, Germany
}

\abstract{
Massive stars end their lives in explosions with kinetic
energies on the order of \(10^{51}\) erg.
Immediately after the explosion has been launched, a region of
low density and high entropy forms behind the ejecta, which is
continuously subject to neutrino heating.
The neutrinos emitted from the remnant at the center,
the protoneutron star (PNS), heat the material above the PNS surface.
This heat is partly converted into kinetic energy, and the material
accelerates to an outflow that is known as the neutrino-driven wind.
For the first time we simulate the collapse, bounce, explosion, and the
neutrino-driven wind phases consistently over more than \(20\) seconds.
Our numerical model is based on spherically symmetric general
relativistic radiation hydrodynamics using spectral three-flavor
Boltzmann neutrino transport.
In simulations where no explosions are obtained naturally,
we model neutrino-driven explosions for low- and intermediate-mass
Fe-core progenitor stars by enhancing the charged current reaction rates.
In the case of a special progenitor star, the \(8.8\) M\(_\odot\)
O-Ne-Mg-core, the explosion in spherical symmetry was obtained without
enhanced opacities.
The post-explosion evolution is in qualitative agreement with static
steady-state and parametrized dynamic models of the neutrino-driven wind.
On the other hand, we generally find lower neutrino luminosities
and mean neutrino energies, as well as a different evolutionary behavior
of the neutrino luminosities and mean neutrino energies.
The neutrino-driven wind is proton-rich for more than \(10\) seconds
and the contraction of the PNS differs from the assumptions made for the
conditions at the inner boundary in previous neutrino-driven wind studies.
Despite the moderately high entropies of about \(100\) k\(_\text{B}\)/baryon
and the fast expansion timescales, the conditions found in our models
are unlikely to favor \(r\)-process nucleosynthesis.
The simulations are carried out until the neutrino-driven wind
settles down to a quasi-stationary state.
About \(5\) seconds after the bounce, the peak temperature inside the PNS
already starts to decrease because of the continued deleptonization.
This moment determines the beginning of a cooling phase dominated
by the emission of neutrinos.
We discuss the physical conditions of the quasi-static PNS evolution
and take the effects of deleptonization and mass accretion from early
fallback into account.
}


\maketitle

\section{Introduction}

Stars more massive than \(8\) M\(_\odot\) run into gravitational collapse
at the end of their evolution, due to pressure loss via the
photodisintegration of heavy nuclei and electron captures.
The collapse halts at nuclear density, typically \(2-4\times10^{14}\)
g/cm\(^3\) depending on the equation of state (EoS).
However, the supersonically infalling material from the outer
core continues to fall into the center.
The core overshoots its equilibrium configuration and bounces back.
A dynamic shock wave forms, which propagates outwards and
continuously loses energy owing to the dissociation of heavy nuclei.
As soon as the shock reaches the neutrinospheres, i.e. the neutrino
energy and flavor dependent spheres of last scattering, additional
electron captures emit a large amount of electron neutrinos.
This burst of electron neutrinos, known as the deleptonization burst,
carries away energy of several \(10^{53}\) erg/s on a timescale of
\(10-20\) ms.
This energy loss turns the expanding shock into a standing
accretion shock (SAS) already about \(5\) ms after bounce.
Due to the high energy in the neutrino radiation field,
neutrino heating between the neutrinospheres and the SAS
has long been investigated as a possible source of reviving the SAS
and for triggering neutrino-driven explosions
(\citet{BetheWilson:1985}, \citet{Janka:2001},
\citet{Janka:etal:2005}, \citet{Mezzacappa:etal:2006}).

Up to now, neutrino-driven explosions in spherical symmetry have
only been obtained for the low-mass \(8.8\) M\(_\odot\) O-Ne-Mg-core
by \citet{Kitaura:etal:2006} and for low- and intermediate-mass Fe-core
progenitor stars by \citet{Sagert:etal:2009} assuming a hadron-quark
phase transition during the early post-bounce phase.
On the other hand, multi-dimensional core collapse models
with spectral neutrino transport have only recently become available.
They demonstrate the complexity of the underlying physical phenomena
such as rotation and the development of fluid instabilities.
Such models have been shown to increase the neutrino heating efficiency
(see for example
\citet{Miller:etal:1993},
\citet{Herant:etal:1994},
\citet{Burrows:etal:1995},
\citet{JankaMueller:1996})
and help to understand aspherical explosions
(see for example
\citet{Bruenn:etal:2006} and \citet{MarekJanka:2009}).
For a review of axially-symmetric neutrino-driven explosions,
see also \citet{Janka:etal:2008a}.

The following dynamical evolution of the PNS and hence the properties
of the neutrino spectra emitted is determined by the mass accretion
and the EoS.
On a timescale of several seconds after the explosion has been launched,
the region between the expanding explosion shock and the PNS at the
center is subject to the formation of the neutrino-driven wind as follows.
Neutrinos continuously diffuse out of the hot PNS and heat the material
on top of the PNS surface before they reach the neutrinospheres.
We define the PNS surface to be the radius of the
energy-integrated electron-neutrinosphere.
The dominant neutrino heating contributions are given by the captures
of electron-neutrinos and electron-antineutrinos at free nucleons.
Matter is heated by neutrinos and the thermal energy is converted
into kinetic energy, which accelerates material on top of the PNS
surface to positive velocities.
This matter outflow is known as the neutrino-driven wind.

In this context, two particular studies are of special importance.
The properties of the neutrino-driven wind as described in
\citet{Woosley:etal:1994} are based on the detailed radiation
hydrodynamics simulation of a \(20\) M\(_\odot\) Fe-core progenitor
applying the numerical model from \citet{WilsonMayle:1993}.
The simulation was carried until about \(18\) seconds after bounce.
Another state-of-the-art model of that time was the explosion
of the O-Ne-Mg-core by \citet{MayleWilson:1988}.
Both investigations were milestones in the research of
core collapse supernovae and are based on detailed neutrino input
physics including neutrino transport, developed by J.~R. Wilson.
The results obtained, in particular the properties of the ejecta
and the neutrino observables such as luminosities and energies,
were considered the standard reference for more than \(10\) years.
Neutrino-driven wind studies used the results as parameters,
where the conditions found indicated the possible site for
the production of heavy elements via the \(r\)-process.
In the simulations discussed in the present paper, we follow a similar
approach as \citet{Woosley:etal:1994} and \citet{MayleWilson:1988},
where we apply the neutrino input physics based on \citet{Bruenn:1985}.
Although the previous work is in qualitative agreement with our findings,
in particular the explosion phase, significant differences occur
in several properties of the neutrino-driven wind.
The entropies per baryon are lower by a factor of 2-3 and
the wind stays proton-rich for more than \(10\) seconds for all our models.
In addition, the neutrino luminosities and mean energies are generally
lower.
The mean neutrino energies decrease with time, where they
remain almost constant in the simulation of \citet{Woosley:etal:1994}.
The largest difference arises in the decreasing difference
between the mean electron neutrino and antineutrino energies found
in our simulations, i.e. the neutrino spectra become more similar with
respect to time.
The difference in the neutrino spectra in \citet{Woosley:etal:1994} remains
large and even increases with time.
Using the results from \citet{Woosley:etal:1994} as reference,
\citet{QianWoosley:1996} analyzed the neutrino-driven wind and formulated
approximate analytical expressions for various properties of the neutrino-driven
wind, e.g. the neutrino heating rate, the electron fraction, the entropy per baryon
and the mass outflow rate.

Based on the static wind equations, the results obtained in parameter studies
(see e.g.
\citet{Duncan:etal:1986},
\citet{Hoffman:etal:1997a},
\citet{Thompson:etal:2001}
and
\citet{ThompsonBurrows:2001})
became known as static steady-state wind models, where
\citet{WoosleyBaron:1992},
\citet{Woosley:etal:1994},
\citet{Takahashi:etal:1994}
and
\citet{Witti:etal:1994}
described the neutrino-driven wind in a radiation-hydrodynamics context.
Of special importance for the neutrino-driven wind investigations
is the impact to the nucleosynthesis.
Most interesting is the possibility to explain the production
of heavy elements via the \(r\)-process due to the high entropies per baryon,
the fast expansion timescales and the low electron fraction of
\(Y_e < 0.5\) in the wind.
\citet{Otsuki:etal:2000}
explored general relativistic effects of the neutrino-driven wind
and investigated the possible impact to the nucleosynthesis.
Recently,
\citet{Wanajo:2006a} and
\citet{Wanajo:2006b}
investigated the neutrino-driven wind with respect to
the \(r\)- and \(rp\)-processes.

The possibility of supersonic wind velocities has been explored
in most of the references.
The supersonically expanding material in the wind collides
with the much slower expanding and denser explosion ejecta.
The material decelerates and a reverse shock forms which is
known as the neutrino-driven wind termination shock
(first observed by \citet{JankaMueller:1995} and
\citet{Burrows:etal:1995}).
Recently, \citet{Arcones:etal:2007} examined the post-bounce phase
of core collapse supernovae of several massive progenitor stars.
Their models were launched from massive progenitor stars
that were previously evolved through the core collapse,
bounce and early post-bounce phases using sophisticated
radiation hydrodynamics based on spectral neutrino transport
in spherical symmetry.
The simulations were then continued applying a simplified
radiation hydrodynamics description (see \citet{Scheck:etal:2006}),
assuming high luminosities to trigger neutrino-driven
explosions in spherical symmetry.
The neutrino-driven wind develops supersonic outflow
and the wind termination shock appears in all their models.
Like most of the present neutrino-driven wind studies,
an interior boundary was assumed instead of simulating
the PNS interior for the PNS contraction and the
diffusion of neutrinos out of the PNS.
However, steady-state wind studies could not predict the important
dynamical features from the presence of the neutrino-driven wind
termination shock, especially the deceleration of the wind material
and the consequent entropy as well as density and temperature
increase during the deceleration.
In this respect, the investigation from \citet{Arcones:etal:2007}
was a milestone in modeling the neutrino-driven wind consistently.
On the other hand, they were focusing on parameters (luminosities
and mean neutrino energies) in agreement with the simulations
of \citet{BetheWilson:1985} and \citet{Woosley:etal:1994}.
Although the dynamics is in general agreement, several properties
of the neutrino-driven wind as well as the neutrino spectra emitted
differ significantly from our findings.

The present paper follows a different approach.
We simulate consistently the dynamical evolution
of the collapse, bounce and post-bounce phases
until the neutrino-driven wind phase for more than \(20\) seconds.
The simulations are launched from the \(8.8\) M\(_\odot\)
O-Ne-Mg-core from Nomoto~(1983,1984,1987)
and the \(10.8\) and \(18\) M\(_\odot\) Fe-core progenitors
from \citet{Woosley:etal:2002}.
Our numerical model is based on general relativistic
radiation hydrodynamics with spectral three-flavor Boltzmann
neutrino transport in spherical symmetry.
The explosion mechanism of massive Fe-core progenitors is
an active subject of research.
To model neutrino-driven explosions for such progenitors in
spherical symmetry, we enhance the electronic charged current
reaction rates artificially which increases the neutrino
energy deposition and revives the SAS.
The mechanism including the tuned neutrino reaction rates
will be further discussed in \S 2 and \S 3.
Such explosion models were investigated with respect to
the nucleosynthesis by Fr\"ohlich~et~al.~(2006a-c).
Here, we report on the formation of the neutrino-driven wind
and the possibility of the wind developing supersonic
velocities and hence the formation of the wind termination shock.
In addition, we will also illustrate the explosion and the
neutrino-driven wind for the O-Ne-Mg core, where the explosion is
obtained in spherical symmetry applying the standard neutrino opacities.
The results are in qualitative agreement with those of
\citet{MayleWilson:1988} and \citet{Kitaura:etal:2006},
who used a different EoS.

We find that for low progenitor masses, the neutrino-driven wind
termination shock will develop, using the tuned neutrino reaction rates.
When the neutrino reaction rates are switched back
to the standard opacities given in \citet{Bruenn:1985},
the neutrino-driven wind develops only a subsonic matter outflow.
For intermediate progenitor masses, the neutrino-driven wind
remains subsonic even with the artificially enhanced
neutrino emission and absorption rates.
Since the neutrino-driven wind depends sensitively
on the emitted neutrino spectra at the neutrinospheres,
we believe accurate neutrino transport and general relativity
in the presence of strong gravitational fields
are essential in order to describe the dynamical evolution.
Furthermore, the accurate modeling of the electron fraction
in the wind is essential for nucleosynthesis calculations,
which can only be obtained solving the neutrino transport equation.
In addition, it is beyond the present computational
capabilities to carry multi-dimensional simulations with
neutrino transport to several seconds after bounce.
Hence, our investigations are performed in spherical symmetry
where we simulate the entire PNS interior rather than
approximating an interior boundary.
We find significant discrepancies in comparison
with the assumptions made in previous wind studies.
Material is found to be proton-rich for more than \(10\) seconds,
where most wind models assume luminosities and mean neutrino
energies such that the neutrino-driven wind becomes neutron-rich.
We question the validity of the approximations made in such wind studies.
We believe that the accurate and consistent modeling of the physical
conditions in the neutrino-driven wind is essential,
especially in order to be able to draw conclusions with respect
to the nucleosynthesis.

The paper is organized as follows.
In \S 2, we will present our spherically symmetric core collapse model.
\S 3 is devoted to the explosion phase of neutrino-driven explosions
in spherical symmetry.
We examine the \(8.8\) M\(_\odot\) progenitor
model from Nomoto~(1983,1984,1987)
using the standard neutrino opacities
and the \(10.8\) and \(18\) M\(_\odot\)
progenitor models from \citet{Woosley:etal:2002}
using artificially enhanced neutrino reaction rates.
In \S 4 we discuss the conditions for the formation of the
neutrino-driven wind and the possibility for the wind to develop
supersonic velocities.
We discuss in \S 5 the electron fraction approximation
used in the literature.
Since a generally neutron-rich neutrino-driven wind is found in many
previous and present wind studies, we illustrate the differences and
investigate why we find a generally proton-rich wind.
\S 6 is dedicated to the long term post-bounce evolution
for more than \(20\) seconds.
In \S 7 we discuss the results and emphasize the main differences
of the present investigation to previous wind studies.
Finally we close with a summary in \S 8.
%

\section{The model}

Our core collapse model, AGILE-BOLTZTRAN, is based on general
relativistic radiation hydrodynamics in spherical symmetry,
using three-flavor (anti)neutrino Boltzmann transport.
For details see Mezzacappa and Bruenn (1993a-c),
\citet{MezzacappaMesser:1999}, Liebend\"orfer et al. (2001a,b)
and \citet{Liebendoerfer:etal:2004}.
For this study we include the neutrino input physics based on
\citet{Bruenn:1985}.
The charged current reactions considered
\begin{equation}
e^{-}  + p \rightleftharpoons n + \nu_e,
\label{eq-charged-current1}
\end{equation}
\begin{equation}
e^{+}  + n \rightleftharpoons p + \overline{\nu}_e,
\label{eq-charged-current2}
\end{equation}
\begin{equation}
e^{-}  + \left<A,Z\right> \rightleftharpoons \left<A,Z-1\right> + \nu_e,
\label{eq-charged-current3}
\end{equation}
are electron and positron captures at free nucleons
as well as electron captures at nuclei. The nuclei are characterized
by an average atomic mass and charge \(\left<A,Z\right>\).
In addition, the standard scattering reactions considered
are iso-energetic neutrino nucleon (\(N\in\{n,p\}\)) and nuclei
(\(N=\left<A,Z\right>\)) scattering,
\begin{equation*}
\nu + N \rightleftharpoons \nu + N,
\end{equation*}
where \(\nu\in\{\nu_e,\nu_{\mu/\tau}\}\) (equivalent for antineutrinos \(\bar{\nu}\)),
and neutrino electron/positron scattering
\begin{equation*}
\nu + e^\pm \rightleftharpoons \nu + e^\pm.
\end{equation*}
The classical neutrino pair process is electron-positron annihilation,
\begin{equation*}
e^- + e^- \rightleftharpoons \nu + \bar{\nu}.
\end{equation*}
The standard neutrino energy \(E\) dependent emissivity \(j(E)\)
and absorptivity \(\chi(E)\) for the charged current reactions
as well as the scattering and pair-reaction rates are given in
\citet{Bruenn:1985} based on \citet{YuehBuchler:1976}
and \citet{SchinderShapiro:1982}.
The additional pair-process nucleon-nucleon-Bremsstrahlung,
\begin{equation*}
N + N \rightleftharpoons N + N + \nu + \bar{\nu},
\end{equation*}
has been implemented into our model according to
\citet{ThompsonBurrows:2001} and is also taken into account.
The emission of (\(\mu/\tau\))-neutrino pairs
via the annihilation of trapped electron-neutrino pairs,
\begin{equation*}
\nu_e + \bar{\nu}_e \rightleftharpoons
\nu_{\mu/\tau} + \bar{\nu}_{\nu/\tau},
\end{equation*}
as well as contributions from nucleon-recoil and weak magnetism
as studied in \citet{Horowitz:2002} are investigated in
\citet{Fischer:etal:2009} and are not taken into account in the
present study of the neutrino-driven wind.

\subsection{Recent improvements of the adaptive grid}

Long-term simulations of the supernova post-bounce phase with
AGILE-BOLTZTRAN lead to a very large contrast of densities,
reaching from $\sim10^{15}$ g/cm$^{3}$ at the center of the
protoneutron star (PNS) to densities on the order of g/cm$^{3}$ and
lower in the outer layers.
The version of AGILE described in \citet{Liebendoerfer:etal:2002} is
not able to resolve such large density contrasts.
If the enclosed mass $a$ is large and the density in one zone very low,
then the evaluation of the mass contained in the zone according to Eq.~(39)
in \citet{Liebendoerfer:etal:2002},
\begin{equation*}
da_{i+\frac{1}{2}}=a_{i+1}-a_{i},
\end{equation*}
is subject to large cancellation so that truncation errors inhibit
the convergence of the Newton-Raphson scheme in the implicitly finite
differenced time step. However, the problem can be avoided by a simple
modification that was first explored in Fisker (2004, priv. comm.).
The state vector of AGILE-BOLTZTRAN is given by the following
set of quantities
\begin{equation}
y = \left(a\,,r\,,u\,,m\,,\rho\,,T\,,Y_{e}\right),
\end{equation}
with enclosed baron mass \(a\),
radius \(r\),
velocity \(u\),
gravitational mass \(m\),
baryon density \(\rho\),
temperature \(T\),
electron fraction \(Y_e\).
In the improved version, the state vector at time \(t^{n}\) is based
on zone masses, \(da_{i+\frac{1}{2}}^{n}\), where the enclosed mass
\begin{equation*}
a_{i}^{n}=\sum_{1}^{i-1}da_{i+\frac{1}{2}}^{n}
\end{equation*}
becomes the derived quantity.

The form of the generic equation (30) in \citet{Liebendoerfer:etal:2002}
applies to the continuity equation,
the momentum equation and the energy equation. If we define
$\delta_{i}=a_{i}^{n+1}-a_{i}^{n}$ as the difference of the enclosed
mass $a_{i}$ between time $t^{n}$and $t^{n+1}$,
Eq. (30) in \citet{Liebendoerfer:etal:2002} becomes
\begin{equation}
\frac{y_{i+\frac{1}{2}}^{n+1}\left(da_{i+\frac{1}{2}}^{n}
+ \delta_{i+1}-\delta_{i}\right)-y_{i+\frac{1}{2}}^{n}da_{i+\frac{1}{2}}^{n}}{dt}
= y_{i+1}^{\textrm{adv}}-y_{i}^{\textrm{adv}}-y_{i+\frac{1}{2}}^{\textrm{ext}}
= 0,
\label{eq:generic.equation}
\end{equation}
where the relative velocity between the fluid and the grid
in the advection term \(y^{\textrm{adv}}\)
is defined by
\begin{equation}
u_{i}^{\textrm{rel}} =
-\frac{a_{i}^{n+1}-a_{i}^{n}}{dt} =
-\frac{\delta_{i}}{dt}.
\label{eq:relativ.velocity}
\end{equation}
The cancellation of large numbers during the Newton-Raphson iterations
of the implicit time step is avoided if the time shifts $\delta_{i}$
are chosen as the unknowns in the state vector when Eqs. (\ref{eq:generic.equation})
and (\ref{eq:relativ.velocity}) are solved. The vector of zone masses
is then updated between the implicit time steps by
\[da_{i+\frac{1}{2}}^{n+1}=
da_{i+\frac{1}{2}}^{n}+\delta_{i+1}-\delta_{i}.\]
This leads to satisfactory convergence of the Newton-Raphson iterations
even in the presence of large density contrasts.

\subsection{The equation of state}

For the present investigation of the neutrino-driven wind,
the baryon EoS from \citet{Shen:etal:1998} for hot and dense nuclear matter
has been implemented for matter in nuclear statistical equilibrium (NSE).
For temperatures below \(T=0.5\) MeV where NSE does not apply, the baryon
EoS combines an ideal gas approximation for a distribution of nuclei,
based on \citet{TimmesArnett:1999} (including ion-ion-correlations),
and a nuclear reaction network using the composition given by the
progenitor model.
Details of the reaction network are given in
\citet{Thielemann:etal:2004} and references therein.
The nuclear abundances are included in the state vector of
AGILE-BOLTZTRAN, which reads as follows
\begin{equation}
y=\left(a\,,r\,,u\,,m\,,\rho\,,T\,,Y_{e}\,,Y_1\,, .\, .\, .\,,Y_N \right).
\end{equation}
For all these quantities, including the nuclear abundances
\(Y_1, ..., Y_N\),
the corresponding advection equations are solved as
described in \citet{Liebendoerfer:etal:2002} \S 3,
but with an improved second order accurate total variation
diminishing advection scheme.
The nuclear reaction network is used in an operator-split manner
in order to calculate the abundance changes due to the source
terms which in turn depend on employed reaction rates.

Due to computational limitations, we restrict ourselves to \(N=19\).
We consider the free nucleons and the \(14\) symmetric nuclei,
from \(^{4}\)He to \(^{56}\)Ni.
In order to model matter with \(Y_e\neq0.5\) to some extent,
we additionally include \(^{53}\)Fe, \(^{54}\)Fe and \(^{56}\)Fe.
The network calculates the composition dynamically
from the progenitor stage until the simulations are stopped.
It is used for an accurate internal energy evolution.
In addition, we can approximately reflect the composition
of the PNS surface for more than \(20\) seconds after bounce,
where nuclei that have been previously in NSE are freezing
out of NSE as the temperature drops rapidly below \(0.5\) MeV
already about \(1\) second after bounce and reach below \(0.01\) MeV
at about \(10\) seconds post-bounce.
In previous studies the simplification of an ideal gas of
Si-nuclei was used for matter which is not in NSE.
This leads to an increasing inaccurate internal energy evolution after
\(500\) ms post-bounce when the explosion shock reaches the Si-layer
of the progenitor and simplifications could not be extended beyond
\(1\) second post-bounce time.
The implementation of the nuclear reaction network now makes it possible
to include more mass (up to and including a large fraction of the He-layer,
depending on the progenitor model)
into the physical domain and follow the dynamical evolution
of the explosion by one order of magnitude longer.

The baryon EoSs for NSE and for non-NSE are coupled to an
electron-positron EoS (including photons),
developed by \citet{TimmesArnett:1999}.

\subsection{Enhanced neutrino emissivity and opacity}

By our choice of a spherically symmetric approach, we implement
the explosion mechanism of massive Fe-core progenitor stars
artificially to trigger neutrino-driven explosions
during the post-bounce evolution after the deleptonization burst
has been launched.
Neutrino heating between the neutrinospheres and the SAS transfers
energy from the radiation field into the fluid.
A part of this energy is converted into thermal energy
which revives the SAS and launches the explosion.
The revival of the SAS and hence the neutrino-driven
explosions take place on a timescale of several \(100\) ms.

During the post-bounce evolution, heavy nuclei continue to fall
onto the SAS and dissociate into free nucleons.
These free nucleons accrete onto the PNS surface.
Hence the dominant neutrino heating contributions behind the SAS
are due to the electronic charged current reactions,
expressions~(\ref{eq-charged-current1}) and (\ref{eq-charged-current2}).
To trigger explosions in spherically symmetric core collapse
simulations of massive Fe-core progenitors, we enhance the
emissivity \(j\) and absorptivity \(\chi\) by a certain factor
(typically \(5-7\)) in the region between the SAS and the neutrinospheres.
This corresponds to matter with entropies above \(6\)
k\(_\text{B}\)/baryon and baryon densities below
\(10^{10}\) g/cm\(^{3}\).
The entropies ahead of the shock are lower and the central densities
of the PNS are higher, such that the artificial heating only
applies to the region between the neutrinospheres and the SAS.
The artificially enhanced reaction rates do not change the neutrino
luminosities and mean neutrino energies significantly
for the electron-neutrinos and electron-antineutrinos.
Furthermore, \(\beta\)-equilibrium is fulfilled since the
reverse reaction rates are obtained via the detailed balance.
However, the timescale for weak-equilibrium to be established is
reduced and hence the electron fraction changes on a shorter timescale.
In combination with the increased neutrino energy deposition,
this leads to a deviation of the thermodynamic variables
in comparison to simulations using the standard opacities
given in \citet{Bruenn:1985},
which will be further discussed in \S 7.
The weak neutrino-driven explosions obtained have explosion energies
of \(6.5 \times 10^{50}\) erg and \(2 \times 10^{50}\) erg for
the \(10.8\) and the \(18\) M\(_\odot\) progenitor model respectively.

\subsection{Explosion energy and mass cut}

The explosion energy estimate is a quantity which changes
during the dynamical evolution of the system.
It is given by the total specific energy of the fluid
in the laboratory frame
\begin{equation}
E_\text{specific}(t,a) = \Gamma e 
  + \frac{2}{\Gamma + 1} \left (\frac{u^2}{2} - \frac{m}{r} \right ),
\end{equation}
which in turn is the sum of the specific internal energy \(e\)
\footnote{The baryon contribution to the internal energy is
composed of a thermal and nuclear part, i.e.
\(e=e_\text{thermal}+e_\text{nuclear}\).
In NSE,  \(e\) is given implicitly via the EoS of hot and dense nuclear matter.
In non-NSE, \(e_\text{nuclear}\) is the binding energy
of the nuclear composition used in the reaction network.},
the specific kinetic energy given by the fluid velocity
\(u=\partial r/\alpha\partial t\) squared
and the specific gravitational binding energy \(m/r\)
with gravitational mass \(m\) and radius \(r\)
(see \citet{Liebendoerfer:etal:2001b}).
\(\Gamma(t,a) = \sqrt{1-2m/r+u^2}\) and \(\alpha(t,a)\)
are the metric functions in a non-stationary
and spherically symmetric spacetime with
coordinate time \(t\), baryon mass \(a\) and the two angular coordinates
(\(\theta,\phi\)) describing a 2-sphere of radius \(r(t,a)\)
(see \citet{MisnerSharp:1964}).
The explosion is determined by the energy of the ejecta.
Integrating \(E_\text{specific}(t,a)\) with respect to the enclosed
baryon mass starting from the progenitor surface \(\text{M}\)
toward the center
\begin{equation}
E_\text{total}(t_0,a_0) = -\int_{\text{M}}^{a_0} E_\text{specific}(t_0,a)\,da ,
\label{eq-Etot}
\end{equation}
gives the total mass-integrated energy, at a given time \(t_0\)
outside a given mass \(a_0\).
The expression (\ref{eq-Etot}) is negative during the collapse,
bounce and the early post-bounce phases
because the progenitor and central Fe-core
are gravitationally bound.
At some time after bounce, expression (\ref{eq-Etot}) becomes positive
in the region between the shock and the neutrinospheres.
It stays negative at large distances and close to the deep gravitational
potential of the PNS, because the outer layers of the progenitor and the
PNS continue to be gravitationally bound.
While the emission of neutrinos cools the PNS interior, neutrino
absorption deposits energy on the timescale \(\tau_\text{heating}\)
on the order of \(100\) ms into the fluid near the neutrinospheres.
This increases the specific internal energy
which matches at later (\(\sim 500\) ms) post-bounce times the
gravitational binding energy at a certain distance toward the center.

On a suggestion by S. Bruenn, we consider the {\em mass cut} as follows
\begin{equation}
a_\text{cut} = a
\left(
\text{max}\left(E_\text{total}(t\gg\tau_\text{heating},a)\right)
\right).
\end{equation}
The material outside of \(a_\text{cut}\) is gravitationally unbound
and will be ejected while the enclosed material will accrete onto
the central PNS.
The explosion energy estimate is defined as the total mass-integrated
energy of the layers outside the mass cut
\begin{equation}
E_\text{expl} = E_\text{total}(t\gg\tau_\text{heating},a_\text{cut}),
\label{eq-Eexpl}
\end{equation}
at late times (\(t\gg\tau_\text{heating}\)) after the explosion
has been launched.
It becomes clear from the above expressions that the explosion energy
estimate is sensitively determined by the balance of internal and
kinetic energies to gravitational binding energy.

From the time post-bounce when the shock reaches low enough densities
and temperatures such that neutrinos decouple from matter completely,
neutrino heating and cooling does not affect the explosion energy
estimate anymore.
The additional energy deposition from the neutrino-driven wind,
which will be discussed further below, might affect the explosion
estimate at later times.
We will illustrate in particular the effect of the formation
of a supersonic neutrino-driven wind and the wind
termination shock to the explosion energy estimate.
Only after the neutrino-driven wind disappears, the final value
of the explosion energy can be obtained.

\subsection{The neutrino observables}

The neutrino radiation hydrodynamics equations are a coupled system
which combines the evolution of the matter properties given by the
state vector \(y\) and the radiation field,
as documented in \citet{Liebendoerfer:etal:2004} and references therein.
The neutrino radiation field is taken into account via the phase-space
distribution function \(f_\nu(t,a,\mu,E)\) for each neutrino flavor
\(\nu=(\nu_e,\bar{\nu}_e,\nu_{\mu/\tau},\bar{\nu}_{\mu/\tau})\).
In spherical symmetry, it depends on the time \(t\), the enclosed baryon
mass \(a\) as well as on the neutrino energy \(E\) and the cosine of the
propagation angle \(\mu=\cos(\theta)\).
The evolution of the neutrino radiation field is taken into account by
solving the Boltztran transport equation for massless fermions.
It determines the phase-space derivative of the specific distribution
function \(F_\nu=f_\nu/\rho\), i.e. the distribution function divided by the
matter density \(\rho\), in a co-moving frame
(see for example Eq.(8) of \citet{Liebendoerfer:etal:2005})
and due to neutrino-matter interactions such as emission and
absorption as well as scattering and pair reactions.

In order to compare simulation results, neutrino observables can be defined.
Commonly used are the neutrino number-luminosities, which is given by the
phase-space integration of the neutrino distribution function as follows
\begin{eqnarray*}
L_n\left([t_0,t_1],a\right) =
4\pi r^2 \rho
\frac{2\pi c}{(hc)^3}
\int_{-1}^{+1} d\mu \int_{0}^{\infty} E^2\,dE\,\,
F_\nu(t,a,\mu,E),
\end{eqnarray*}
which is the number of neutrinos
\footnote{The integration with respect to \(\mu\) is performed
separately for in- and out-ward direction,
according to the transport coefficients.}
of energy \(E\) passing through the mass coordinate \(a\)
for a given time-interval \([t_0,t_1]\) taken in a co-moving frame
at position \(r(t,a)\).
Equivalently, the energy-luminosity can be defined as follows
\begin{eqnarray*}
L_e\left([t_0,t_1],a\right) &\equiv& L_\nu\left([t_0,t_1],a\right) \\
&=& 4\pi r^2 \rho
\frac{2\pi c}{(hc)^3}
\int_{-1}^{+1} \mu\,d\mu \int_{0}^{\infty} E^3\,dE\,\,
F_\nu(t,a,\mu,E),
\end{eqnarray*}
for each neutrino flavor, i.e.
\((\nu_e,\bar{\nu}_e,\nu_{\mu/\tau},\bar{\nu}_{\mu/\tau})\).
Additionally useful quantities are the mean neutrino and root-mean-squared
(rms) neutrino energies, defined as follows
\begin{eqnarray*}
\left<E_\nu (t,a)\right> &=&
\frac
{\int_{-1}^{+1} d\mu \int_{0}^{\infty} E^3\,dE\,\,F_\nu(t,a,\mu,E)}
{\int_{-1}^{+1} d\mu \int_{0}^{\infty} E^2\,dE\,\,F_\nu(t,a,\mu,E)},
\\
\left<E_\nu (t,a)\right>_\text{rms} &=&
\sqrt
{
\frac
{\int_{-1}^{+1} d\mu \int_{0}^{\infty} E^4\,dE\,\,F_\nu(t,a,\mu,E)}
{\int_{-1}^{+1} d\mu \int_{0}^{\infty} E^2\,dE\,\,F_\nu(t,a,\mu,E)}
}.
\end{eqnarray*}
We will use these observables, i.e. the energy-luminosities and the mean
and root-mean-squared energies, to illustrate the dynamical evolution of
the radiation field as well as for comparisons with previous studies.

\subsection{The electron fraction}

The proton-to-baryon ratio is essential for the composition of the
ejecta, which is obtained via detailed post processing
nucleosynthesis calculations.
In the absence of muons or tauons, the proton-to-baryon ratio
is given by the electron fraction as follows
\begin{equation}
Y_e = Y_{e^{-}} - Y_{e^{+}} = Y_p,
\end{equation}
which is equal to the number of protons and defines the total number
of charges per baryon.
The change of the electron fraction is given by the balance of
emitted and absorbed electrons (positrons) and electron-neutrinos
(antineutrinos) at free nucleons and nuclei.
Weak-equilibrium is achieved if
\begin{equation}
\mu_{e^-} + \mu_p = \mu_n + \mu_{\nu_e},
\end{equation}
\begin{equation}
\mu_{e^+} + \mu_n = \mu_p + \mu_{\bar{\nu}_e},
\end{equation}
where \(\mu_i\) are the chemical potentials for electron and positron
(\(\mu_{e^\pm}\)), proton (\(\mu_p\)), neutron (\(\mu_n\)) and
electron-neutrino (\(\mu_{\nu_e}\)) and electron-antineutrino
(\(\mu_{\bar{\nu}_e}\)).
The time-derivative of the electron fraction, \(\dot{Y}_e\),
is given by the phase-space integration of the emissivities \(j_\nu\) and
the absorptivities \(\chi_\nu\) for electron-neutrinos and
electron-antineutrinos as follows
\begin{equation*}
\dot{Y_e} =
-\frac{2 \pi}{(hc)^3}\frac {m_B\,c}{\rho}
 \int_{-1}^{+1} d\mu\int_{0}^{\infty} E^2 dE
\,\,\times
\end{equation*}
\begin{equation}
\times\,\,
\left (
\left ( j_{\nu_e} - \tilde{\chi}_{\nu_e} f_{\nu_e}
\right ) -
\left (
j_{\bar{\nu_e}} - \tilde{\chi}_{\bar{\nu_e}} f_{\bar{\nu_e}}
\right )
\right )
\label{eq-ye}
\end{equation}
where \(m_B\) is the baryon mass, \(\rho\) is the matter density
and \(\tilde{\chi} = j + \chi\).
The emissivities \(j_\nu(E)\) and absorptivities \(\chi_\nu(E)\)
depend on the neutrino energy.
They are the reaction rates for the electronic charged current
reactions, which are calculated following \citet{Bruenn:1985}
and depend on temperature and density.
Eq.~(\ref{eq-ye}) is found by combining the equation of lepton number
conservation with the phase-space integrated Boltzmann transport equation
(see Mezzacappa and Bruenn (1993a)).
To follow the dynamical evolution of the electron fraction via Eq.~(\ref{eq-ye}),
knowledge of the neutrino distribution functions is required for which
neutrino transport is necessary.
%

\section{Explosions in spherical symmetry}

Progenitor stars more massive than \(9\) M\(_\odot\) develop extended
Fe-cores at the end of stellar evolution.
The explosion mechanism of such Fe-core progenitors
is an active subject of research.
In the following section, we will investigate the neutrino-driven
explosions of the \(10.8\) and \(18\) M\(_\odot\) Fe-core progenitors
from \citet{Woosley:etal:2002} in spherical symmetry by enhancing
the electronic charged current reaction rates artificially.
Further below, we will investigate the explosion phase of the
\(8.8\) M\(_\odot\) O-Ne-Mg-core from Nomoto~(1983,1984,1987),
where the explosion is obtained using the standard neutrino opacities
as introduced in \S~2.
\begin{figure}[ht]
\begin{center}
\includegraphics[width=0.7\columnwidth]{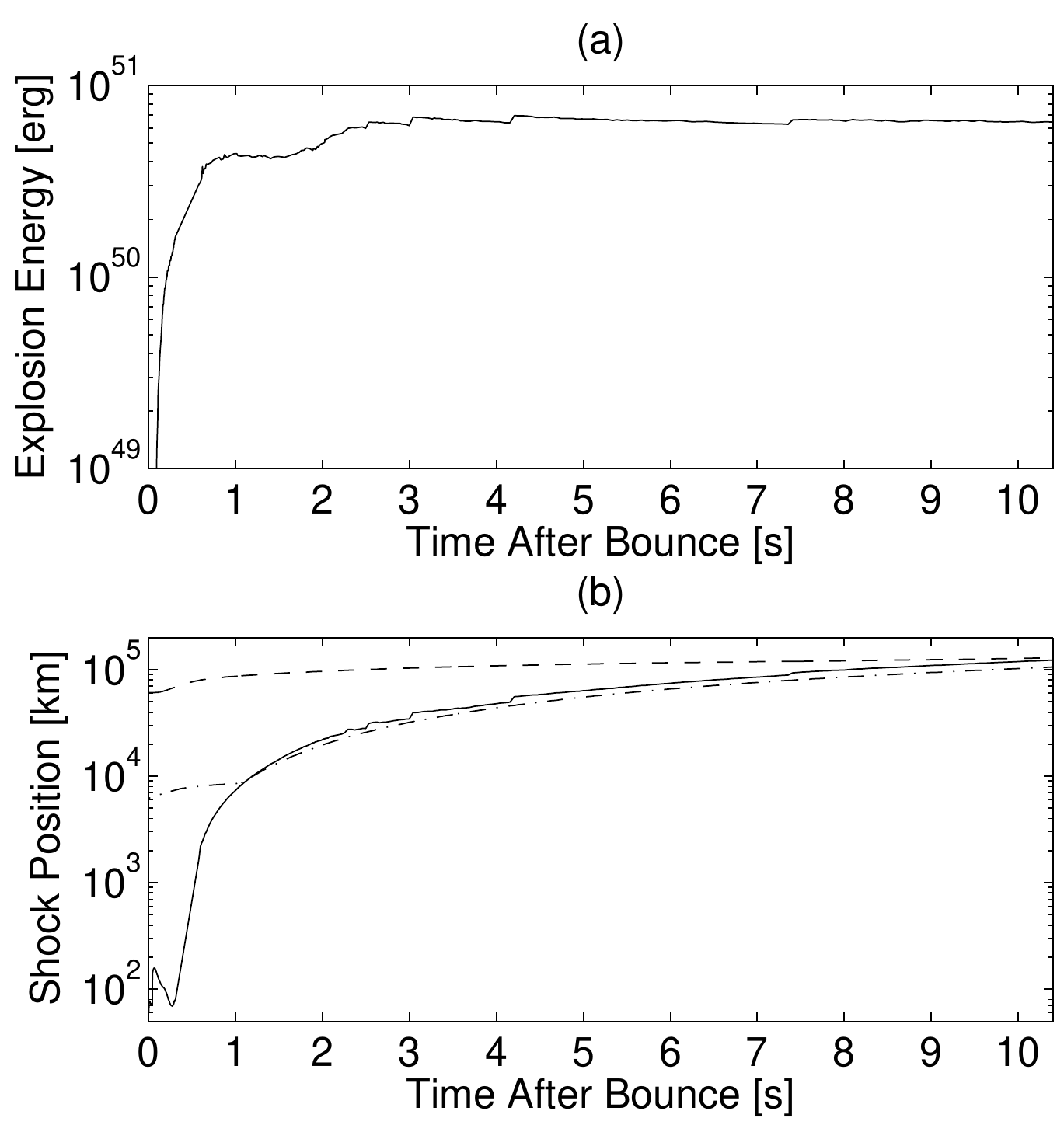}
\caption{Explosion energy estimate and shock position
with respect to time after bounce for the
\(10.8\) M\(_\odot\) progenitor model from \citet{Woosley:etal:2002}.
In addition, graph (b) illustrates the position of the
He-layer (dashed line) and the O-layer (dash-dotted line).}
\label{fig-explosion-h10a}
\end{center}
\end{figure}

\subsection{Neutrino-driven explosions of Fe-core progenitors}

The dynamical behavior of the explosion energy estimate and the shock position
are illustrated in Figs.~\ref{fig-explosion-h10a}~(a) and (b) respectively
with respect to time after bounce.
The figures illustrate the explosion phase and the long term evolution
up to \(10\) seconds after bounce.
After achieving a convergent value between \(600\) ms and \(2\)
seconds post-bounce of \(4.5\times10^{50}\) erg, the explosion energy
estimate is lifted slightly to about \(6.5\times10^{50}\) erg.
This effect coincides with the additional mass outflow obtained
in the neutrino-driven wind and the appearance of the reverse shock,
which will be discussed further below.
In simulations with a less intense (subsonic) neutrino-driven wind,
this effect is negligible and the explosion energy can be obtained
already after about \(1\) second post-bounce.

\begin{figure*}[htp!]
\begin{center}
\includegraphics[width=0.64\columnwidth]{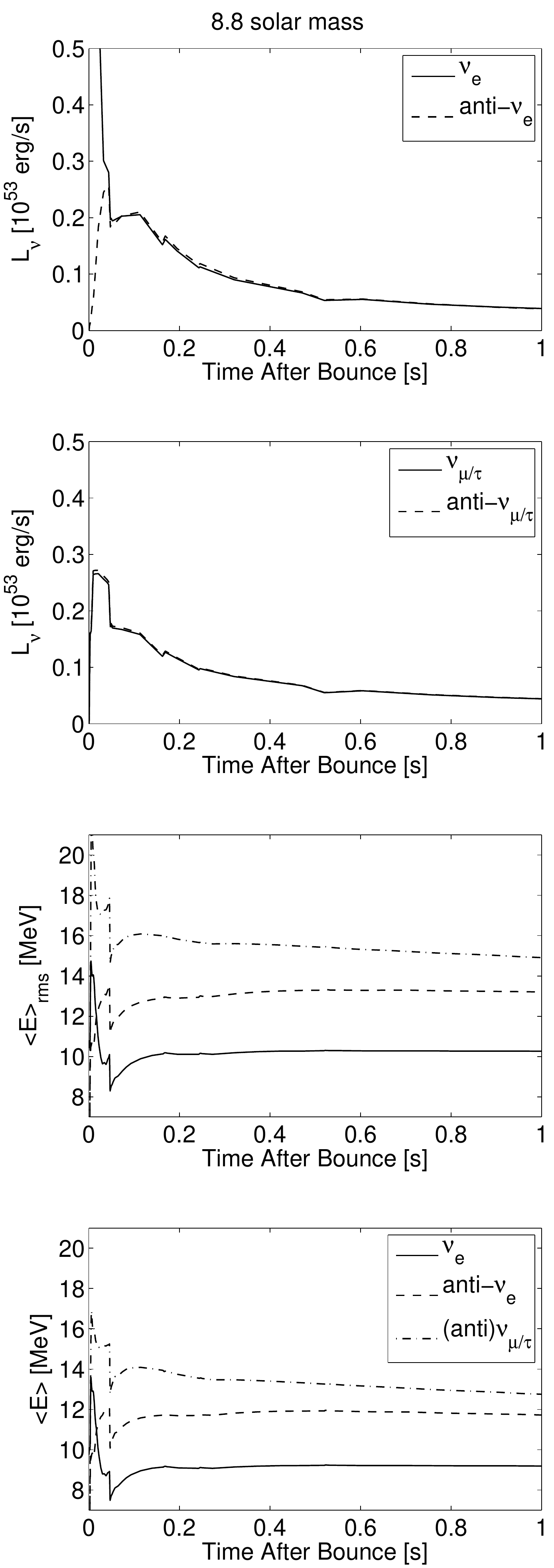}
\includegraphics[width=0.64\columnwidth]{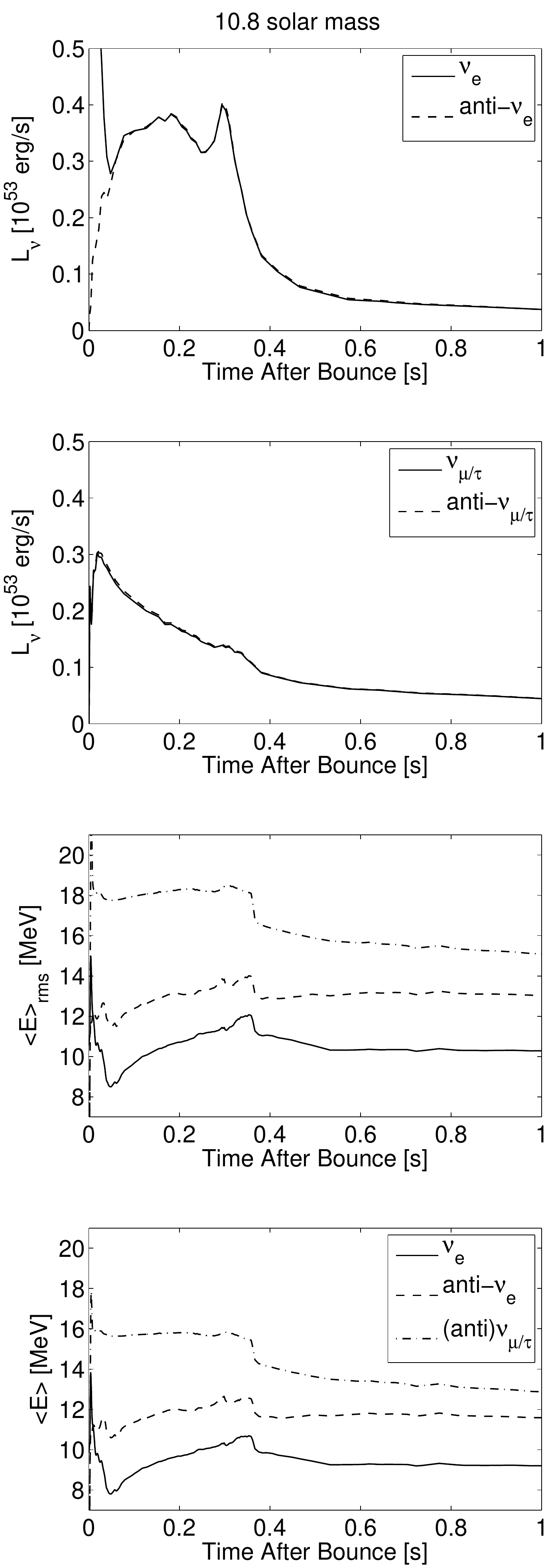}
\includegraphics[width=0.64\columnwidth]{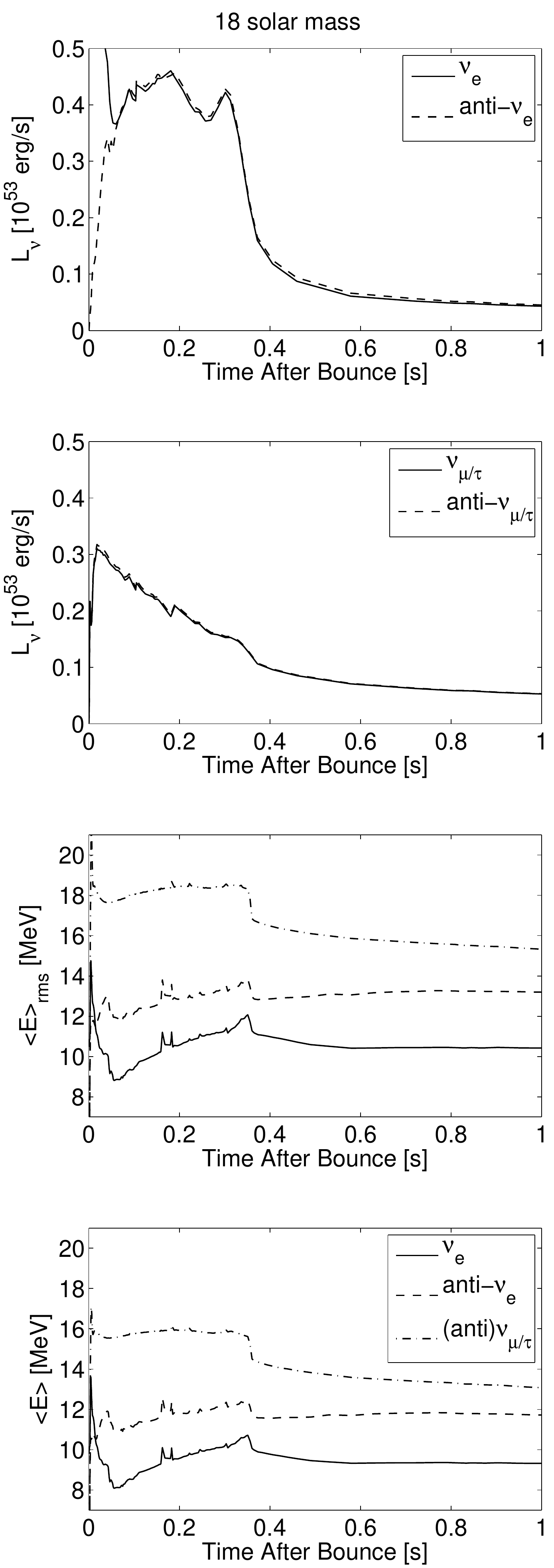}
\caption{Neutrino luminosities and energies with respect to time after
bounce for the $8.8$ M$_\odot$ O-Ne-Mg-core progenitor model from
Nomoto~(1983,1984,1987) (left panel) and the $10.8$ M$_\odot$ and
$18$ M$_\odot$ Fe-core progenitor models from \citet{Woosley:etal:2002}
(middle and right panels respectively),
measured in a co-moving frame at $500$ km distance.}
\label{fig-lumin-expl}
\end{center}
\end{figure*}

The neutrino luminosities and the mean as well as rms neutrino energies
are shown in Fig.\ref{fig-lumin-expl} for the \(10.8\) (middle panel)
and the \(18\) (right panel) M\(_\odot\) progenitor model with respect
to time after bounce.
Note that the more compact PNS of the \(18\) M\(_\odot\) progenitor
model results in generally higher neutrino luminosities.
The oscillating shock position and the consequent contracting and
expanding neutrinospheres during the neutrino heating phase of the
$10.8$ and $18$ M$_\odot$ progenitor models on a timescale of several
$100$ ms are reflected in the electron-flavor neutrino luminosities,
which correspondingly increase and decrease respectively.
During the heating phase, the mean neutrino energies of the
electron-(anti)neutrinos increase from about \(8\) (\(10\)) MeV after
bounce to about \(12\) (\(14\)) MeV until the explosion is launched.
The mean neutrino energy of the (\(\mu/\tau\))-neutrinos remains
constant at about \(18\) MeV during the heating phase.
The mean neutrino energies are generally lower than the rms-energies
but follow the same behavior.
The explosions for both models are launched after about $350$ ms
post-bounce, which defines the neutrino heating timescale for the energy
deposition in the gain region to revive the SAS.
Matter is accelerated to positive velocities and the SAS turns into
the dynamic explosion shock.
The resulting neutrino spectra from these artificially induced explosions
in spherical symmetry are in general agreement with the neutrino spectra
from axially-symmetric neutrino-driven core collapse supernova models
that explode without artificially modified reaction rates
(see \citet{MarekJanka:2009}).
The explosion shock continuously propagates through the remaining
domain of the progenitor star.
After the explosions have been launched, the electron flavor neutrino
luminosities decay exponentially.
Furthermore the jumps in the neutrino energies after \(350\) ms post-bounce
for the \(10.8\) and \(18\) M\(_\odot\) progenitor models are due to the
shock propagation over the position of \(500\) km, where the
observables are measured in a co-moving reference frame.

\subsection{The O-Ne-Mg-core}

\begin{figure*}[ht]
\begin{center}
\includegraphics[width=1.8\columnwidth]{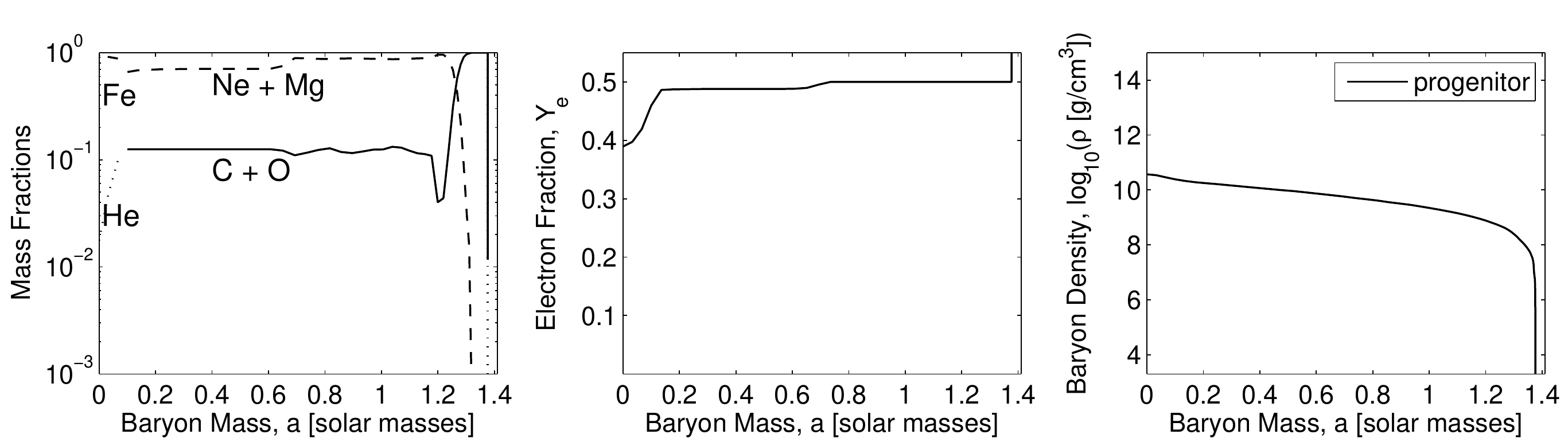}\\
\includegraphics[width=1.8\columnwidth]{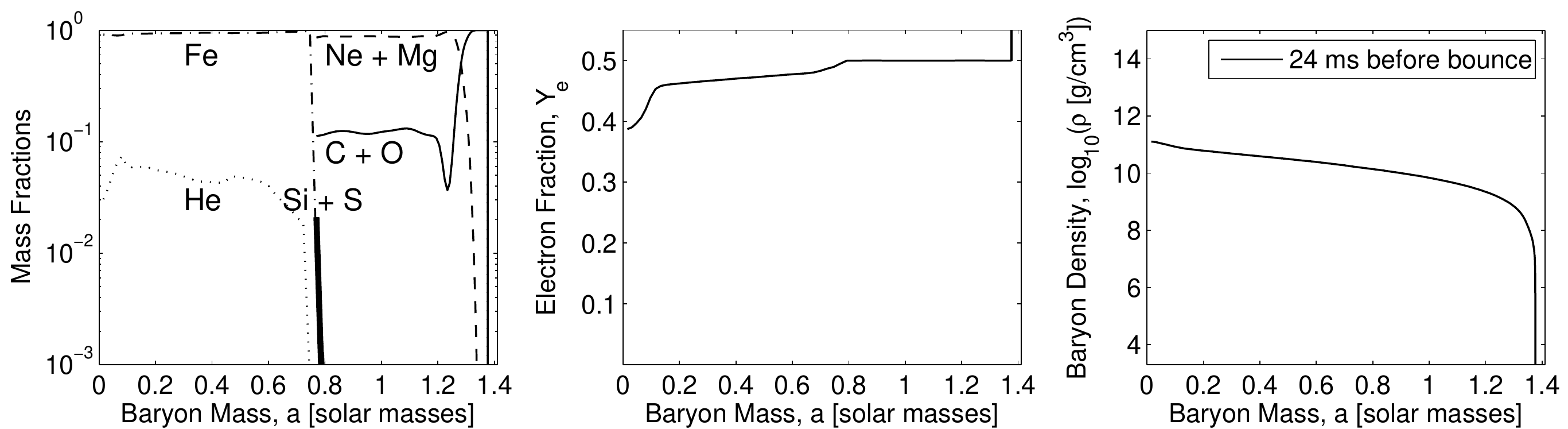}\\
\includegraphics[width=1.8\columnwidth]{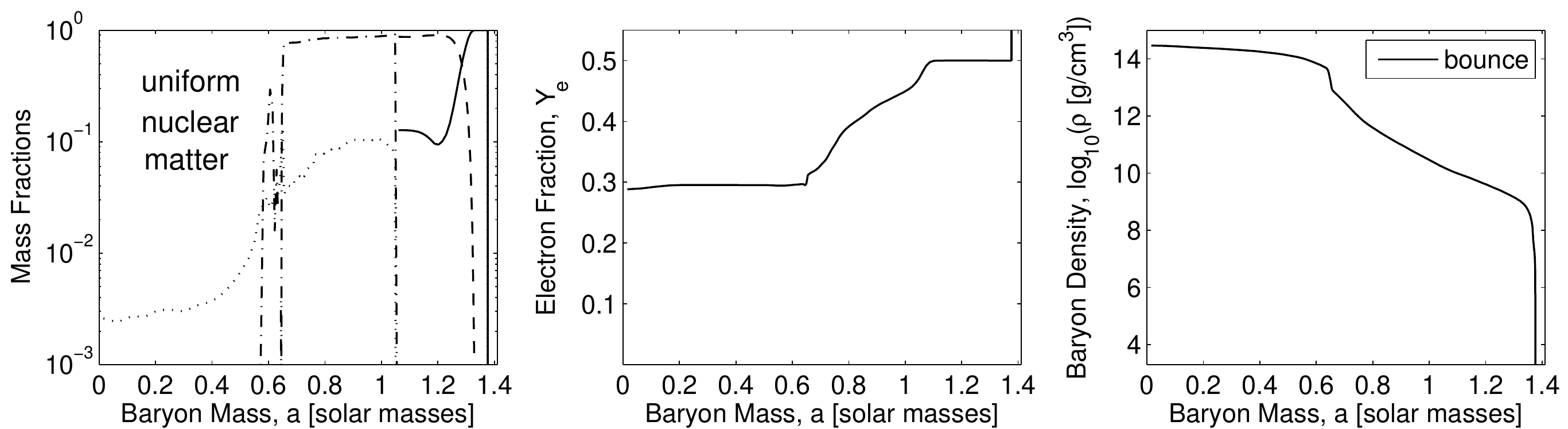}
\caption{Evolution of the $8.8$ M$_\odot$ progenitor model
from Nomoto~(1983,1984,1987) during the core collapse phase
(top: progenitor configuration, middle: $24$ ms before bounce,
bottom: at bounce).
The composition (left panels) are as follows: 
C+O (thin solid line), Ne+Mg (dashed line),
Fe+Ni (dash-dotted line), He (dotted line),
Si+S (thick solid line).}
\label{fig-composition-onemg}
\end{center}
\end{figure*}

\begin{figure*}[ht]
\centering
\subfigure[Before and at bounce.]{
\label{fig-velocity-onemg}
\includegraphics[width=0.8\columnwidth]{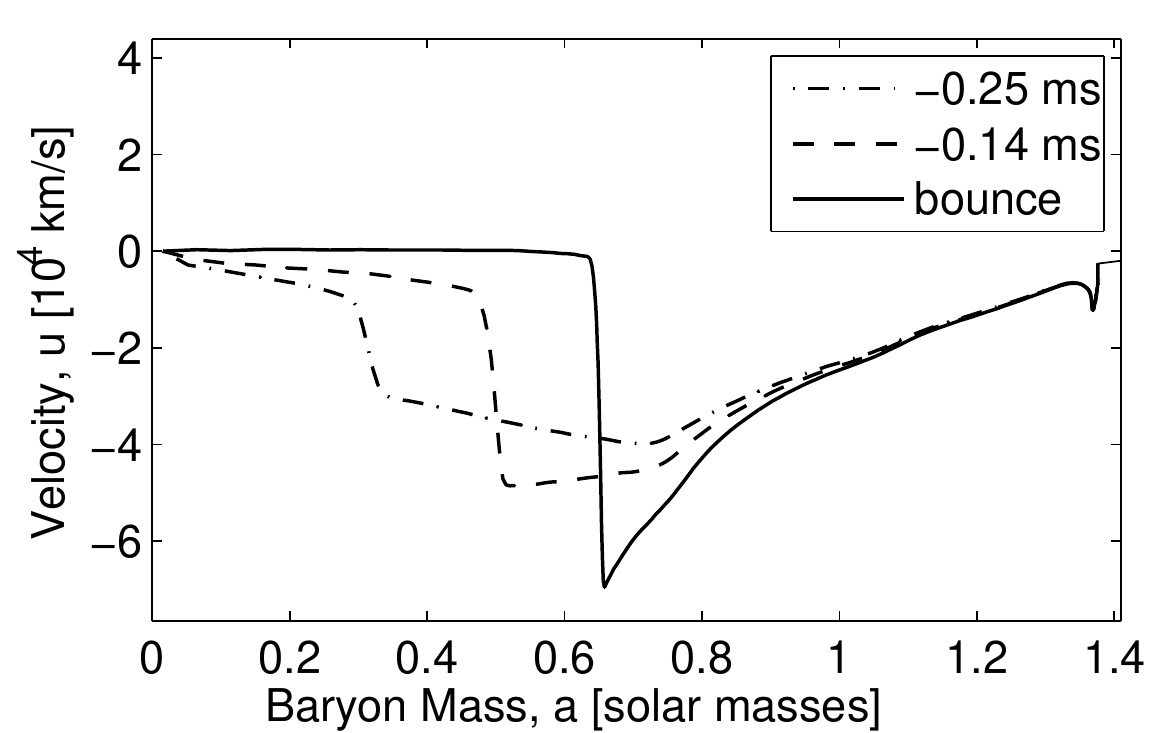}}
\hspace{5mm}
\subfigure[During the explosion.]{
\label{fig-velocity-onemg-expl}
\includegraphics[width=0.8\columnwidth]{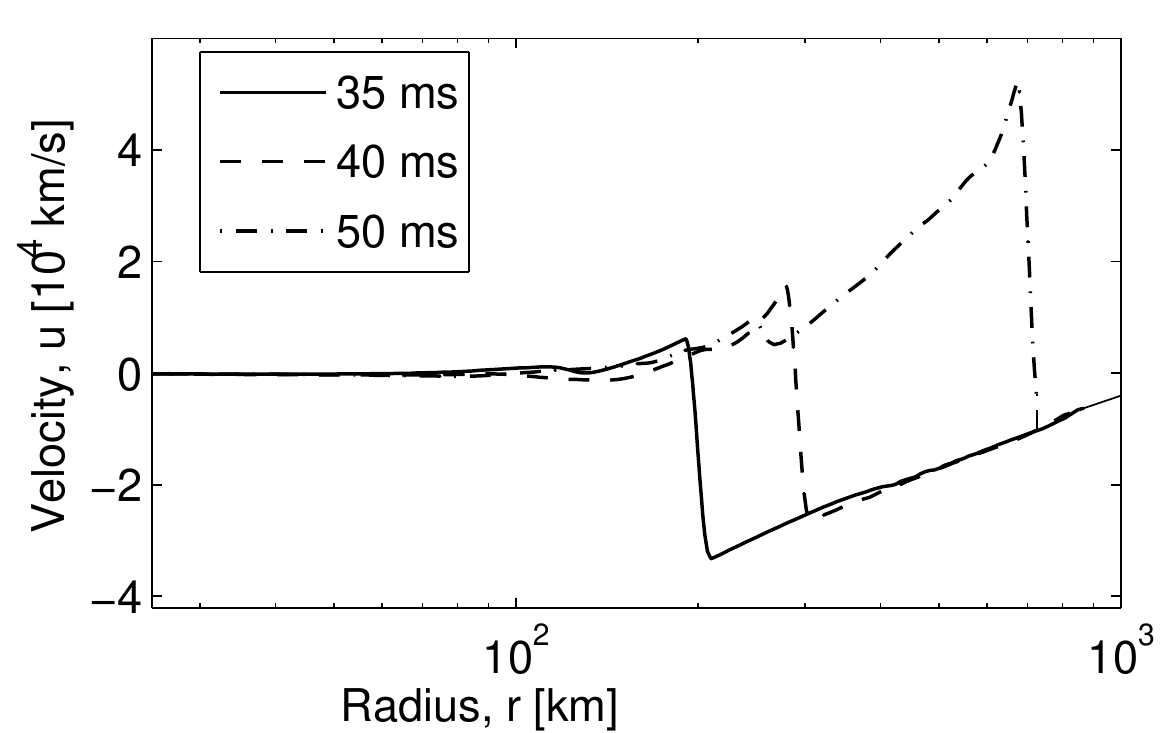}}
\caption{Radial velocity profiles with respect to the
baryon mass (a) and with respect to the radius (b) for the
\(8.8\) M\(_\odot\) progenitor model from Nomoto~(1983,1984,1987)}
\end{figure*}

A special star is the \(8.8\) M\(_\odot\) progenitor model from
Nomoto~(1983,1984,1987).
The central thermodynamic conditions at the end of stellar evolution
are such that only a tiny fraction of about \(0.15\) M\(_\odot\) of
Fe-group nuclei are produced, where nuclear statistical equilibrium (NSE)
applies (see Fig.~\ref{fig-composition-onemg} (a) top panel).
Instead, the central composition is dominated by \(^{16}\)O,
\(^{20}\)Ne and \(^{24}\)Mg nuclei.
Because temperature and density increase during the collapse, these
nuclei are burned into Fe-group nuclei and the NSE regime increases
(see Fig.~\ref{fig-composition-onemg} middle panel).
The core continues to deleptonize, which can be identified at the
decreasing \(Y_e\) in Fig.~\ref{fig-composition-onemg}.
We use our nuclear reaction network as described in \S 2.2
to calculate the dynamically changing composition,
based on the abundances provided by the progenitor model.
The size of the bouncing core of \(M_\text{core}\simeq 0.65\) M\(_\odot\)
is significantly larger in comparison with the previous studies by
\citet{Kitaura:etal:2006} and \citet{Liebendoerfer:2004},
illustrated in Fig.~\ref{fig-velocity-onemg} at different velocity
profiles before and at bounce.
This is because we do not take the improved electron capture rates from
\citet{Hix:etal:2003} and \citet{Langanke:etal:2003} into account,
which are based on the capture of electrons at the distribution of heavy nuclei.
It results in a lower central electron fraction at bounce and a
consequently more compact bouncing core of about \(\simeq 0.1\) M\(_\odot\),
in comparison to the standard rates given in \citet{Bruenn:1985}.
The remaining difference is most likely due to the large asymmetry energy
of the EoS from \citet{Shen:etal:1998} applied to the present study.

This progenitor is not only a special case due to the incomplete
nuclear burning at the end of stellar evolution but also due to the
steep density gradient which separates the dense core from the He- and
H-rich envelope at \(1.376\) M\(_\odot\),
see Fig.~\ref{fig-composition-onemg} (c).
There, the density drops over \(13\) orders of magnitude
which makes it difficult to handle numerically.

\begin{figure*}[ht]
\centering
\includegraphics[width=0.65\columnwidth]{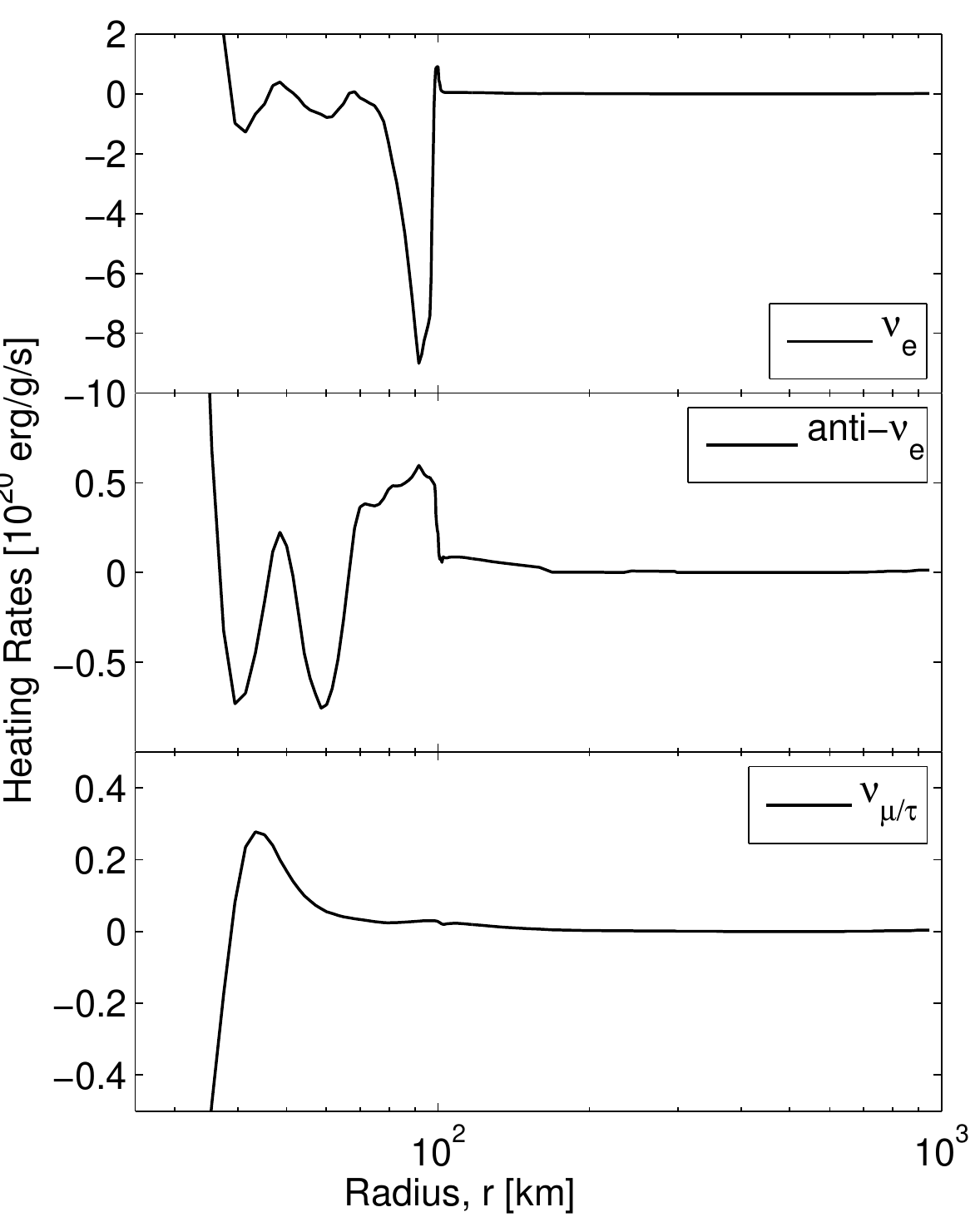}
\includegraphics[width=0.65\columnwidth]{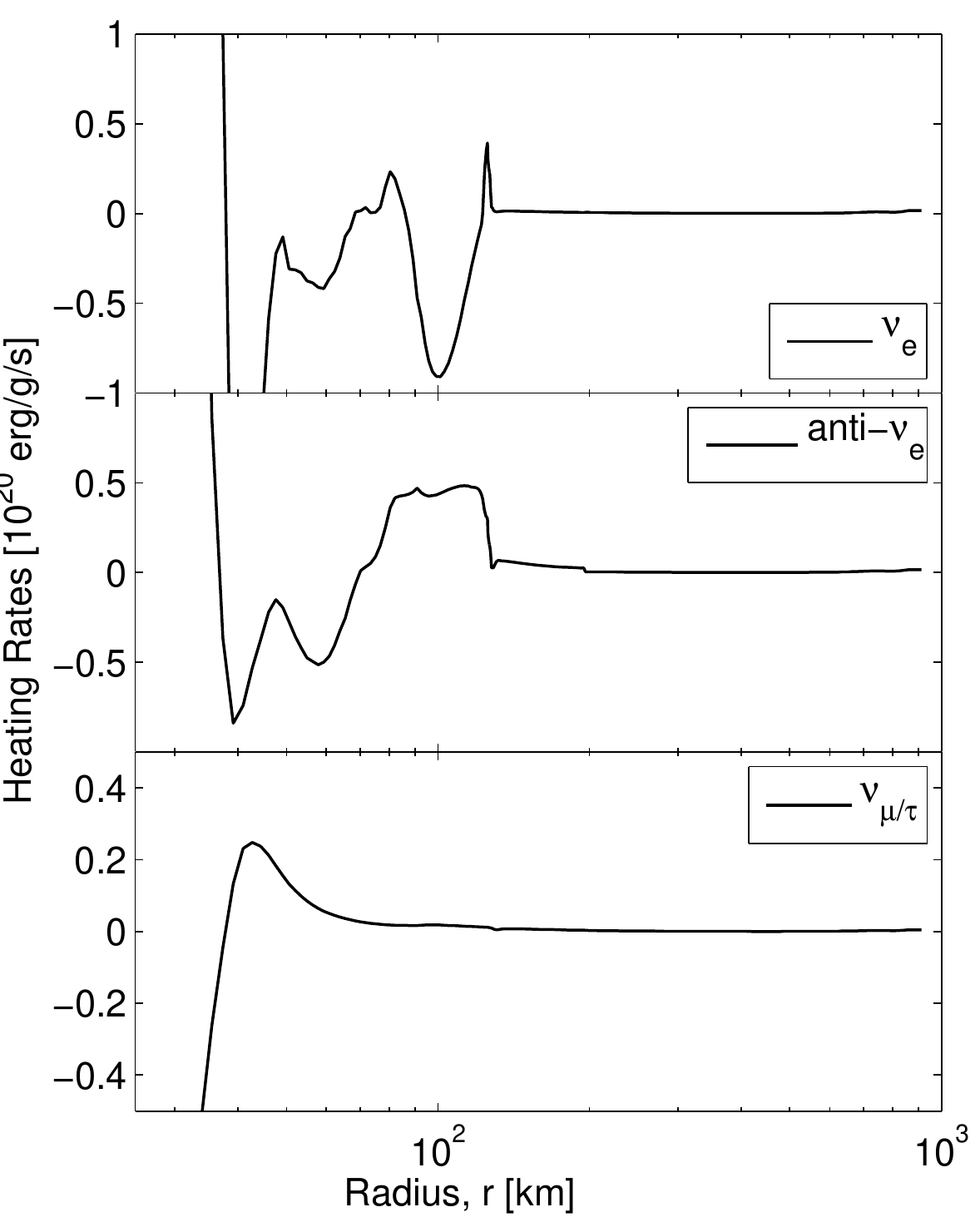}
\includegraphics[width=0.65\columnwidth]{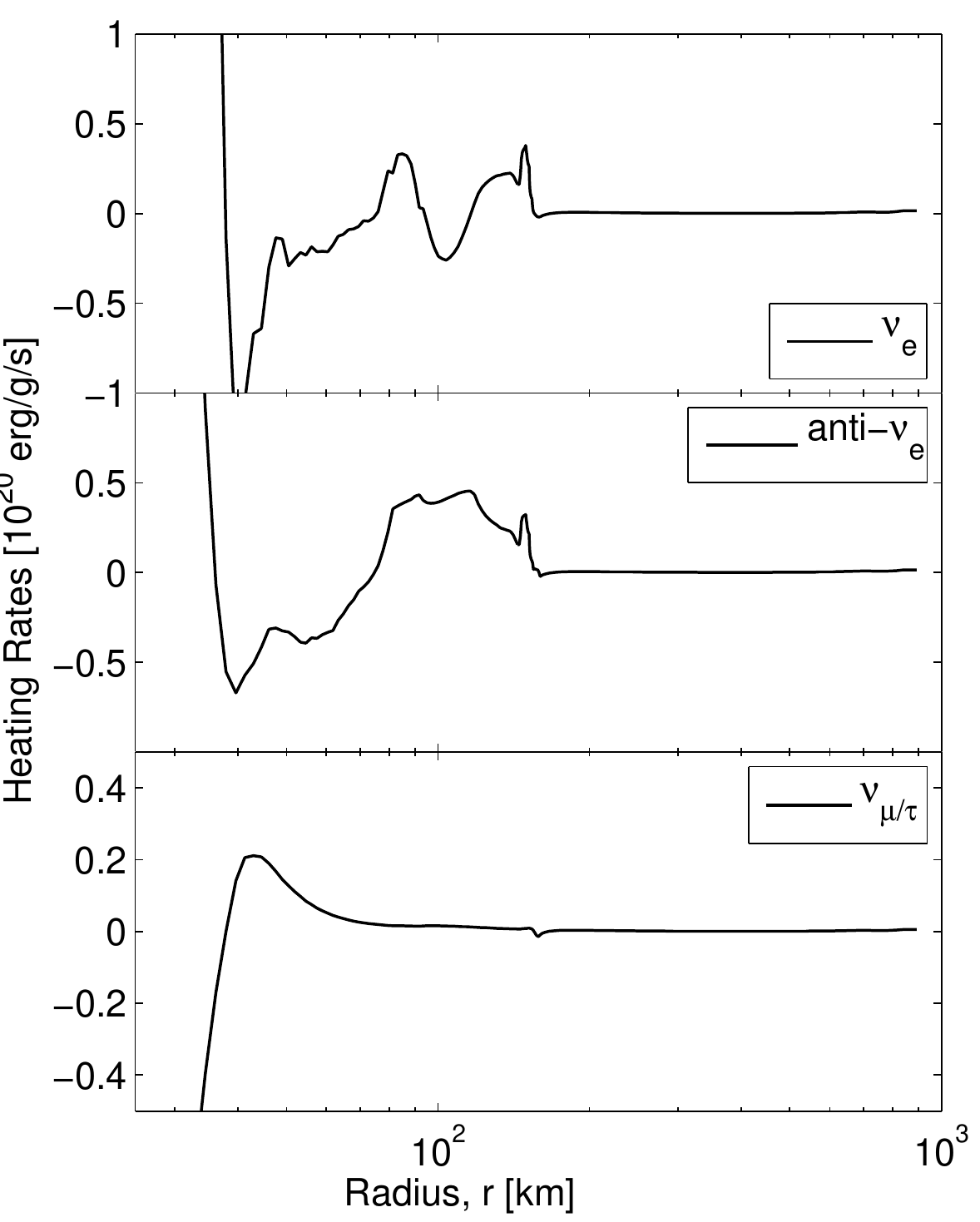}\\
\includegraphics[width=0.65\columnwidth]{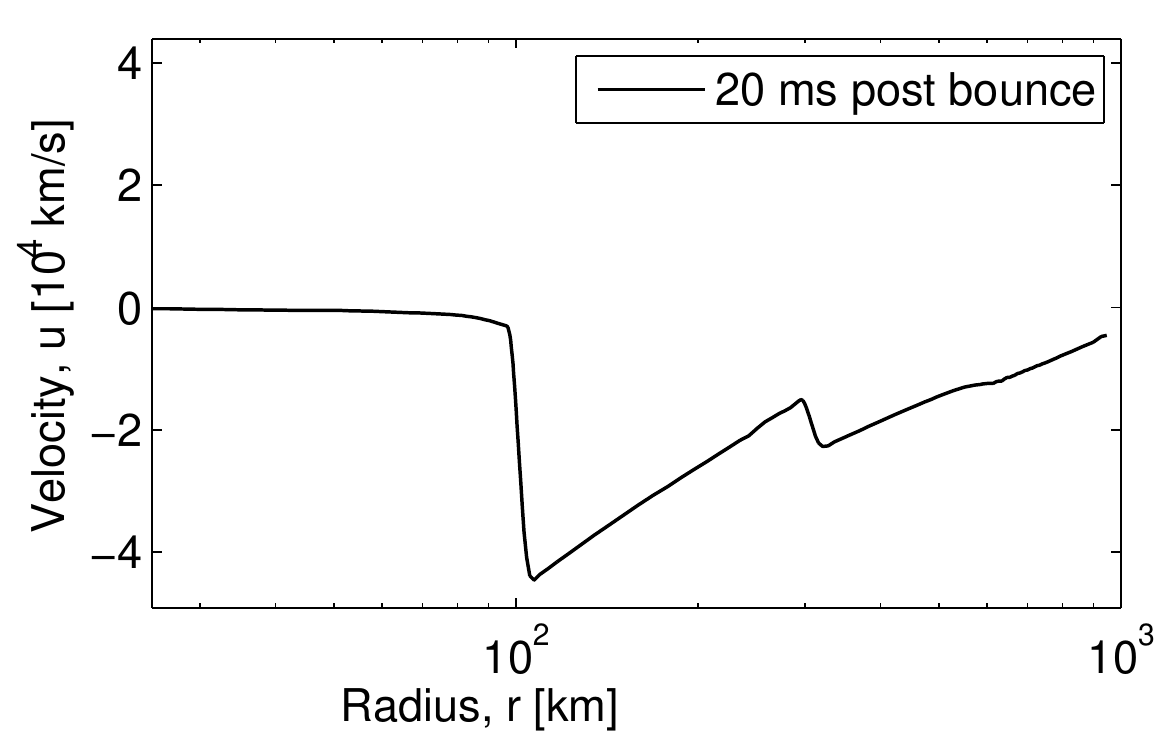}
\includegraphics[width=0.65\columnwidth]{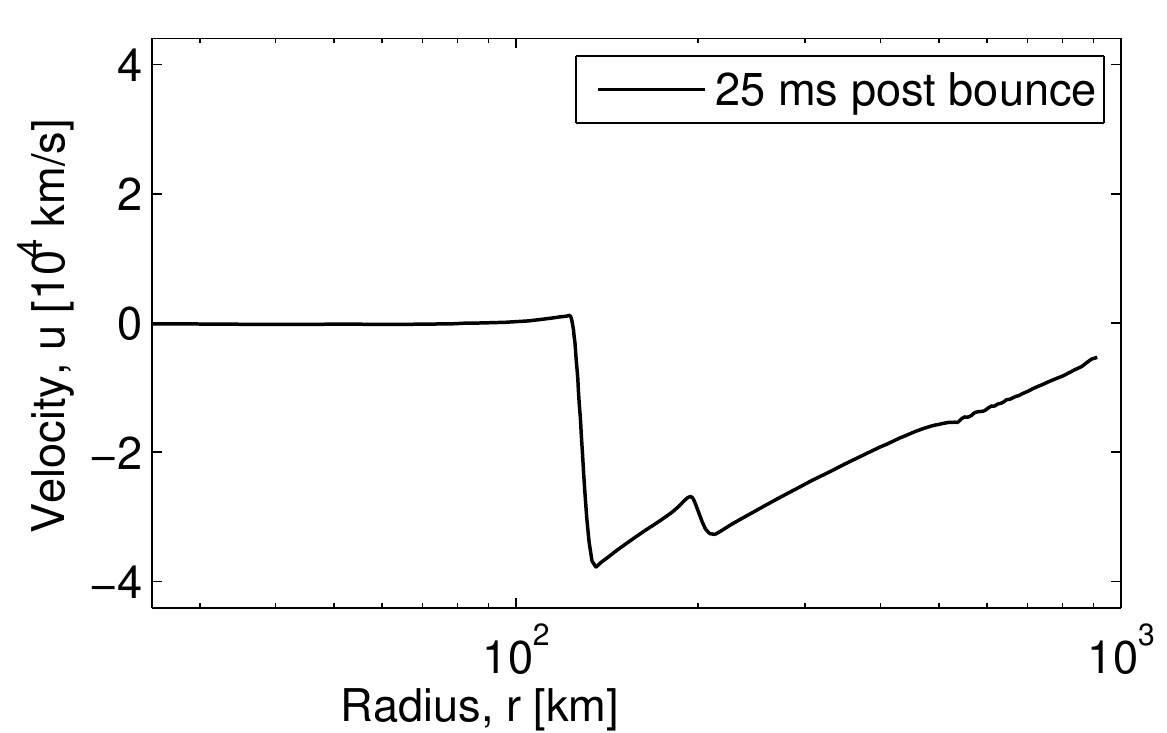}
\includegraphics[width=0.65\columnwidth]{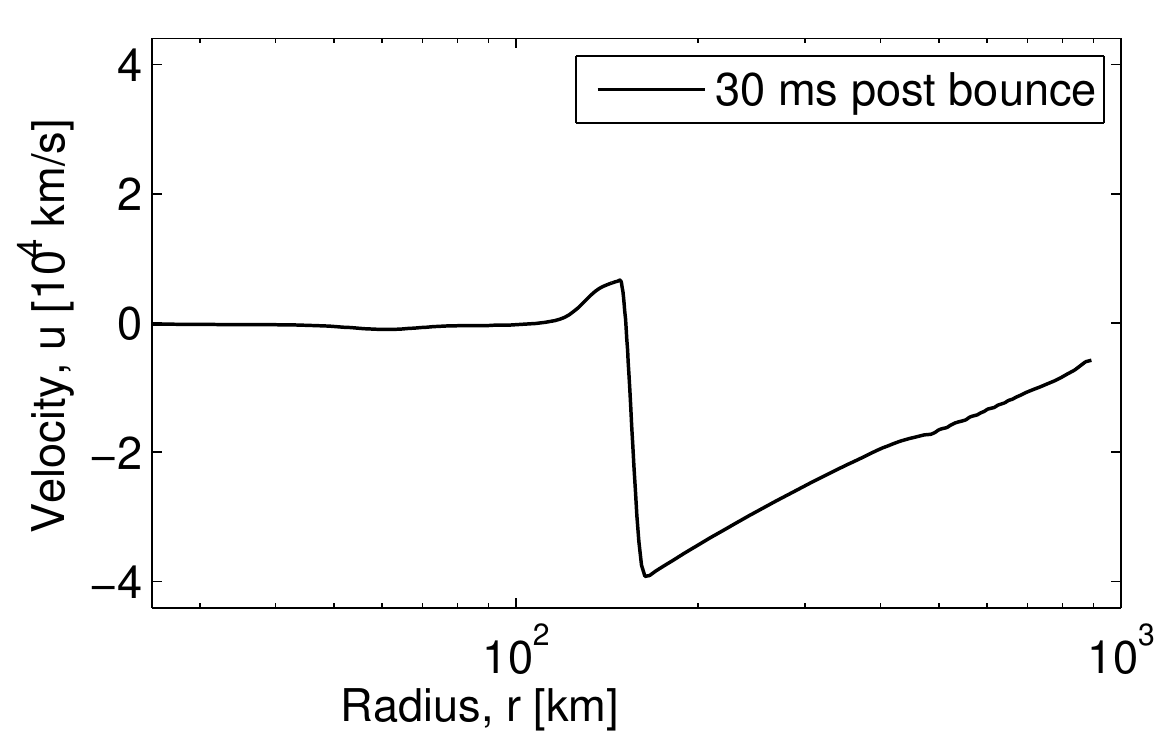}
\caption{Heating (\(>0\)) and cooling (\(<0\)) rates for the
\(8.8\) M\(_\odot\) progenitor model from Nomoto~(1983,1984,1987)
during the explosion phase at \(20\) ms (left panel), \(25\) ms (middle panel) 
and \(30\) ms (right panel) after bounce.
For a better comparison, the velocities are plotted
for the same post-bounce times.}
\label{fig-heatrate-onemg}
\end{figure*}

The low density of the mass outside the O-Ne-Mg-core makes it possible
to obtain the explosion in spherical symmetry supported via neutrino heating.
Neutrino cooling in the region of dissociated nuclear matter causes the
expanding shock front to turn into the SAS with no significant matter outflow.
\(\nu_e\)-cooling dominates over \(\bar{\nu}_e\)-heating  by one order of
magnitude.
Only at the dissociation line of infalling heavy nuclei, the neutrino energy
deposition drives the SAS slowly to larger radii, for illustration see the
heating(cooling) rates and velocity profile in Fig.~\ref{fig-heatrate-onemg}
(left panel) at \(20\) ms post-bounce.
However, the cooling of \(\nu_e\) still contributes to a large amount
at \(25\) ms post-bounce over the heating of \(\bar{\nu}_e\) and
\(\nu_{\mu/\tau}\) in Fig.~\ref{fig-heatrate-onemg}
(middle panel) behind the SAS.
Only directly at the shock a low net-heating rate remains.
Hence the influence of the neutrinos to the explosion itself
is of minor importance.
More important is the region of C-O-burning which produces Ne and Mg.
The hydrodynamic feedback to this thermodynamic transition
can be identified already during the collapse phase of the progenitor
core at the velocity profiles in Fig.~\ref{fig-velocity-onemg}
at about \(1.35-1.374\) M\(_\odot\).
As material is shock heated post-bounce, the transition layer
where Ne and Mg nuclei are burned into NSE propagates together
with the expanding shock wave outwards.
In other words, the Ne-Mg-layer of the progenitor
is converted directly into NSE.
Furthermore, the transition (discontinuity) from C-O-burning
is falling quickly towards the SAS.
It was found to be at about \(350\) km at \(20\) ms post-bounce
and at about \(200\) km at \(25\) ms post-bounce,
illustrated at the velocity profiles (bottom) in
Fig.~\ref{fig-heatrate-onemg} (left-right panels).
At about \(30\) ms post-bounce, the entire Ne-Mg-layer is converted
into NSE due to the temperature increase obtained via shock heating.
Hence, C and O nuclei are burned directly into NSE.

In contrast to more massive Fe-core progenitors where O-burning
produced an extended Si-S-layer, the amount of \(^{28}\)Si and
\(^{32}\)S is less than \(1\%\) at the end of nuclear burning
for the O-Ne-Mg-core discussed here
(see Fig.~\ref{fig-composition-onemg} (a) middle panel).
This low fraction of Si and S is already converted into NSE
during the initial collapse phase, due to the rapid density and
temperature increase of the contracting core.
Hence, C- and O-nuclei are burned directly into NSE
during the post-bounce evolution.
This sharp transition is related to a jump in the density
and the thermodynamic variables.
As the SAS propagates over this transition along the decreasing density,
the shock accelerates to positive velocities
(see Fig.~\ref{fig-heatrate-onemg} right panel).
The consequent explosion is hence driven due to the shock propagation
over the infalling transition between two different thermonuclear
regimes rather than by pure neutrino heating, illustrated at the
velocity profiles in Fig.~\ref{fig-velocity-onemg-expl}.
Although \citet{Kitaura:etal:2006} approximated nuclear
reactions during the evolution of the O-Ne-Mg-core progenitor,
the results of their explosion dynamics
are in qualitative agreement with our findings.

The more massive Fe-core progenitors show the same thermo- and
hydrodynamic features as discussed above for the O-Ne-Mg-core
due to the transition between different thermonuclear regimes.
However, the differences are smaller because C-O-burning produces
an extended layer composed of \(^{28}\)Si and \(^{32}\)S.
The transition of Si-burning into NSE is much smoother than
the transition of C-O-burning into NSE.
Furthermore, due to the more massive Si-S and C-O-layers for the Fe-core
progenitors, the transitions take more time on the order of seconds to fall
onto the SAS.
The presence of neutrino heating becomes important for the more massive
Fe-core progenitors to drive the SAS to large radii on a longer timescale.
The effects of the shock propagation across the transition between
different thermonuclear regimes has been pointed out in
\citet{Bruenn:etal:2006} with respect to the explosion dynamics
in axially-symmetric simulations of massive Fe-core progenitors.
In our spherically symmetric models, we cannot confirm the driving
force of explosions of Fe-core progenitors to be the shock propagation
across different thermonuclear regimes.
We find that the explosions are initiated due to the deposition,
although enhanced, of neutrino energy.
The shock is accelerated additionally when crossing different
thermonuclear regimes due to the density jumps at the transitions.

\subsection{Comparison of the neutrino spectra}

Striking is the agreement in the mean neutrino energies between
all different progenitor models (including the O-Ne-Mg-core and the
Fe-core progenitors) during the explosion phase, although the neutrino
emissivities and opacities are enhanced for the Fe-core progenitor models
(see Fig.~\ref{fig-lumin-expl}).
The explosion phase for the O-Ne-Mg-core lasts only until about
\(40\) ms post-bounce, which is significantly shorter in comparison
to the more massive Fe-core progenitors.
Furthermore, the luminosities are also lower by a factor of \(2\).
For all models, the electron antineutrino luminosity is higher than the
electron neutrino luminosity on a timescale of \(200\) ms after the
explosions have been launched.
This slight difference reduces again at later times where the
electron neutrino luminosity becomes again higher than the electron
antineutrino luminosity.
However, after the explosions have been launched the behaviors of the
luminosities are in qualitative agreement for all models.
The same holds for the mean neutrino energies which increase
continuously during the neutrino heating phase.
The electron (anti)neutrinos have rms energies of about \(12\)
(\(14\)) MeV where as after the explosions have been launched,
rms energies of about \(11\) (\(13\)) MeV are obtained.
The values remain constant on the timescale of \(1\) second post-bounce.
The (\(\mu/\tau\))-neutrinos have rms energies of about \(18\) MeV
during the neutrino heating phase and about \(15\) MeV after
the explosion has been launched.
These differences in the mean neutrino energies and luminosities
during the neutrino heating, initial and proceeding explosion phases
are in correspondence with the electron fraction of the material,
as will be illustrated in the following section.

\subsection{The electron fraction of the early ejecta}

\begin{figure}[ht]
\begin{center}
\includegraphics[width=0.98\columnwidth]{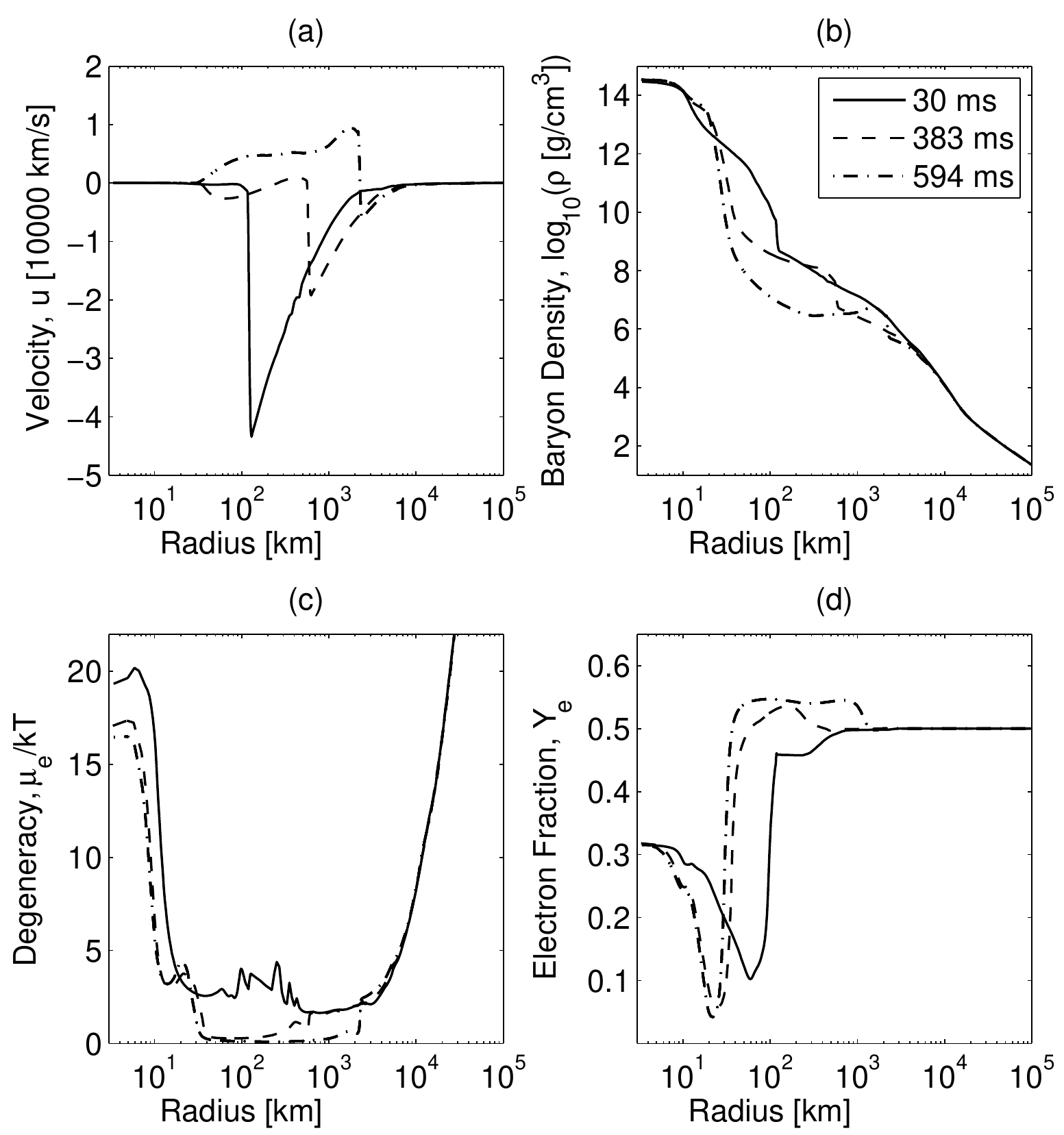}
\caption{Selected hydrodynamic variables during the initial explosion
phase at three different post-bounce times for the \(10.8\)
M\(_\odot\) progenitor model from \citet{Woosley:etal:2002}.}
\label{fig-hydroplot-expl}
\end{center}
\end{figure}

During the neutrino heating phase, the neutrino spectra are mainly
determined by mass accretion at the neutrinospheres.
Neutron-rich nuclei from the progenitor star with an electron
fraction of \(Y_e \simeq 0.45\) are falling onto the oscillating SAS
and dissociate into free nucleons and light nuclei,
see Fig.~\ref{fig-hydroplot-expl} (d).
These nucleons accrete then slowly onto the PNS surface at the center.
Due to the increased electron-degeneracy behind the SAS
in Fig.~\ref{fig-hydroplot-expl} (c), weak-equilibrium is established
at a lower value of the electron fraction of \(Y_e \leq 0.15\).

As soon as the SAS is revived and propagates outward,
see the velocity and density profiles in Fig.~\ref{fig-hydroplot-expl}
(a) and (b), the electron degeneracy behind the expanding shock
is reduced and weak-equilibrium is established at a higher value
of the electron fraction of \(Y_e > 0.56\)
\footnote{
The EoS from \citet{Shen:etal:1998}
is limited to a maximum electron fraction of \(Y_{e}\geq0.564\).
The EoS has been extended by G\"ogelein (2007, priv. comm.)
to model asymmetric nuclear matter
with an electron fraction above \(0.564\).
We assume an ideal gas of free nucleons and light nuclei,
which is a sufficient assumption for the conditions
found in the region of the extremely proton-rich ejecta.
}.
The capture rates for electron-neutrinos at neutrons are favored
over electron-antineutrino captures at protons.
This slight difference results in an electron and hence proton excess.
Consequently the explosion ejecta are found to be initially proton-rich.
This behavior of the electron fraction was found for all our
explosion models, for the \(10.8\) and \(18\) M\(_\odot\) Fe-core
progenitors with artificially enhanced opacities and for the
O-Ne-Mg-core using the standard opacities.
Such explosion models were investigated with respect to the
nucleosynthesis in general and with respect to the
\(\nu p\)-process by Fr\"ohlich et~al. (2006a-c).

One of the main goals of the present investigation is to determine
the behavior of the electron fraction for the initially proton-rich ejecta
on a long timescale on the order of \(10\) seconds, in a consistent manner.
We explore the question if the material ejected in the neutrino-driven wind
becomes neutron-rich and which are the conditions (e.g. entropy per baryon,
expansion timescale) obtained in the neutrino-driven wind.
These aspects are of special relevance for the composition of the
ejecta, which is determined via explosive nucleosynthesis analysis,
in particular in order to be able to draw conclusions with respect
to a possible site for the production of heavy elements via the \(r\)-process.
Therefore, the continued expansion of the explosion ejecta must be
simulated, for which the inclusion of a large physical domain
of the progenitor up to the He-layer is required.
Furthermore, since the electronic charged current reaction rates and
the neutrino fluxes determine the electron fraction,
the PNS contraction at the center and hence the contraction of the
neutrinospheres are essential.
%

\section{The neutrino-driven wind}

In this section we investigate the post explosion evolution
with special focus on the properties of the ejecta, in particular
the electron fraction.
We explore the problem if the initially proton-rich ejecta
become neutron-rich at later times on the order of \(10\) seconds and
if the conditions might indicate a possible site for the 
nucleosynthesis of heavy nuclei via the \(r\)-process.
This has been assumed in static steady-state as well as parametrized
dynamic wind models, based on the results obtained by
\citet{Woosley:etal:1994}.
The ejected material in their simulations does never become proton-rich,
the electron fraction was found to continuously decrease with time
after the explosion has been launched.
Thus, the ejecta were investigated in a region where
the conditions are favorable for the \(r\)-process.
The stellar models applied to the present investigation of the neutrino
driven wind are the \(8.8\) M\(_\odot\) O-Ne-Mg-core and the  \(10.8\) and
\(18\) M\(_\odot\) Fe-core progenitors, where for the latter two cases the
explosions are obtained using the artificially enhanced opacities as
described in \S 2.3.

\begin{figure*}[ht]
\begin{center}
\includegraphics[width=1.7\columnwidth]{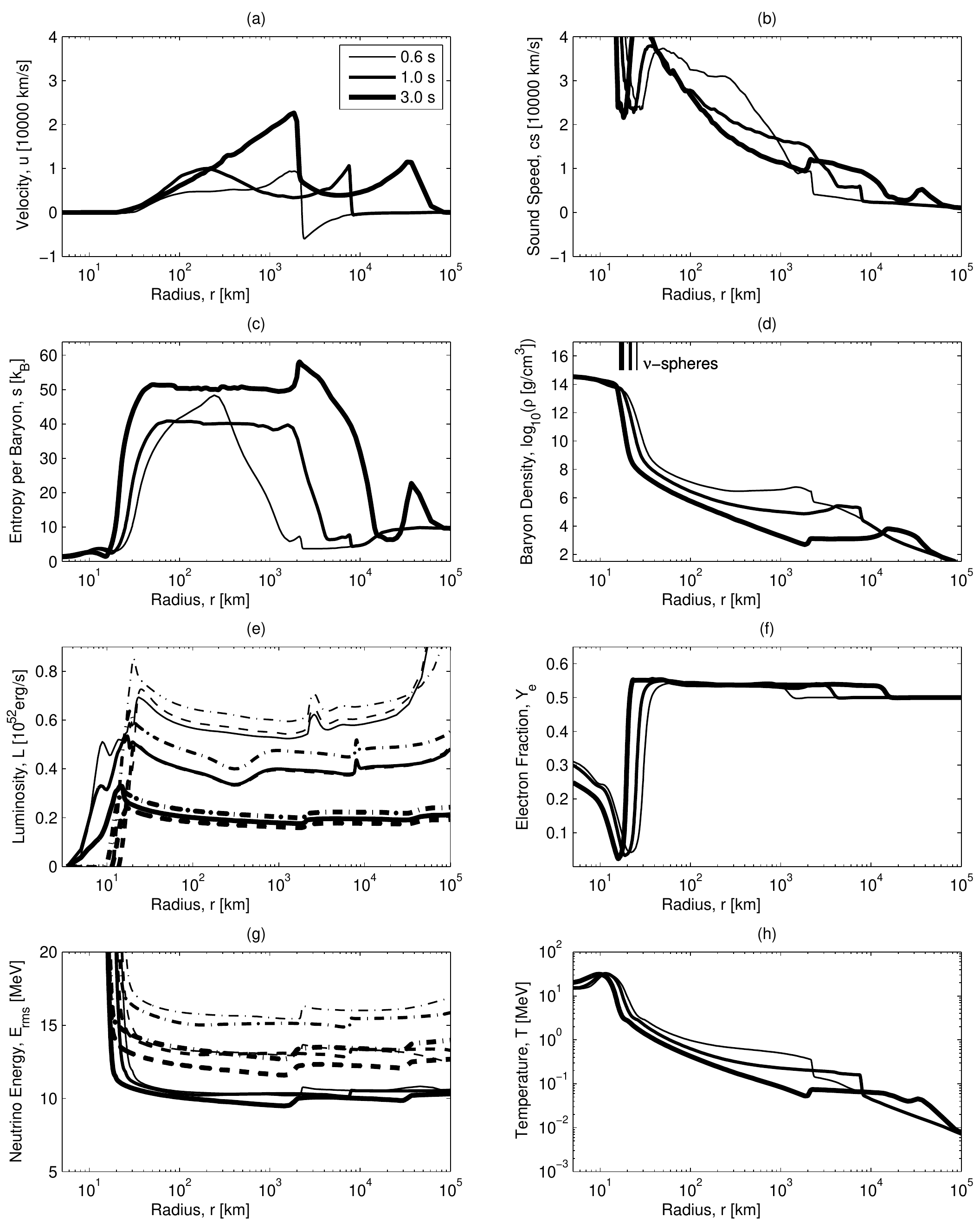}
\caption{Selected hydrodynamic variables during the formation of the
neutrino-driven wind at three different post-bounce times for the
\(10.8\) M\(_\odot\) progenitor model from \citet{Woosley:etal:2002}.
In addition, graphs (e) and (g) show the neutrino luminosities and
rms neutrino energies
(solid lines: \(\nu_e\),
dashed lines: \(\bar{\nu}_e\),
dash-dotted lines: \(\nu_{\mu/\tau}\)).
For this progenitor model the neutrino-driven wind becomes
supersonic, using the enhanced opacities.}
\label{fig-fullstate-wind-h10a}
\end{center}
\end{figure*}

\begin{figure*}[ht]
\begin{center}
\includegraphics[width=1.7\columnwidth]
{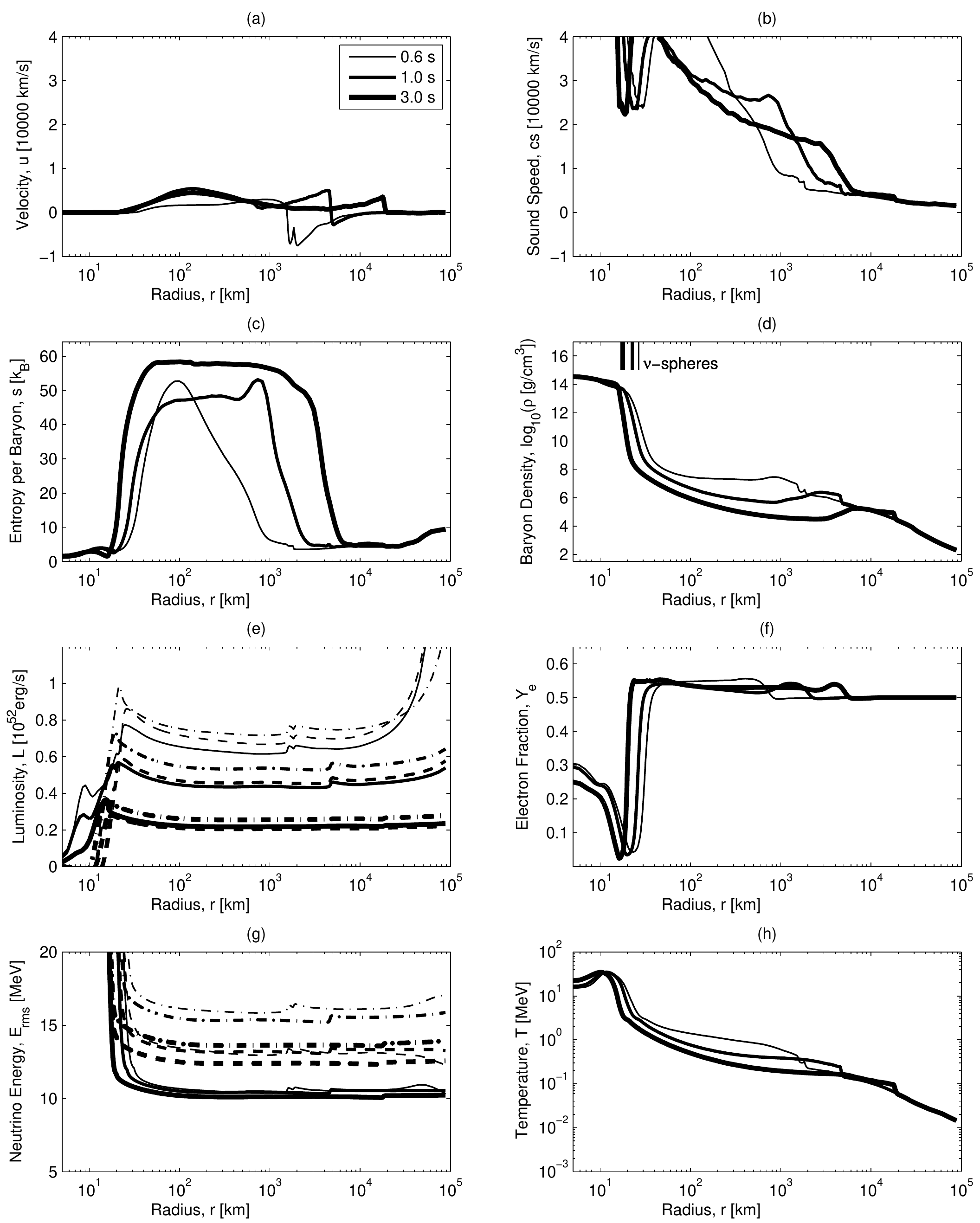}
\caption{The same configuration as Fig.~\ref{fig-fullstate-wind-h10a}
for the \(18\) M\(_\odot\) progenitor model from \citet{Woosley:etal:2002}.
The neutrino-driven wind remains subsonic for this progenitor model,
even using the enhanced opacities.}
\label{fig-fullstate-wind-h18b}
\end{center}
\end{figure*}

After the explosions have been launched, the region between the expanding
explosion shock and the neutrinospheres cools rapidly and the density
decreases continuously as illustrated in Figs.~\ref{fig-fullstate-wind-h10a}
and \ref{fig-fullstate-wind-h18b} (d) and (h).
In order to determine the evolution of the electron fraction \(Y_e\),
the non-local neutrino fluxes are as important as the local neutrino
reaction rates.
Since the PNS and hence the neutrinospheres contract continuously
due to the deleptonization, the degeneracy increases and matter at
the PNS surface is found to be extremely neutron-rich with \(Y_e\le 0.1\)
(see Figs.~\ref{fig-fullstate-wind-h10a}
and \ref{fig-fullstate-wind-h18b} (d) and (f)).


\begin{figure}[ht]
\centering
\includegraphics[width=0.95\columnwidth]
{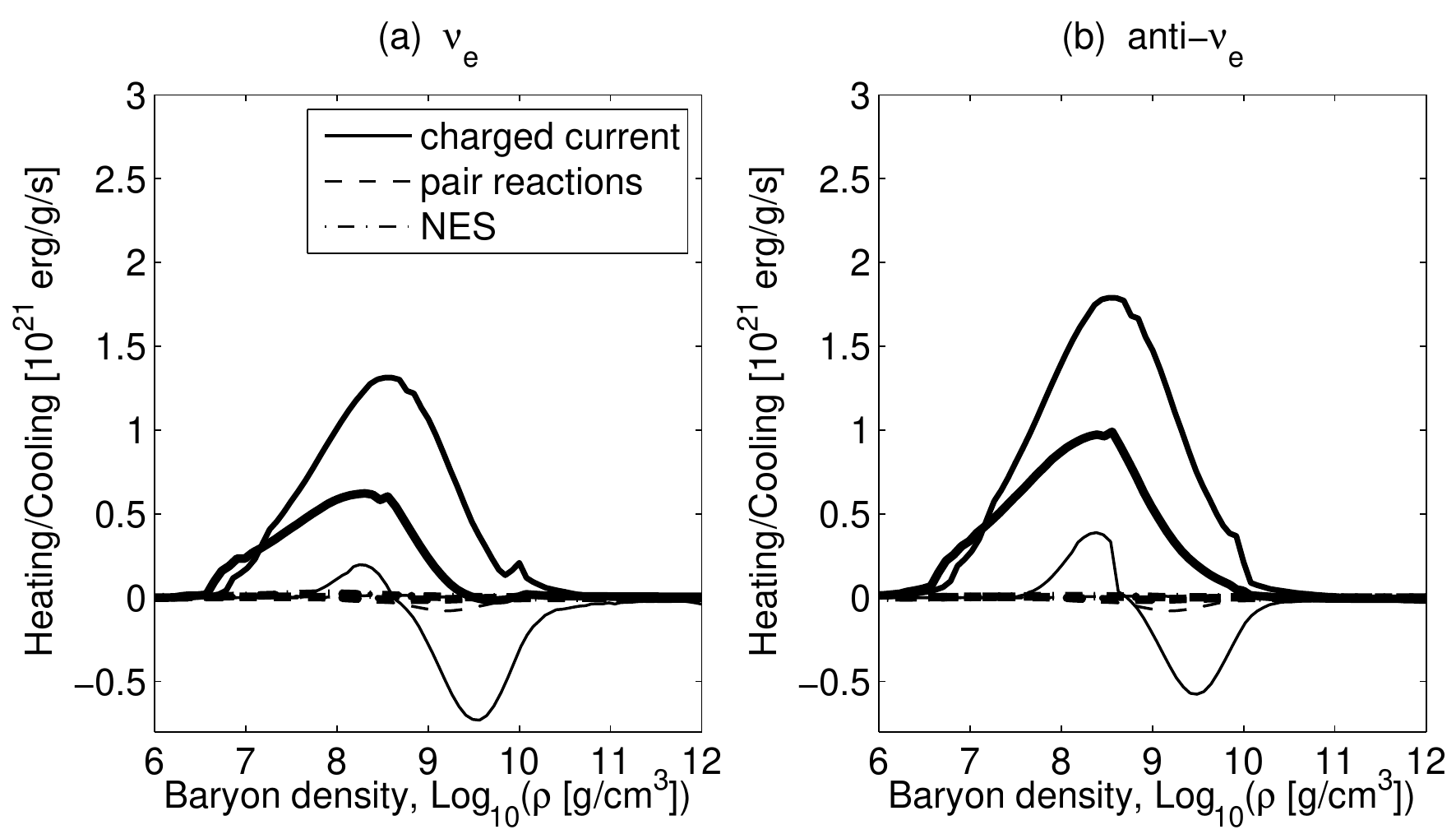}
\caption{Heating and cooling rates with respect to the baryon density
at three different post-bounce times 0.5 second (thin lines), 1 second
(intermediate lines), 5 seconds (thick lines) for the \(10.8\) M\(_\odot\)
progenitor model from \citet{Woosley:etal:2002}.}
\label{fig-heatplot-h10a}
\end{figure}

Independent of the progenitor model, the region on top of the PNS surface
is continuously subject to neutrino heating during the post explosion phase.
The dominant heating sources are the absorption of electron-(anti)neutrinos
at free nucleons, due to the high fraction of free nucleons
(dissociated nuclear matter) present in the region on top of the PNS,
as shown in Fig.~\ref{fig-heatplot-h10a}.
The neutrino pair production and thermalization processes have a
negligible contribution to the heating outside the neutrinospheres.
In order to compare the heating and cooling rates in Fig.~\ref{fig-heatplot-h10a},
we plot the quantities with respect to the baryon density.
While neutrino cooling is still dominantly present at \(\sim500\) ms post-bounce
(thin lines in Fig.~\ref{fig-heatplot-h10a}), at later times after \(\sim1\) second
post-bounce (thick lines in Fig.~\ref{fig-heatplot-h10a})
neutrino cooling vanishes and only heating is found in the density
domain of interest, i.e. between \(10^{7}-10^{12}\) g/cm\(^{3}\).
Figs.~\ref{fig-fullstate-wind-h10a} (d) and \ref{fig-fullstate-wind-h18b} (d)
show the conditions for the contracting PNSs at the center via the radial
baryon density profiles and the electron-neutrinospheres.
The region of interest where the neutrino-driven wind develops
corresponds to the density domain of \(10^7 - 10^{11}\) g/cm\(^3\).
The degeneracy of the early ejecta favors proton-rich matter where
a high electron fraction of \(Y_e\simeq0.54\) is obtained.
Hence, the absorption of electron-antineutrinos at free protons
dominates over electron-neutrino absorption at free neutrons.
The corresponding radial neutrino luminosities and rms energies are
shown in Figs.~\ref{fig-fullstate-wind-h10a} and
\ref{fig-fullstate-wind-h18b} (e) and (g).
In addition, for the first time we were able to follow the deleptonization
burst from core bounce for several seconds over a large physical domain
including several \(10^{5}\) km of the progenitor star.
The outrunning luminosity peak can be identified in the luminosities in
Figs.~\ref{fig-fullstate-wind-h10a} and \ref{fig-fullstate-wind-h18b} (e)
at \(0.6\) seconds after bounce at a distance between \(5\times10^{4}\)
and \(10^{5}\) km, leaving the computational domain between
\(1-2\) seconds post-bounce.

After the explosions have been launched, the continued energy transfer
from the neutrino radiation field into the fluid outside the
neutrinospheres as illustrated in Fig.~\ref{fig-heatplot-h10a}
drives the matter entropies to high values, shown in
Figs.~\ref{fig-fullstate-wind-h10a} and
\ref{fig-fullstate-wind-h18b} (c).
The heat deposition at the PNS surface accelerates matter
to positive velocities, see Figs.~\ref{fig-fullstate-wind-h10a} and
\ref{fig-fullstate-wind-h18b} (a), after \(\simeq 1\) second post-bounce.
This matter outflow is known as the neutrino-driven wind,
which proceeds adiabatically at larger radii.
This is consistent with the constant radial entropy per baryon
profiles in the graphs (c).
Furthermore, the rapidly decreasing luminosities reach
values below \(5 \times 10^{51}\) erg/s already \(1\) second
after bounce
(see Figs.~\ref{fig-fullstate-wind-h10a} and
\ref{fig-fullstate-wind-h18b} (e)).
The luminosities continue to decrease and reach values below
\(1 \times 10^{51}\) erg/s at \(10\) seconds after bounce.
The mean neutrino energies also decrease constantly where values
below \(10\) MeV for the electron-flavor neutrinos and below \(12\)
MeV for the (\(\mu/\tau\))-neutrinos are obtained
(see Figs.~\ref{fig-fullstate-wind-h10a} and
\ref{fig-fullstate-wind-h18b} (g)).

\begin{figure*}[ht]
\begin{center}
\includegraphics[width=1.7\columnwidth]{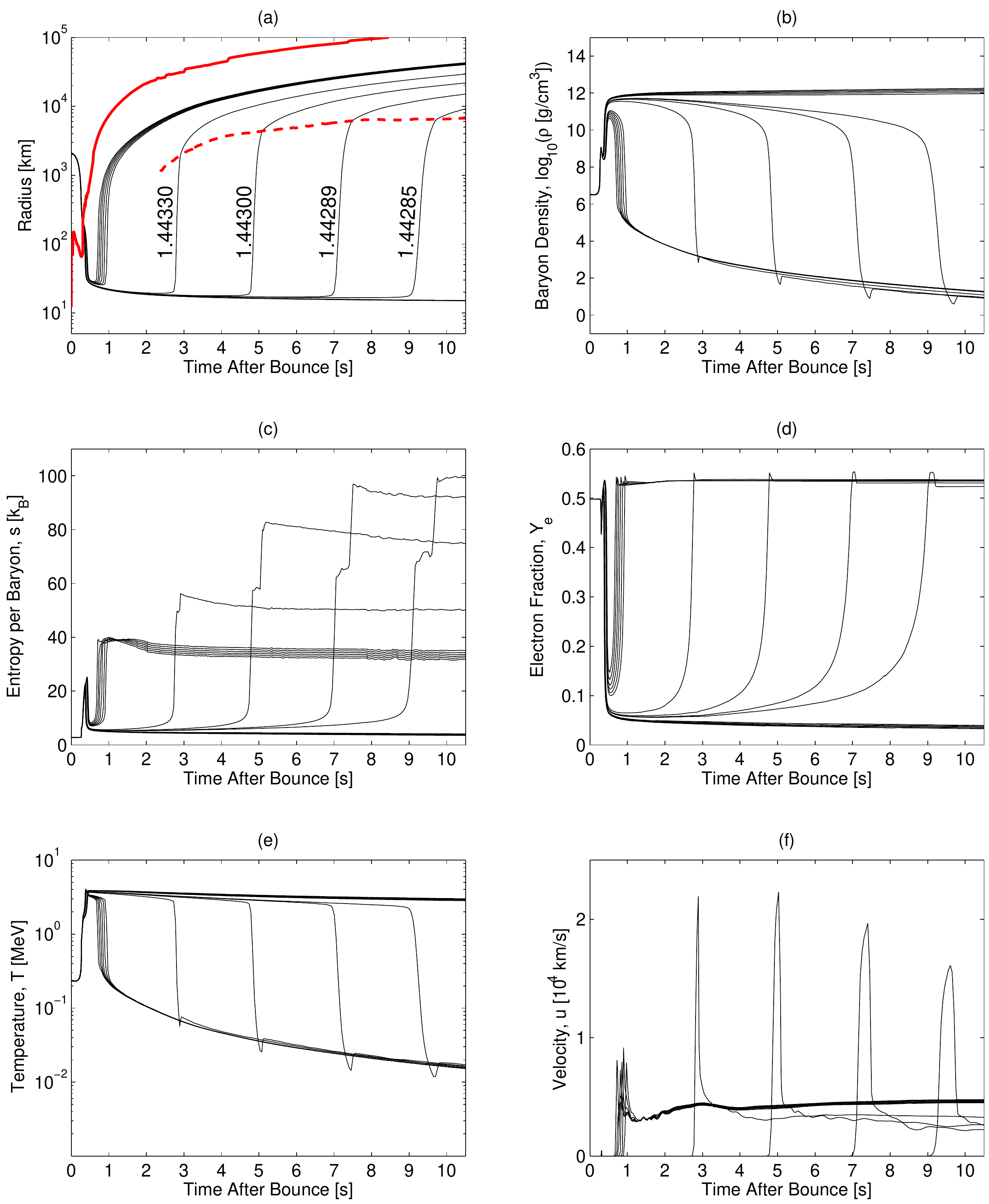}
\caption{Evolution of selected mass elements in the neutrino-driven wind
(as listed in graph (a) from \(1.44285-1.44450\) M\(_\odot\) baryon mass)
for the \(10.8\) M\(_\odot\) progenitor model from \citet{Woosley:etal:2002}
where the enhanced opacities are used.
Graph (a) shows in addition the position of the expanding explosion shock
(red solid line) and the position of the wind termination shock
(red dashed line).}
\label{fig-shellplots-h10a}
\end{center}
\end{figure*}

Several previous wind studies achieved supersonic matter outflow velocities
for the neutrino-driven wind due to assumed high luminosities.
In any case, the accelerated material of the neutrino-driven wind
collides with the slower and subsonically expanding explosion ejecta.
Comparing Figs.~\ref{fig-fullstate-wind-h10a} and
\ref{fig-fullstate-wind-h18b}, the more compact wind region of the
\(18\) M\(_\odot\) progenitor model produces a less pronounced
neutrino-driven wind in comparison to the \(10.8\) M\(_\odot\) progenitor model.
The densities of the wind region are higher up to two orders of magnitude
and the temperatures are higher by a factor of \(2\).
The resulting velocities of the neutrino-driven wind outflow are lower
by a factor of \(2\) and stay below \(10^4\) km/s.
Hence, the neutrino-driven wind remains subsonic for all times for the
\(18\) M\(_\odot\) progenitor model (see Fig.~\ref{fig-fullstate-wind-h18b} (a))
where the wind develops supersonic velocities for the \(10.8\) M\(_\odot\)
progenitor model (see Fig.~\ref{fig-fullstate-wind-h10a} (a)).
In the case of a supersonic neutrino-driven wind, this leads to the
formation of the reverse shock known as the wind termination shock.
The formation of the wind termination shock
of the \(10.8\) M\(_\odot\) progenitor model
is illustrated in Fig.~\ref{fig-shellplots-h10a}
and will be discussed in the following paragraph.
In order to analyze the dynamical evolution and the consequences of the
formation of the reverse shock, steady-state wind models cannot be used.
Radiation hydrodynamics is required in order to describe
the dynamical effects consistently.
Our results, obtained using general relativistic radiation hydrodynamics
based on spectral three-flavor Boltzmann neutrino transport,
are in qualitative agreement with the detailed parametrized 
investigation by \citet{Arcones:etal:2007}.

During the initial and subsonic wind expansion,
the matter entropies in Fig.~\ref{fig-shellplots-h10a} (c)
increase slowly from \(4\) to \(5-10\) k\(_\text{B}\)/baryon
and the densities in  Fig.~\ref{fig-shellplots-h10a} (b) and
temperatures in Fig.~\ref{fig-shellplots-h10a} (e) decrease
on a long timescale over several seconds.
Furthermore, the reduced degeneracy in the wind increases
the electron fraction shown in Fig.~\ref{fig-shellplots-h10a} (d)
slowly on the same timescale.
When the material is accelerated supersonically with velocities
of several \(10^4\) km/s up to radii of a several \(10^3\) km
(see Figs.~\ref{fig-shellplots-h10a} (a) and (f)), 
the entropies increase from s \(\simeq 5-10\) k\(_\text{B}\)/baryon
to s \(\simeq 40-60\) k\(_\text{B}\)/baryon on a short timescale of
the order of \(100\) ms.
During this rapid expansion, the density and temperature decrease
drastically from \(10^{10}\) g/cm\(^3\) to \(10^{4}-10^{2}\) g/cm\(^3\)
and from \(3\) MeV to \(0.1-0.01\) MeV respectively
(see Figs.~\ref{fig-shellplots-h10a} (b) and (e)).
It also corresponds to a rapid decrease of the degeneracy which in turn
is reflected in a rapid increase of the electron fraction of the
accelerated material on top of the PNS surface,
from \(Y_e \simeq 0.1\) to \(Y_e \simeq 0.56\)
(see Fig.~\ref{fig-shellplots-h10a} (d)).
Furthermore, the supersonically expanding neutrino-driven wind collides
with the explosion ejecta as can bee seen in Fig.~\ref{fig-shellplots-h10a}
(a) (solid red line) at radii of several \(10^4\) km.
Consequently, the previously accelerated material decelerates behind the
explosion ejecta as can be seen in the velocities in 
Fig.~\ref{fig-shellplots-h10a} (f).
This phenomenon becomes significant after about \(2\) seconds post-bounce
and corresponds to the formation of the reverse shock,
i.e. the wind termination shock.
(see Fig.~\ref{fig-shellplots-h10a} (a) dashed red line at radii of several
\(10^3\) km).
It causes an additional entropy increase to the final values of
s \(\simeq 50 - 100\) k\(_\text{B}\)/baryon.
During the rapid deceleration on the same short timescale on the order of
\(100\) ms, the densities in Fig.~\ref{fig-shellplots-h10a} (b) and temperatures
in Fig.~\ref{fig-shellplots-h10a} (e) increase again slightly,
where the degeneracy increases
and hence the electron fraction reduces slightly to values of \(Y_e \simeq 0.54\).
The following dynamical evolution is given by the subsonic and adiabatic
expansion of the explosion ejecta on a longer timescale on the order of seconds.
The density and temperature decrease slowly where the entropies of
s \(\simeq 50 - 100\) k\(_\text{B}\)/baryon
and the electron fraction of about \(Y_e = 0.54\) remain constant.
The latter aspects are essential for the nucleosynthesis analysis of the ejecta.
It can be understood in the sense that the neutrino reaction rates freeze out
and the matter conditions correspond to the neutrino free streaming regime.

\begin{figure*}[ht]
\begin{center}
\includegraphics[width=1.7\columnwidth]{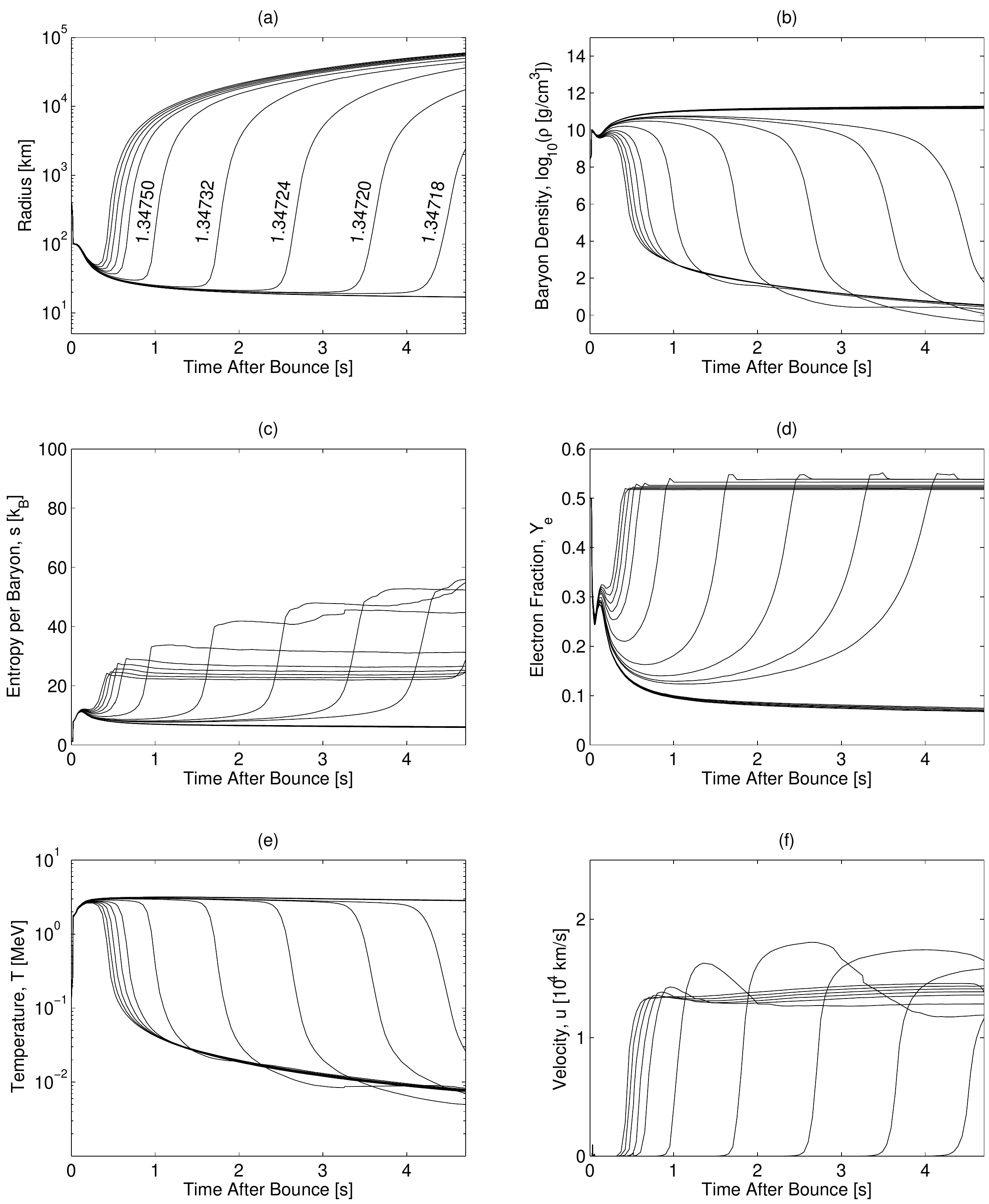}
\caption{Evolution of selected mass shells in the neutrino-driven wind
(as listed in graph (a) from \(1.34718 - 1.34750\) M\(_\odot\) baryon mass)
for the \(8.8\) M\(_\odot\) progenitor model from Nomoto~(1983,1984,1987)
where the standard emissivities and opacities given in \citet{Bruenn:1985} are used.
The graphs show the same configurations as Fig.~\ref{fig-shellplots-h10a}.}
\label{fig-shellplot-n08c}
\end{center}
\end{figure*}

Note that the strong neutrino-driven wind for the \(10.8\) M\(_\odot\) progenitor
model is obtained using the enhanced opacities as introduced in \S 2.3.
We additionally illustrate selected properties of the neutrino-driven
wind for the \(8.8\) M\(_\odot\) progenitor model in
Fig.~\ref{fig-shellplot-n08c} where a strong neutrino-driven wind was obtained
using the standard emissivities and opacities given in \citet{Bruenn:1985}.
This is due to the low density of the region between the neutrinospheres
at the PNS surface and the expanding explosion shock, where neutrino heating
via the standard rates and energy from nuclear burning are sufficient
to drive a strong supersonic matter outflow.
Matter entropies increase to \(s\simeq 10\) k\(_\text{B}\)/baryon during
the initial acceleration of the wind and the densities and temperatures
decrease slowly on a timescale of seconds.
The properties during the initial acceleration observed are similar
to those of the more massive \(10.8\) M\(_\odot\) Fe-core progenitor.
The same holds for the acceleration to supersonic velocities.
The timescale is reduced to \(100\) ms where the entropies increase
rapidly to \(s\simeq 20-50\) k\(_\text{B}\)/baryon
(see Fig.~\ref{fig-shellplot-n08c} (c))
and due to the reduced degeneracy the electron fraction increases
from \(Y_e=0.1\) at the PNS surface to \(Y_e=0.56\)
(see Fig.~\ref{fig-shellplot-n08c} (d)).
Density and temperature decrease to \(10-100\) g/cm\(^3\) and \(0.001\)
MeV respectively (see Fig.~\ref{fig-shellplot-n08c} (b) and (e)).
The difference to the more massive \(10.8\) M\(_\odot\) Fe-core progenitor
is due to the lower mass enclosed between the PNS surface and the
expanding explosion ejecta.
For the more massive \(10.8\) M\(_\odot\) Fe-core
progenitor in Fig.~\ref{fig-shellplots-h10a} (f), the previously accelerated
material collides with the explosion ejecta already after a few \(100\) ms.
Here the supersonic wind expands on a much longer timescale up to
several seconds before it collides with the explosion ejecta
(see Fig.~\ref{fig-shellplot-n08c} (f)).
During this adiabatic expansion, entropy and electron fraction remain constant.
The fast material collides with the much slower expanding explosion ejecta
so that the material is decelerated and the reverse shock appears.
This is again similar to the formation of the reverse shock for the
more massive \(10.8\) M\(_\odot\) Fe-core progenitor as discussed above.
Matter entropies increase to \(s=100\) k\(_\text{B}\)/baryon
(see Fig.\ref{fig-shellplot-n08c} (c)), density and temperature increase
sightly (see Fig.\ref{fig-shellplot-n08c} (b) and (e))
and the electron fraction reduces slightly to
\(Y_e\simeq 0.52-0.54\) due to the increased degeneracy
(see Fig.\ref{fig-shellplot-n08c} (d)).
The following evolution is determined by the adiabatic expansion of the
explosion ejecta during which the entropy and electron fraction remain constant.
\begin{figure*}[ht]
\centering
\includegraphics[width=0.95\columnwidth]{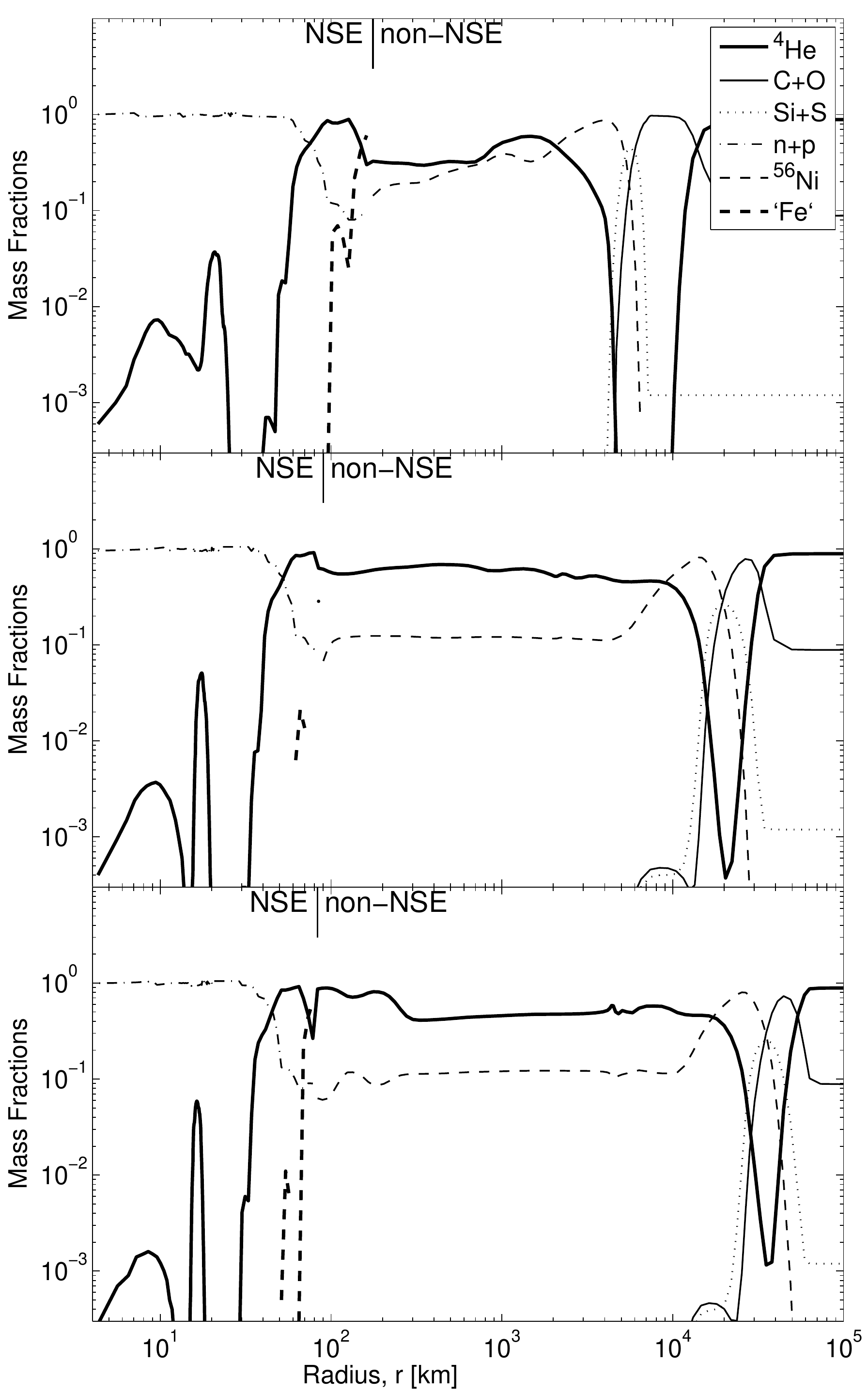}
\includegraphics[width=0.95\columnwidth]{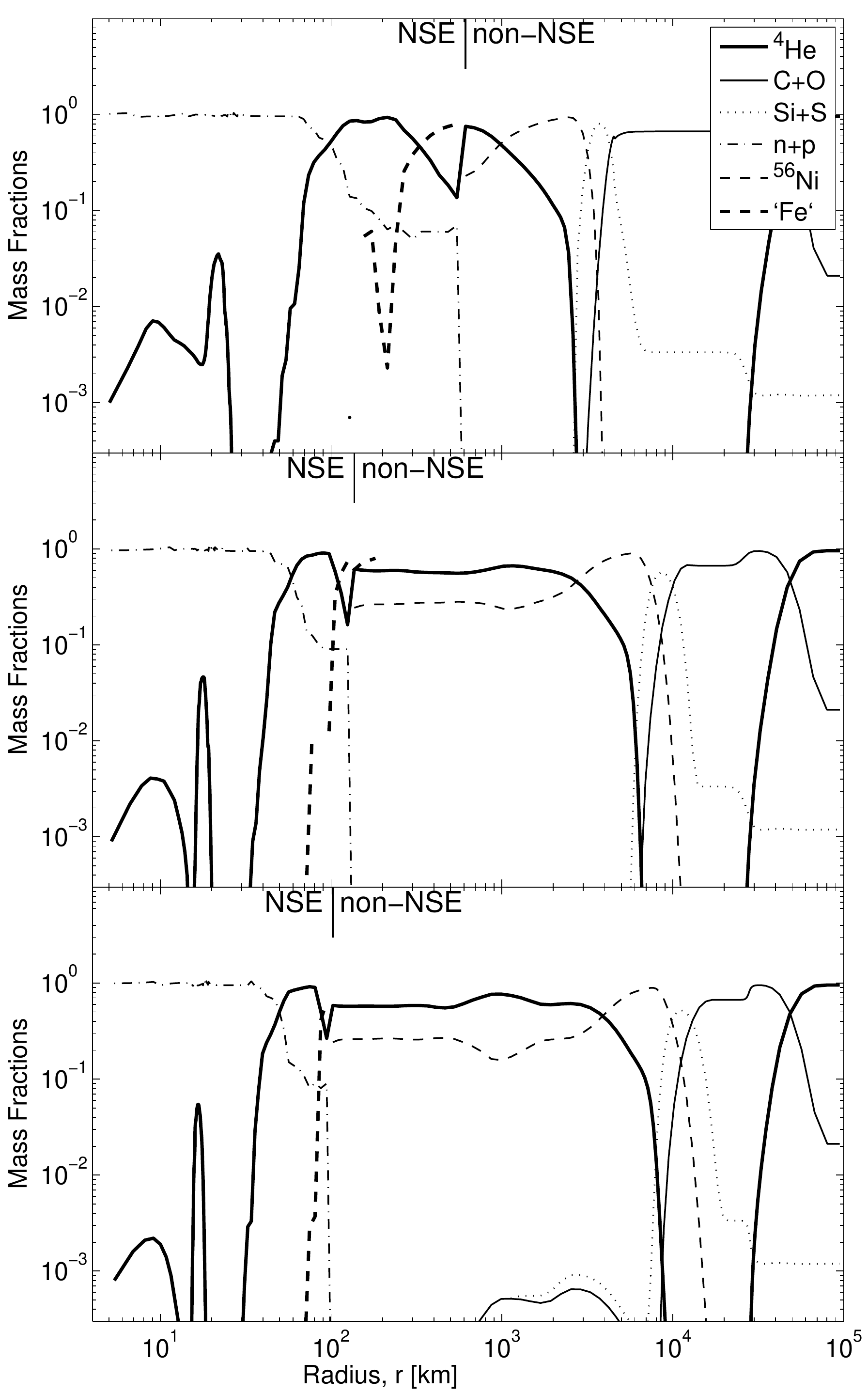}
\caption{Selected radial mass fraction profiles during the neutrino
driven wind phase for the \(10.8\) M\(_\odot\) (left panel) and the
\(18\) M\(_\odot\) (right panel) progenitor model from
\citet{Woosley:etal:2002} at \(1\) second (top), \(3\) seconds (middle)
and \(5\) seconds (bottom) post-bounce.
The vertical lines represent the separation of NSE (EoS for hot and
dense nuclear matter) where heavy nuclei are represented by a single
Fe-group nucleus 'Fe' with average atomic mass and charge
and non-NSE (nuclear reaction network) where the most abundant Fe-group
element is \(^{56}\)Ni, at temperatures of \(\simeq 0.5\) MeV.}
\label{fig-abundance}
\end{figure*}

In the following paragraph, we will discuss the composition
of the neutrino-driven wind region to some extent.
This is possible due to the recently implemented nuclear reaction network.
It includes the free nucleons and the symmetric nuclei from
\(^{4}\)He to \(^{56}\)Ni plus \(^{53}\)Fe, \(^{54}\)Fe and \(^{56}\)Fe.
The initial composition is given by the progenitor model.
Mostly \(^{28}\)Si and \(^{30}\)S are shock-heated and
burned to Fe-group nuclei due to the temperature and density jump
during the initial expansion of the explosion shock (see Fig.~\ref{fig-abundance}
and compare with Figs.~\ref{fig-fullstate-wind-h10a}
and \ref{fig-fullstate-wind-h18b} (d) and (f)).
The high fraction of these Fe-group nuclei reduces behind the
explosion shock due to photodisintegration,
indicated by the region of low density and high entropy
in Figs.~\ref{fig-fullstate-wind-h10a}
and \ref{fig-fullstate-wind-h18b} (b) and (c).
This produces a high fraction of \(\alpha\)-particles,
which in our model represent light nuclei.
The region of \(\alpha\)-particle domination
behind the expanding explosion shock increases with time.
This behavior is illustrated in Fig.~\ref{fig-abundance}
for both Fe-core progenitor models under investigation.
The position of the explosion shock coincides with the maximum of the
mass fraction of Fe-group nuclei (in particular \(^{56}\)Ni).
In addition, density and temperature of the neutrino-driven wind
on top of the PNS surface decrease continuously with time.
The low temperatures and densities in that region do not justify
the assumption of NSE beyond \(\sim1\) second after bounce,
where temperatures reach values below \(0.5\) MeV.
Instead, our nuclear reaction network is used to determine
the composition in that region.
The decreasing density and temperature and the presence of a
high fraction of free nucleons favor the freeze out of light nuclei.
Finally, the entire region between the expanding explosion shock
and the PNS surface is found to be dominated in our simulations by
\(\alpha\)-particles. 
In Fig.~\ref{fig-abundance}, the radii of the NSE to non-NSE transitions
are indicated by vertical lines.
The slight mismatch between the abundances between the heavy 'Fe'-group
nuclei (the representative heavy nucleus with average atomic mass
and charge in NSE) and \(^{56}\)Ni (non-NSE) as well as between the
\(\alpha\)'s is due to the different nuclear models used for the two regimes.
While in NSE the EoS for hot and dense nuclear matter assumes \(^{56}\)Fe
as the most stable nucleus due to the lowest mass per nucleon
for low temperatures and densities,
the nuclear reaction network applied in non-NSE calculates the
composition dynamically based on tabulated reaction rates.
%

\section{Comparison with previous wind studies}

\subsection{The proton-to-baryon ratio of the wind}

The most fundamental approximations made in previous wind studies is the
simplified description of the radiation-hydrodynamics equations,
see for example \citet{Duncan:etal:1986} and \citet{QianWoosley:1996}.
More crucial is the absence of neutrino transport.
Neutrino heating and cooling is calculated based on
parametrized neutrino luminosities and mean energies.
Hence, such models explore the neutrino-driven wind by varying
the neutrino luminosities and energies, where the simplified
radiation-hydrodynamics equations are solved
(see for example \citet{Thompson:etal:2001}).
Since neutrino transport is neglected, the evolution equation for
the electron fraction Eq.~(\ref{eq-ye}) cannot be solved consistently
because the neutrino distribution functions are unknown.
In the following paragraph, we will discuss the assumptions made
for the evolution of the electron fraction in the neutrino-driven wind
which go back to \citet{QianWoosley:1996}.

Applying the theory of weak interactions based on the reaction rates
\(\lambda_{ij}\) for the reaction partners \((i,j)\), i.e. electron and
positron as well as electron neutrino and antineutrino captures,
the evolution equations for the electron and positron fractions
can be written as follows 
\begin{equation}
\frac{dY_{e^{-}}}{dt} =
- \lambda_{e^{-}p} Y_{e^{-}} Y_{p} 
+ \lambda_{\nu_{e}n} Y_{\nu_{e}} Y_{n},
\label{eq-dyedt}
\end{equation}
\begin{equation}
\frac{dY_{e^{+}}}{dt} =
- \lambda_{e^{+}n} Y_{e^{+}} Y_{n} 
+ \lambda_{\bar{\nu}_{e}p} Y_{\bar{\nu}_{e}} Y_{p}.
\label{eq-dypdt}
\end{equation}
These expressions can be combined to calculate the evolution of the
total number of charges, using the relations \(Y_{p}=Y_{e}\) and
\(Y_{n} = 1-Y_{e}\),
\begin{eqnarray*}
\frac{dY_e}{dt} &=&
\frac{d}{dt}\left(Y_{e^-}-Y_{e^+}\right) \\
&=&
  \lambda_{e^{+}n} Y_{e^{+}}
+ \lambda_{\nu_{e}n} Y_{\nu_{e}} \\
&-&
\left (
\lambda_{e^{-}p} Y_{e^{+}} + \lambda_{e^{+}n} Y_{e^{+}} +
\lambda_{\nu_en} Y_{\nu_e} + \lambda_{\bar{\nu}_ep} Y_{\bar{\nu}_e}
\right )
Y_{e},
\end{eqnarray*}
assuming fully dissociated nuclear matter.
This expression is approximated in a crucial but powerful way
(\citet{QianWoosley:1996} Eq.(73)), ignoring contributions from electron
and positron captures as well as the decoupling of radiation from matter
and the angular dependency of the neutrino distribution function on the
distance from the energy-dependent neutrinospheres, as follows
\begin{eqnarray}
Y_{e} &\simeq&
\frac{
  \lambda_{e^{+}n} Y_{e^{+}}
+ \lambda_{\nu_{e}n} Y_{\nu_{e}}
}
{
   \lambda_{e^{-}p} Y_{e^{+}}
+  \lambda_{e^{+}n} Y_{e^{+}}
+  \lambda_{\nu_en} Y_{\nu_e} 
+  \lambda_{\bar{\nu}_ep} Y_{\bar{\nu}_e}
} \\
&\simeq&
\frac{\lambda_{\nu_{e}n} Y_{\nu_{e}}}
{\lambda_{\nu_en} Y_{\nu_e} + \lambda_{\bar{\nu}_ep} Y_{\bar{\nu}_e}}.
\label{eq-yeproxy1}
\end{eqnarray}
This approximation of the electron fraction was further simplified and
expressed in terms of the neutrino luminosities \(L_{\nu}\) and
\(\left<\epsilon_{\nu}\right>\), which is the ratio of mean-square energy
to average energy, and the well known rest mass difference between neutron
and proton \(Q=m_{n}-m_{p}=1.2935\)~MeV, as follows
\begin{equation}
Y_{e} \simeq
\left (
1 +
\frac{L_{\bar{\nu}_{e}}}
{L_{\nu_{e}}}
\frac
{
\left<\epsilon_{\bar{\nu}_{e}}\right> - 
2Q + \frac{1.2\,Q^{2}}{\left<\epsilon_{\bar{\nu}_{e}} \right >}
}
{
\left<\epsilon_{\nu_{e}}\right> + 
2Q + \frac{1.2\,Q^{2}}{\left<\epsilon_{\nu_{e}} \right >}
}
\right )^{-1},
\label{eq-yeproxy2}
\end{equation}
which is used in previous static steady-state and parametrized
dynamic studies of the neutrino-driven wind.
\begin{figure*}[ht]
\centering
  \subfigure[Different electron fraction approximations
  (dash-dotted line: Eq.~\ref{eq-yeproxy1}, dashed line: Eq.~\ref{eq-yeproxy2})
   in comparison with Boltzmann transport (solid line).]{
    \includegraphics[width=0.8\columnwidth]{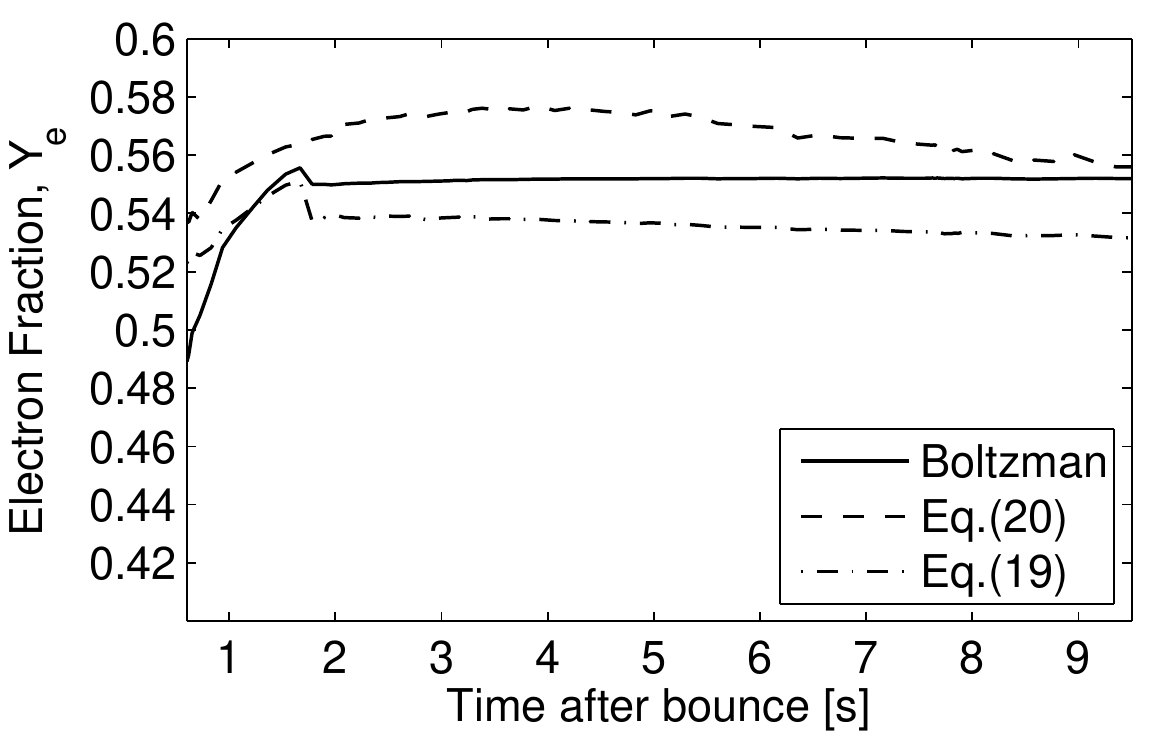}
    \label{fig-yeproxy-10}}
  \hspace{5mm}
  \subfigure[Evolution of the electron fraction according to
    Eq.~(\ref{eq-yeproxy2}). The three lines correspond to different
    assumptions for the electron antineutrino mean energy.]{
    \includegraphics[width=0.8\columnwidth]{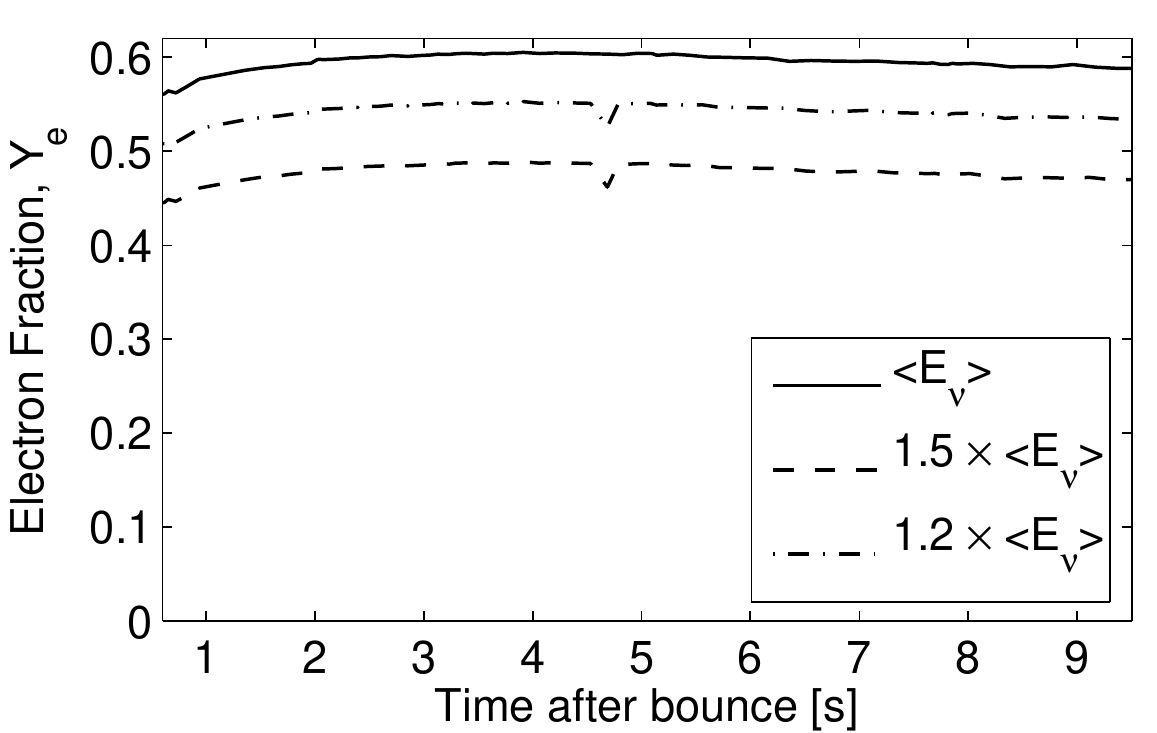}
    \label{fig-yeproxy-10-lumin}}
  \caption{The Electron fraction approximations
  at a distance of \(10\) km outside the electron-neutrinosphere
  for the \(10.8\) M\(_\odot\) progenitor model from \citet{Woosley:etal:2002}.}
\end{figure*}

Fig.~\ref{fig-yeproxy-10} compares the electron fraction behavior at a
distance of \(10\) km outside the electron-neutrinosphere, from Boltzmann
neutrino transport (solid line) with the approximations
Eq.~(\ref{eq-yeproxy1}) based on the neutrino capture rates
(dashed line) and Eq.~(\ref{eq-yeproxy2}) based on the luminosities
and mean neutrino energies (dash-dotted lines).
The approximations are in qualitative agreement with Boltzmann transport.
The differences on the longer timescale are most likely due to the presence
of light and heavy nuclei which are not taken into account explicitly
in the approximations.
They change the number of free nucleons available for the reactions in 
Eq.~(\ref{eq-dyedt}) and (\ref{eq-dypdt}).
All descriptions agree qualitatively in the prediction of a generally
proton-rich material in the wind, based on the neutrino spectra
obtained via Boltzmann transport.

\subsection{The neutrino observables in the wind}

\begin{figure*}[ht]
\begin{center}
\includegraphics[width=0.65\columnwidth]{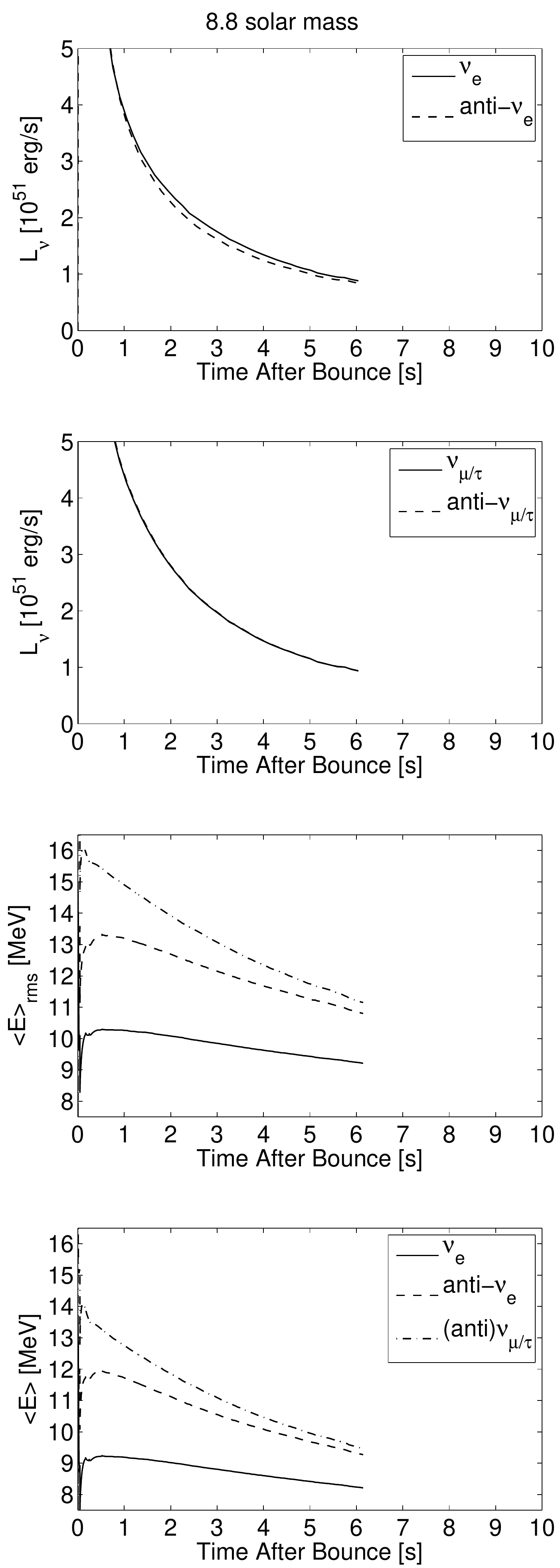}
\includegraphics[width=0.65\columnwidth]{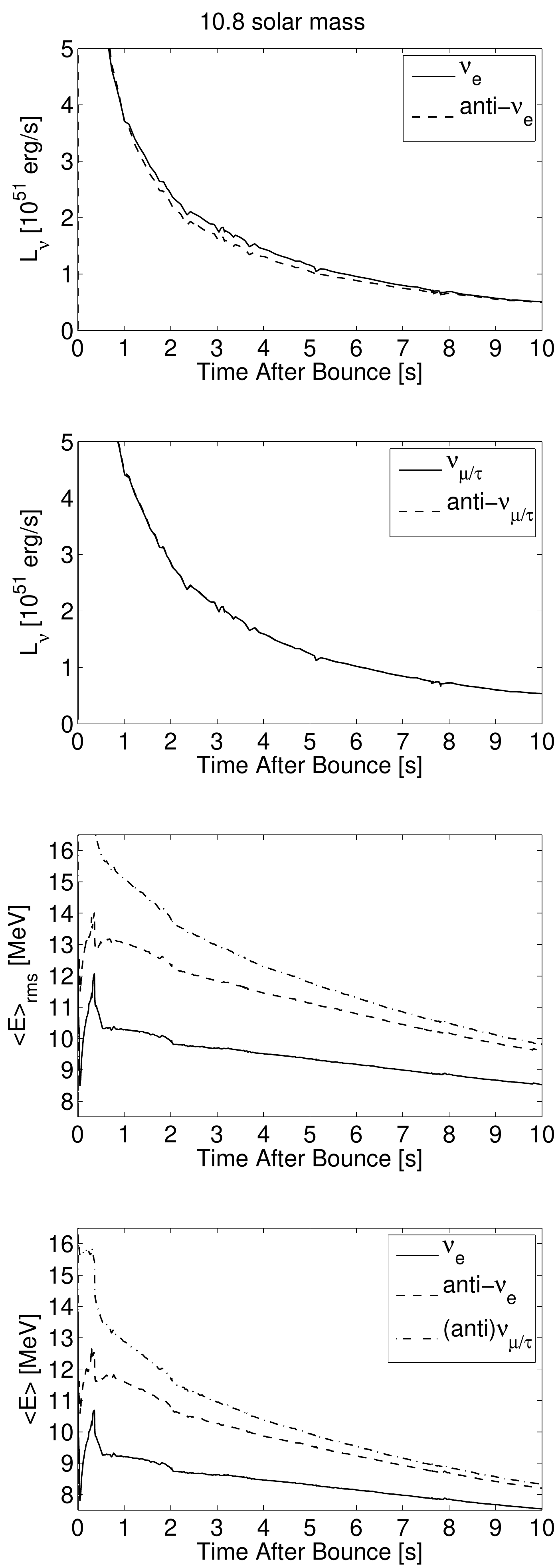}
\includegraphics[width=0.65\columnwidth]{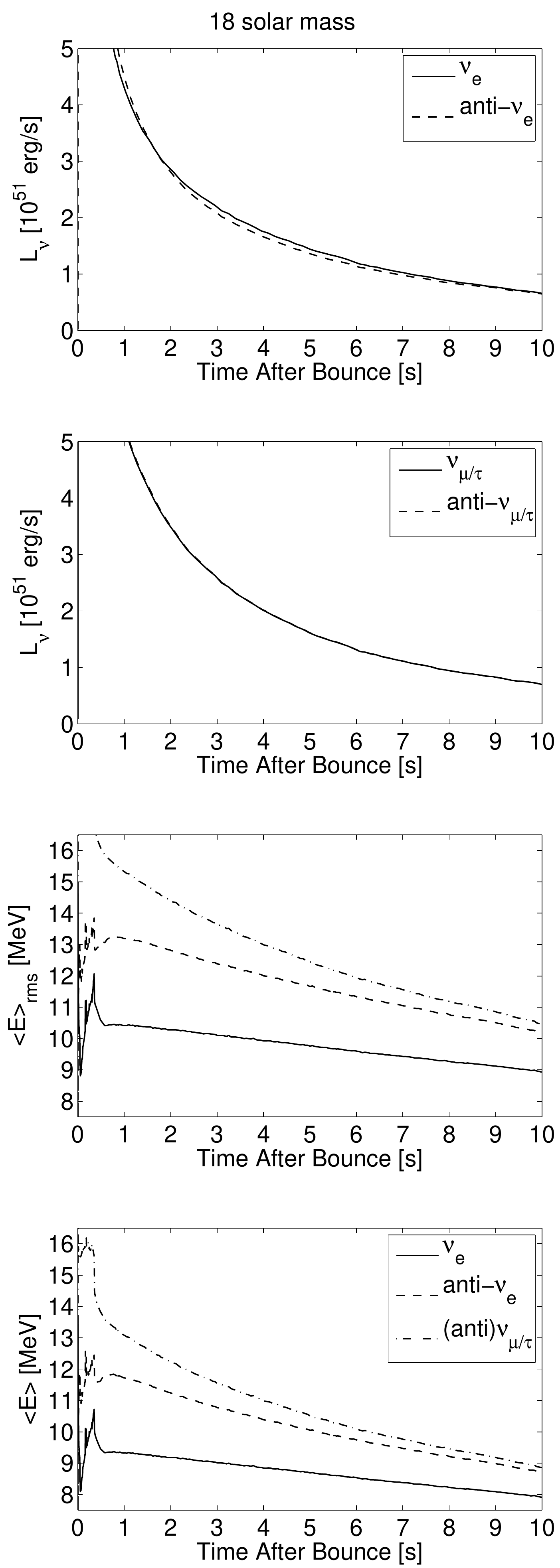}
\caption{Neutrino luminosities and mean energies with respect to time
after bounce for the \(8.8\) M\(_\odot\) O-Ne-Mg-core from
Nomoto~(1983,1984,1987) (left panel) and the \(10.8\) M\(_\odot\) (middle panels)
and \(18\) M\(_\odot\) (right panel) Fe-core progenitor models from
\citet{Woosley:etal:2002}, measured in the co-moving reference frame
at a distance of \(500\) km.}
\label{fig-lumin-n08c-h10a-h18b}
\end{center}
\end{figure*}

Comparing the neutrino spectra in Fig.~\ref{fig-lumin-n08c-h10a-h18b} with the
spectra assumed in previous static steady-state and dynamic wind studies
(see for example \citet{Thompson:etal:2001} and \citet{Arcones:etal:2007}),
we find two major differences:
{\em One}, the neutrino luminosities and mean neutrino energies
assumed are significantly higher than those we find and
{\em two}, the assumed behavior with respect to time is different.

The commonly used assumptions made in static steady-state and parametrized
dynamic wind studies go back to the detailed investigation from
\citet{Woosley:etal:1994}, who performed core collapse simulations based on
sophisticated input physics.
They investigated the neutrino-driven explosion of a \(20\) M\(_\odot\)
progenitor star and followed the evolution for \(18\) seconds post-bounce
into the neutrino-driven wind phase.
In their simulations the electron (anti)neutrino luminosities decreased from 
initially \(4\times 10^{52}\) (\(3\times 10^{52}\)) erg/s at the onset of
the explosion to \(6\times 10^{51}\) (\(7.5\times 10^{50}\)) erg/s
at \(10\) seconds after bounce, where strictly
\(L_{\bar{\nu}_e}>L_{\nu_e}\) after the onset of the explosion.
The difference between the neutrino and antineutrino luminosities remained
small and constant with respect to time up to \(3\) seconds post-bounce
and increased only significantly after \(4-5\) seconds post-bounce,
after which the difference reached its maximum of \(1.5\times10^{50}\) erg/s
at the end of the simulation at about \(18\) seconds post-bounce.
The electron flavor neutrino luminosities in our models follow
a different behavior.
They reach \(1\times 10^{51}\) erg/s at about \(5\), \(6\) and \(8\) seconds
post-bounce for the \(8.8\), \(10.8\) and \(18\) M\(_\odot\)
progenitor models respectively.
The higher electron flavor neutrino luminosities for the more massive
progenitors are in correlation with the more massive PNSs and the hence
larger number of neutrinos emitted.
However, the difference between electron-neutrino and electron-antineutrino
luminosities found in the present investigation is significantly lower
than the difference in \citet{Woosley:etal:1994}.
During the initial explosion phase until about \(300\) ms after the onset of
the explosion, the electron antineutrino luminosity is slightly higher than
the electron neutrino luminosity by about \(1\times 10^{50}\) erg/s which in
our models explains the electron fraction of \(Y_e>0.5\) of the
early explosion ejecta.
After about \(900\) ms post-bounce, the luminosities can hardly be distinguished
where during the initial neutrino-driven wind phase after about \(1\) second
after bounce the electron neutrino luminosity becomes higher than the electron 
antineutrino luminosity by about \(1\times 10^{50}\) erg/s.
This difference reduces again at later times at about \(6\) seconds post-bounce
and the electron flavor neutrino luminosities become more and more similar
(see Fig.~\ref{fig-lumin-n08c-h10a-h18b}).

Even more different are the values and the behavior of the mean neutrino
energies, see Fig.~\ref{fig-lumin-n08c-h10a-h18b} and compare with
Fig.~2 of \citet{Woosley:etal:1994}.
They found (\(\mu/\tau\))-neutrino energies of about \(35\)~MeV
which remained constant with respect to time.
Their electron-antineutrino energies increased slightly from about \(20\)~MeV
to \(22\)~MeV where the electron-neutrino energies decrease from
\(14\)~MeV to \(12\)~MeV.
This increasing difference between the electron neutrino and antineutrino
spectra favored neutron-rich material, which was consistent with their
findings of \(Y_e<0.5\) for the material ejected in the neutrino-driven
wind in \citet{Woosley:etal:1994}.
We cannot confirm these results for the mean neutrino energies
nor the evolution of the spectra.
In contrast, all mean neutrino energies decrease with respect to time
for all our models.
This is a consequence of lepton number and energy loss of the central
PNS where the neutrinos diffuse out.
The electron (anti)neutrino energies decrease from about
\(10\)~(\(12\))~MeV at the onset of the explosion to about
\(8.5\)~(\(9\))~MeV and the (\(\mu/\tau\))-neutrino energies
decrease from \(16\)~MeV to \(10\)~MeV at the end of the simulations.
Hence, not only the mean energies decrease also the difference between
the electron flavor neutrino spectra decreases.
The reason for the neutrino spectra to become more similar with respect to
time is related to the evolution of the thermodynamic properties at the
neutrinospheres, and will be discussed in the following subsection.

\subsection{The PNS contraction}

\begin{figure*}[ht]
\begin{center}
\includegraphics[width=0.95\textwidth]{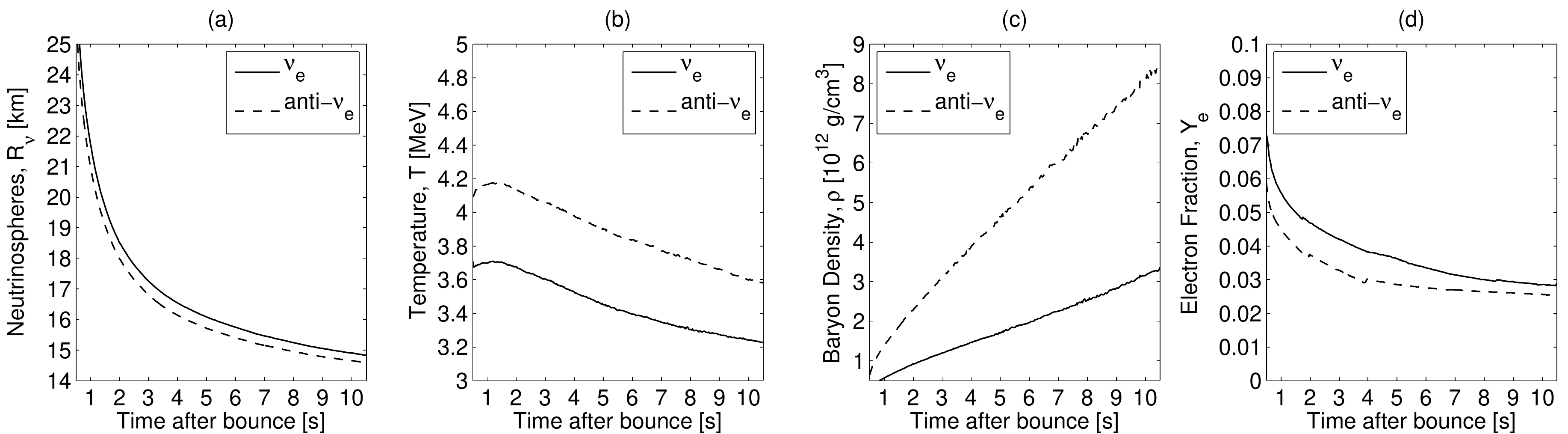}
\caption{Evolution of the neutrinosphere radii in graph (a) and temperature
and density at the corresponding neutrinospheres in graphs (b) and (c)
respectively for the \(10.8\)~M\(_\odot\) progenitor model from
\citet{Woosley:etal:2002}.}
\label{fig-nuspheres-h10a-wind}
\end{center}
\end{figure*}

The behavior of the neutrino spectra and hence the evolution and the
properties of the neutrinospheres is related to the PNS contraction.
The contraction is caused by a continuous deleptonization and translates
to a continued steepening of the density gradient at the PNS surface.
Hence, the neutrinosphere radii for the electron flavor neutrinos
move closer together with time.
The evolution of the neutrinosphere radii for both electron neutrino
and antineutrino are illustrated in Fig.~\ref{fig-nuspheres-h10a-wind}
(a) for the \(10.8\)~M\(_\odot\) progenitor model.
Their difference reduces from \(740\)~m at about \(1\)~second post-bounce
to \(370\) m at about \(5 \)~seconds post-bounce and further to
\(260\)~m at about \(10\)~seconds post-bounce.

\begin{figure*}[ht]
\begin{center}
\includegraphics[width=0.95\textwidth]{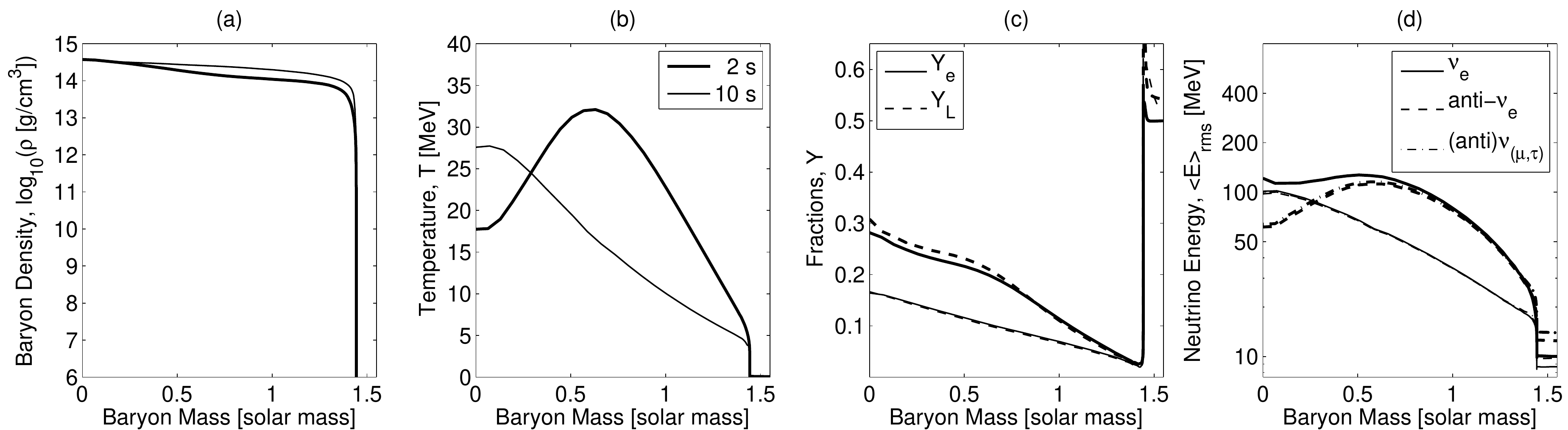}
\caption{Radial PNS profiles at two different post-bounce times
(thick lines: 2 seconds, thin lines: \(10\) seconds),
for the \(10.8\)~M\(_\odot\) progenitor model from
\citet{Woosley:etal:2002}.}
\label{fig-delept-h10a}
\end{center}
\end{figure*}

This contraction behavior has consequences for the neutrino spectra,
which are determined during the neutrino-driven wind phase by
diffusion rather than by mass accretion.
Hence, the electron flavor neutrino luminosities can be determined as follows
\begin{equation}
L_\nu = \frac{1}{4}\,4\pi\,r^2\,u_\nu\,\vert_{R_{\nu}},
\end{equation}
where \(u_\nu\propto T^4\) is the thermal black body spectrum for
ultra-relativistic fermions with temperature \(T\).
The matter temperatures at the neutrinospheres decrease with
respect to time as shown in Fig.~\ref{fig-nuspheres-h10a-wind} (b).
This is due to the continued loss of lepton number and energy,
carried away by the diffusing neutrino radiation field as
illustrated in Fig.~\ref{fig-delept-h10a} for the \(10.8\) M\(_\odot\)
from \citet{Woosley:etal:2002}.
The lepton number decreases from \(Y_L\simeq0.3\) at \(2\) seconds post
bounce to \(Y_L\simeq0.18\) at \(10\) seconds post-bounce,
see Fig.~\ref{fig-delept-h10a} (c).
The additionally reduced mean neutrino energies (on average), from
\(\left<E\right>_\text{rms}\simeq150\) MeV to
\(\left<E\right>_\text{rms}\simeq50\) MeV (see Fig.~\ref{fig-delept-h10a}
(d)), and the consequent reduced temperature-peak inside the PNS
(see Fig.~\ref{fig-delept-h10a} (b)) cause the contraction of the outer
layers of the PNS.
This can be identified via the density increase in
Fig.~\ref{fig-delept-h10a} (a).
Note in addition to the maximum temperature decrease at the outer layers
of the PNS, the central temperature increases from \(18\) MeV to \(23\)
MeV on the post-bounce times between \(2\) and \(10\) seconds.
This is caused by the contraction of the deleptonizing outer layers
of the PNS which compresses the central part.
The evolution of the reducing temperature at the neutrinospheres is shown
in Fig.~\ref{fig-nuspheres-h10a-wind} (b).
In combination with the loss of leptons number, it explains the decreasing
electron flavor neutrino luminosities and mean neutrino energies with
respect to time.
Furthermore, the temperature difference decreases with respect to time
from \(0.467\)~MeV at about \(1\)~second post-bounce to \(0.362\)~MeV
at about \(10\)~seconds post-bounce.
Consequently the neutrino spectra become more similar with respect to time.
As illustrated in Fig.~\ref{fig-lumin-n08c-h10a-h18b},
the difference in the electron flavor neutrino luminosities and
mean neutrino energies decreases for all models under investigation.
In addition, Fig.~\ref{fig-nuspheres-h10a-wind} (c) and (d) illustrate
the evolution of the baryon density and the electron fraction at the
corresponding neutrinospheres.
It becomes additionally clear from the electron fraction approximation
Eq.(~\ref{eq-yeproxy2}), that it is not the absolute values for the mean
neutrino and antineutrino energies that determine whether matter becomes
neutron- or proton-rich but their ratio.

Since this difference is small in our simulations, with initially 
\(\left<E_{\nu_{e}}\right>_{\text{rms}}\simeq 10\)~MeV 
and \(\left<E_{\bar{\nu}_{e}}\right>_{\text{rms}}\simeq 13\)~MeV
(at about \(1\)~second post-bounce)
and only \(\left<E_{\nu_{e}}\right>_{\text{rms}}\simeq 9\)~MeV
and \(\left<E_{\bar{\nu}_{e}}\right>_{\text{rms}}\simeq 11\)~MeV
(at later times at \(10\) seconds post-bounce),
the values found for the electron fraction of \(Y_{e}>0.5\)
(solid line in Fig.~\ref{fig-yeproxy-10-lumin} for the \(Y_e\)-approximation
based on the luminosity and mean neutrino energies)
clearly illustrate that the accelerated matter in the neutrino
driven wind stays proton-rich for more than \(10\)~seconds.
This is in qualitative agreement with Boltzmann transport
as discussed above and shown in Fig.~\ref{fig-yeproxy-10}.
Hence we find Eq.(~\ref{eq-yeproxy2}) to be a good approximation
to model the electron fraction in the wind.
On the other hand, most of the previous studies select the neutrino
luminosities and mean energies to investigate a neutron-rich neutrino
driven wind.
In order to test the appearance of \(Y_{e}<0.5\) under such conditions,
we increase the difference between the mean neutrino and antineutrino
energies by hand.
We evaluate expression (\ref{eq-yeproxy2}) shown in
Fig.~\ref{fig-yeproxy-10-lumin} at \(10\)~km outside the
electron-neutrinosphere for \(1.2\) (dashed line) and \(1.5\)
(dash-dotted line) times larger electron-antineutrino mean energies.
For the first value, \(Y_e\) decreases but matter remains slightly
proton-rich, where for the latter value matter becomes neutron-rich.
Indeed, the larger the difference between neutrino and antineutrino
spectra are, the lower becomes the electron fraction in the wind.
Note that the luminosities and electron-neutrino energies
remained unmodified for this experiment.
Such an increase of the energy difference between neutrinos and
antineutrinos could perhaps be related to the uncertainty of the EoS
for nuclear matter, which will be discussed in the following paragraph.

The assumed PNS radii in previous wind studies reach about \(10\)~km
shortly (\(\leq 1\) second) after the onset of the explosion.
We define the radius of the PNS as the position of the
electron-neutrinosphere at the steep density gradient at the PNS surface.
The approximated inner boundary of the physical domain in most wind
models is close to but still inside this radius.
The position of the neutrinospheres and the contraction of the PNSs
found in the present paper differ significantly from the assumptions made
in most previous wind studies.
We find PNS radii of about \(40\)~km at the time of the explosion
and \(20\)~km at about \(2\)~seconds after bounce
(see Fig.~\ref{fig-nuspheres-h10a-wind} (a)).
During the later evolution, the PNS contraction slows down.
The PNS profile and hence the position of the neutrinospheres as well as
the contraction behavior itself is given implicitly by the EoS for
hot and dense nuclear matter as well as the PNS deleptonization.
For the stiff EoS from \citet{Shen:etal:1998} and both the
\(10.8\) and \(18\)~M\(_\odot\) progenitors, the PNSs reach radii
of \(14.5-15\)~km only at about \(10\)~seconds after bounce
(see Fig.~\ref{fig-nuspheres-h10a-wind}).
The larger radii of the neutrinospheres result in lower neutrino
luminosities and mean energies and a lower difference between
neutrino and antineutrino spectra in comparison to the assumptions
made in most previous wind models.
This is in agreement with \citet{Arcones:etal:2007} who additionally
assume PNS radii of \(15\)~km and find conditions that differ more
from previous wind studies.
They obtained significantly higher values for the electron fraction.
To summarize, this effects and the different behavior of the neutrino 
spectra assumed in the previous wind studies leads to different
matter properties of the neutrino-driven wind.
A detailed comparison study of fast and slow contracting PNSs
with respect to the neutrino-driven wind, e.g. applying EoSs with
different compressibilities and asymmetry energies,
would be necessary in the context of radiation hydrodynamics
simulations using spectral Boltzmann neutrino transport.
\begin{SCfigure*}
\centering
\includegraphics[width=1.5\columnwidth]{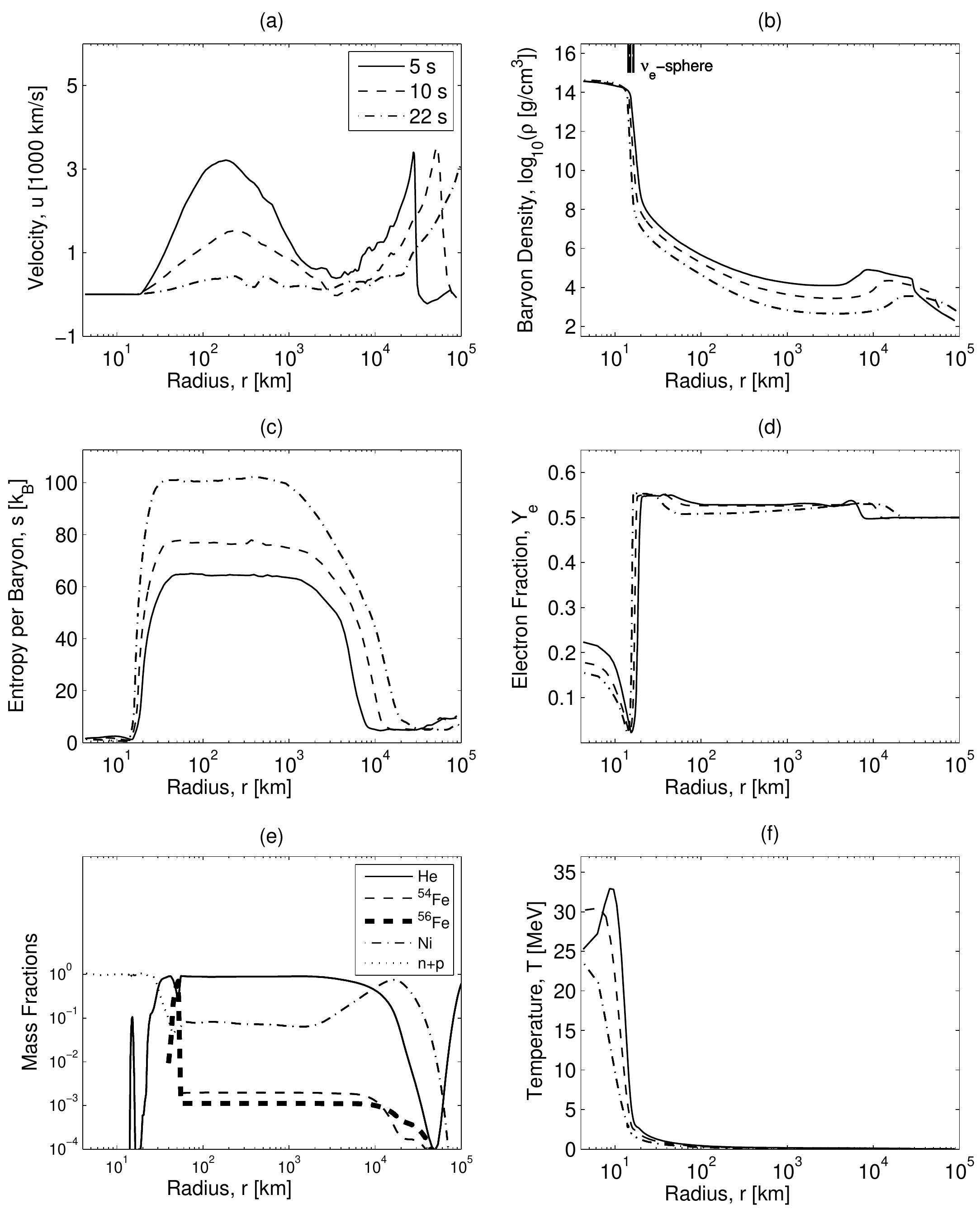}
\caption{Radial profiles of selected hydrodynamic variables
for the \(18\) M\(_\odot\) progenitor model at three different post-bounce times,
illustrating the disappearance of the neutrino-driven wind and the PNS
cooling and contraction.
Graph (e) illustrates the composition at \(22\)~seconds post-bounce.}
\label{hydroplot-postwind-h18b}
\end{SCfigure*}
%

\section{Long term post-bounce evolution}

During the neutrino-driven wind phase, the neutrino luminosities and mean
neutrino energies decrease continuously, which leads to a constant decrease
in the net-heating rates.
At luminosities below \(10^{51}\) erg/s (see Fig.~\ref{fig-lumin-n08c-h10a-h18b}),
the supersonic matter outflow for the \(10.8\) M\(_\odot\) progenitor model
descends into a subsonic expansion.
The wind termination shock turns again into a subsonic neutrino-driven wind.
At later times, the neutrino-driven wind settles down to a quasi-stationary
state with no significant matter outflow,
illustrated at the example of the \(18\) M\(_\odot\) progenitor model
in Fig.\ref{hydroplot-postwind-h18b} (a).
The explosion shock continues to expand and the material enclosed inside
the mass cut accretes onto the PNS at the center.
In combination with the deleptonization, this leads to the
continuous PNS contraction.
However, the contraction proceeds on a timescale of seconds and hence
the PNS can be considered in a quasi-stationary state.
The dense and still hot and lepton-rich PNS at the center is surrounded
by a low density and high entropy atmosphere, composed of light and heavy nuclei.
See for example the abundances of the \(18\) M\(_\odot\) progenitor
for the post-bounce time of \(22\) seconds in
Fig.~\ref{hydroplot-postwind-h18b} (e).
The region at sub-saturation densities where light nuclei are present
belongs to the inhomogeneous matter phase where clusters, known as pasta-
and spaghetti-phases, are predicted to dominate the EoS.
However, the EoS from \citet{Shen:etal:1998} approximates these effects
by the presence of light nuclei represented in our model by
\(\alpha\)-particles.

The internal temperature profile of the PNS is not constant.
The central region of the PNS did not experience shock heating immediately
after the Fe-core bounce, since the initial shock forms at the edge of the
bouncing core.
Its mass scales roughly with \(Y_{e}^2\) and is typically around values
of \(0.5-0.6\) M\(_\odot\) for low- and intermediate-mass Fe-core progenitors.
Hence, the central temperature after bounce is given by the thermodynamic
conditions at bounce.
The temperature changes only during the post-bounce evolution due to
compressional heating and the diffusion of neutrinos.
The shock heated material inside the PNS shows significantly
higher temperatures than at the center.
The temperature decreases again towards the PNS surface where the matter
is less dense (for the illustration of the radial temperature profile inside the
PNS as well as the dynamical evolution of temperature and density,
see Fig.\ref{hydroplot-postwind-h18b} (f) and (b) at selected
post-bounce times between \(5-22\)~seconds).
The neutrinos diffuse continuously out of the PNS and carry away energy.
The central electron fraction reduces from \(Y_e \simeq 0.25\) at the
onset of the explosion to \(Y_e \simeq 0.15\) at \(22\)~seconds after
bounce (see Fig.\ref{hydroplot-postwind-h18b} (d)).
It relates to a temperature decrease from about \(35\)~MeV initially
(at \(3\)~seconds post-bounce) to \(23\)~MeV at about \(22\)~seconds post-bounce. 
This corresponds to the initial and neutrino dominated cooling phase.
Unfortunately the achieved temperatures are not representative since
important neutrino reactions, such as the direct and modified Urca processes,
are not yet taken into account.
\begin{figure*}[ht]
\begin{center}
\includegraphics[width=1.95\columnwidth]{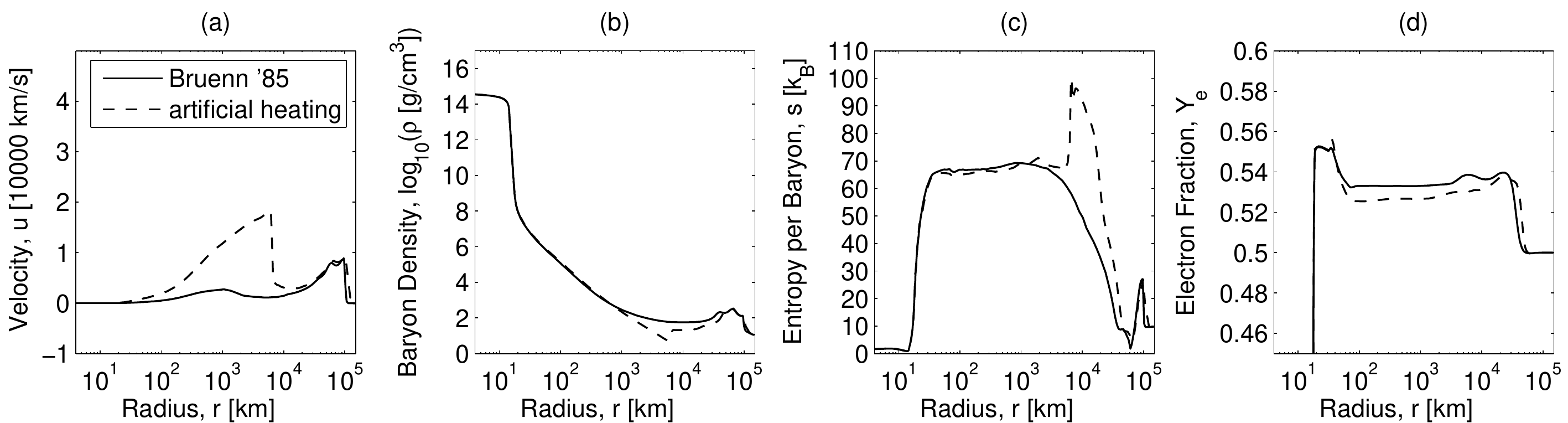}\\
\includegraphics[width=1.95\columnwidth]{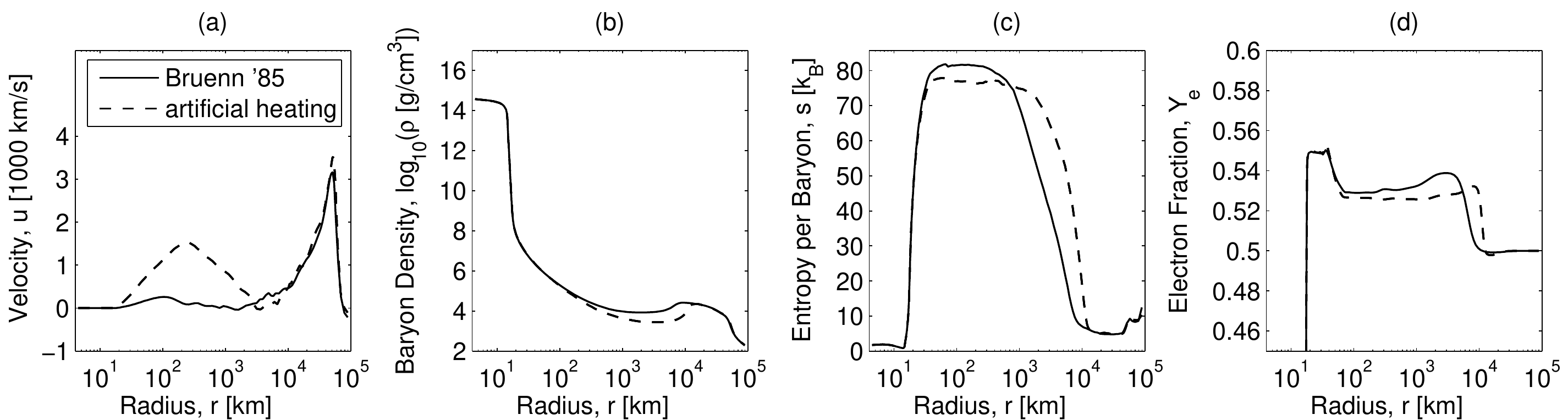}
\caption{Comparing selected hydrodynamic variables using the standard
reactions rates based on \citet{Bruenn:1985} (solid lines) and the
artificially enhanced rates (dashed lines) for the \(10.8\) M\(_\odot\)
(top) and the \(18\) M\(_\odot\) progenitor model (bottom).}
\label{fig-heating}
\end{center}
\end{figure*}
%

\section{Discussion}

The neutrino-driven wind was found to occur in all three progenitor models
under investigation,
the \(8.8\) M\(_\odot\) O-Ne-Mg-core and the
\(10.8\) and \(18\) M\(_\odot\) Fe-core progenitor models.
Because the neutrino-driven explosions for the Fe-core progenitors are
launched using artificially enhanced neutrino reaction rates, one may ask
about the impact of these modified rates to the neutrino-driven wind.
Therefore we performed additional runs for which we switch back to the
standard opacities given in \citet{Bruenn:1985} after the explosions
have been launched.
The times when we switch back is about \(500\) ms after bounce,
chosen such that the dynamics of the explosion ejecta does not change
anymore significantly due to neutrino heating.
However, the lower opacities translate to a significantly lower net-heating
by a factor of \(5-6\) in the region on top of the PNS where the neutrino-driven
wind develops.
The energy deposition is still sufficient to drive the neutrino-driven wind
but the matter velocities are lower by a factor of \(2-5\) in comparison
to the wind velocities using the enhanced reaction rates
(see Fig.~\ref{fig-heating} (a)).
The main effect of the artificially enhanced reaction rates and the hence
increased neutrino heating to the dynamics is clearly the stronger
neutrino-driven wind.
For the \(10.8\) M\(_\odot\) progenitor model and with the enhanced heating,
the wind even develops supersonic velocities (as discussed above in \S~4)
in Fig.~\ref{fig-heating} (a) (top panel).
The supersonic wind collides with the explosion ejecta where matter decelerates
and hence the reverse shock forms, which additionally increases the
entropy in the wind (see Fig.~\ref{fig-heating}(c) (top panel)).
This additional entropy increase is absent in the simulations using the
standard opacities, where the wind stays subsonic.
This is also the case for the \(18\) M\(_\odot\) progenitor model
(Fig.~\ref{fig-heating}, bottom panel), with and without the enhanced
opacities.
The neutrino-driven wind of the O-Ne-Mg-core is illustrated in
Fig.~\ref{fig-shellplot-n08c} using the standard rates based on
\citet{Bruenn:1985}.
The formation of a supersonic neutrino-driven wind could be confirmed
including the formation of the wind termination shock.
Hence, one may speculate whether only low-mass progenitors develop strong
neutrino-driven winds, while for more massive progenitors the influence
of the winds to the matter properties of the ejecta becomes small.
The progenitor dependency of the neutrino-driven wind is related to
the density of the envelope surrounding the PNS after the explosion
has been launched, which is significantly higher for more massive
progenitors and hence the neutrino-driven wind is weaker.

The agreement of the time evolution of the mean neutrino energies between all
three progenitor models under investigation (using the enhanced and standard
opacities) in Fig.~\ref{fig-lumin-n08c-h10a-h18b} is striking.
The impact of the artificial heating to the neutrino observables
and hence to the electron fraction in the wind is less pronounced.
The influence on the composition of the wind is illustrated via the
electron fraction in Fig.~\ref{fig-heating} (d).
Using the standard rates, the wind stays slightly more proton-rich.
Increasing the charged current reaction rates allows \(\beta\)-equilibrium
to be established on a shorter timescale.
In addition, matter stays slightly more proton-rich for the
the less intense neutrino-driven wind, which develops for the
Fe-core progenitors using the standard neutrino opacities.
The additional electron fraction decrease in the neutrino-driven wind
for the models using the enhanced neutrino reactions is found due to
the higher degeneracy obtained in the stronger deceleration behind the
explosion ejecta, and is therefore a dynamic effect.
However, the findings of generally proton-rich ejecta as well as the
generally proton-rich neutrino-driven wind does not change.
The corresponding densities and entropies per baryon in the wind
are shown in Fig.~\ref{fig-heating} (b) and (c).
The effects of the artificial heating are slightly lower entropies per baryon.
The higher matter outflow velocities in the wind region using the artificial
heating results additionally in lower densities, shown in Fig.~\ref{fig-heating} (b).

The artificially increased charged current reaction rates cannot be
justified by physical uncertainties of the rates themselves.
Similar to the high luminosities assumed in \citet{Arcones:etal:2007},
they could be seen as a lowest order attempt to take the effects
of multi-dimensional phenomena into account.
For example, known fluid instabilities increase the neutrino energy
deposition efficiency.
Convection allows matter to stay for a longer time in the neutrino heating
region (see \citet{Herant:etal:1994}, \citet{JankaMueller:1996}).
Present axially symmetric core collapse models of massive Fe-core
progenitor stars (even non-rotating) predict bipolar explosions
(see \citet{Janka:etal:2008a}).
The deviation from a spherical description and hence the deformation
of the SAS due to fluid instabilities takes place during the neutrino
heating phase on a timescale of several \(100\) ms after bounce.
In multi-dimensional models, the luminosities are powered by a
significantly higher mass accretion because the up-streaming neutrino
heated matter is accompanied by large down-streams of cold material.
These higher luminosities may power a strong (even supersonic) neutrino-driven
wind behind the explosion ejecta, while the neutrino-driven wind may remain
absent in the angular directions of the accreting material which will
not be ejected.
\begin{figure*}[htp!]
\centering
\includegraphics[width=1.7\columnwidth]{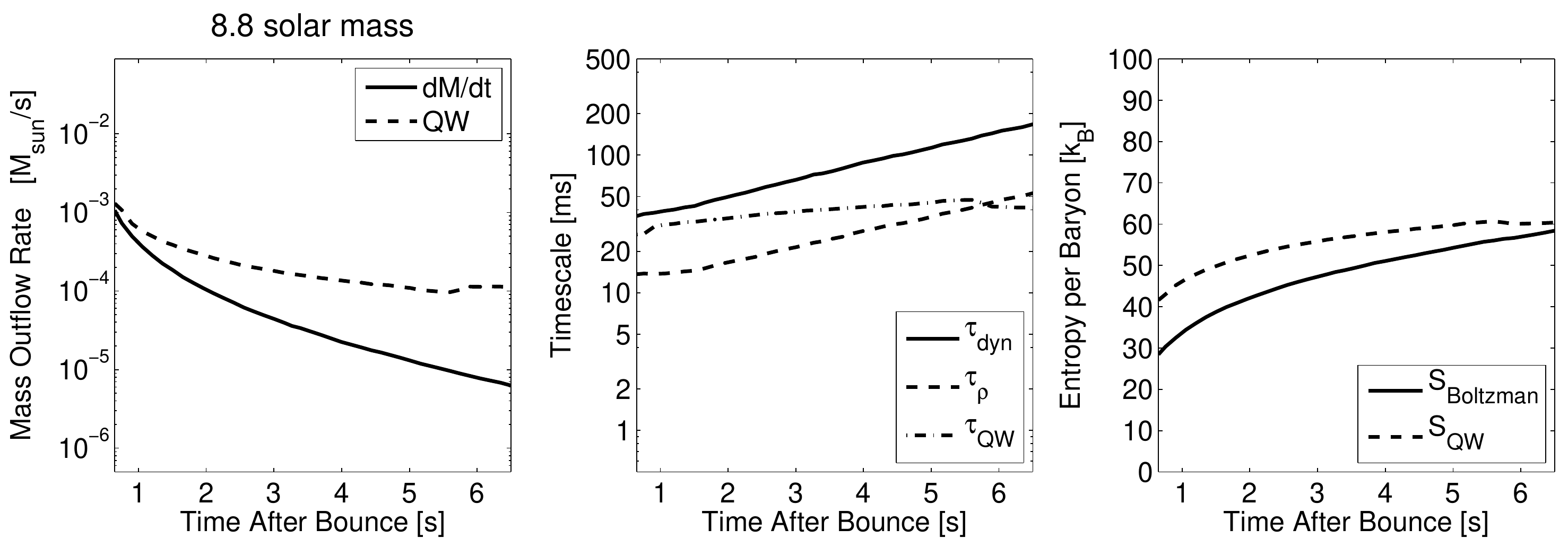}\\
\includegraphics[width=1.7\columnwidth]{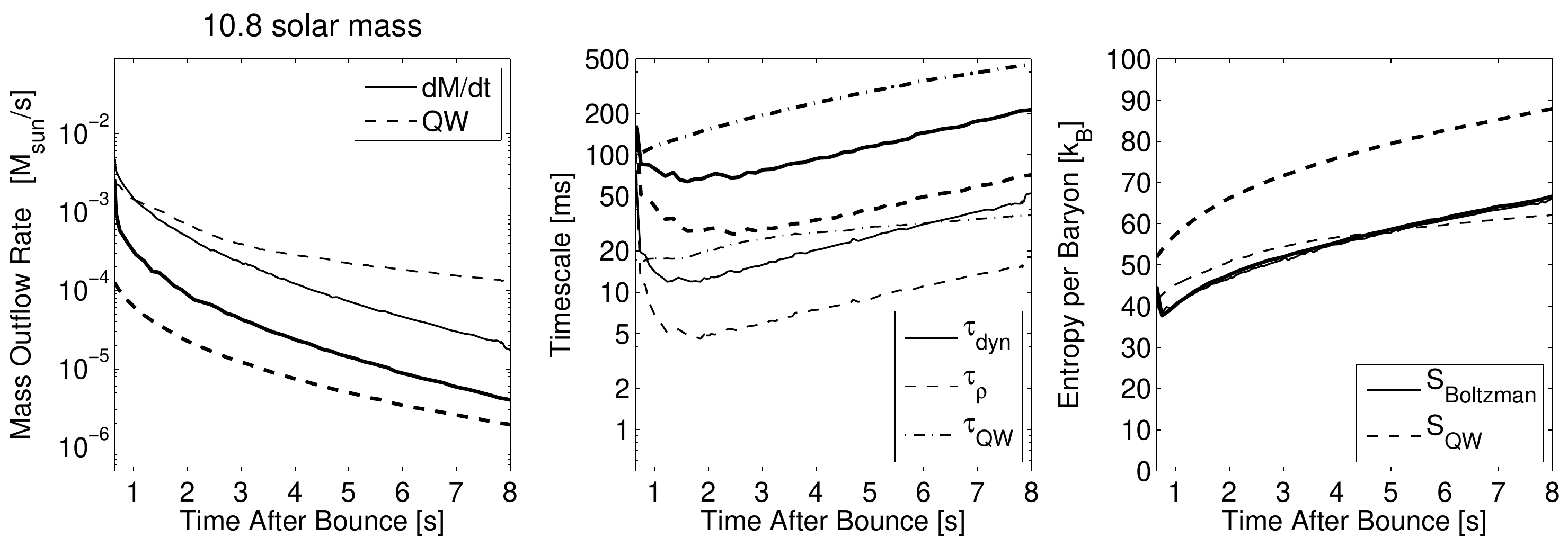}\\
\includegraphics[width=1.7\columnwidth]{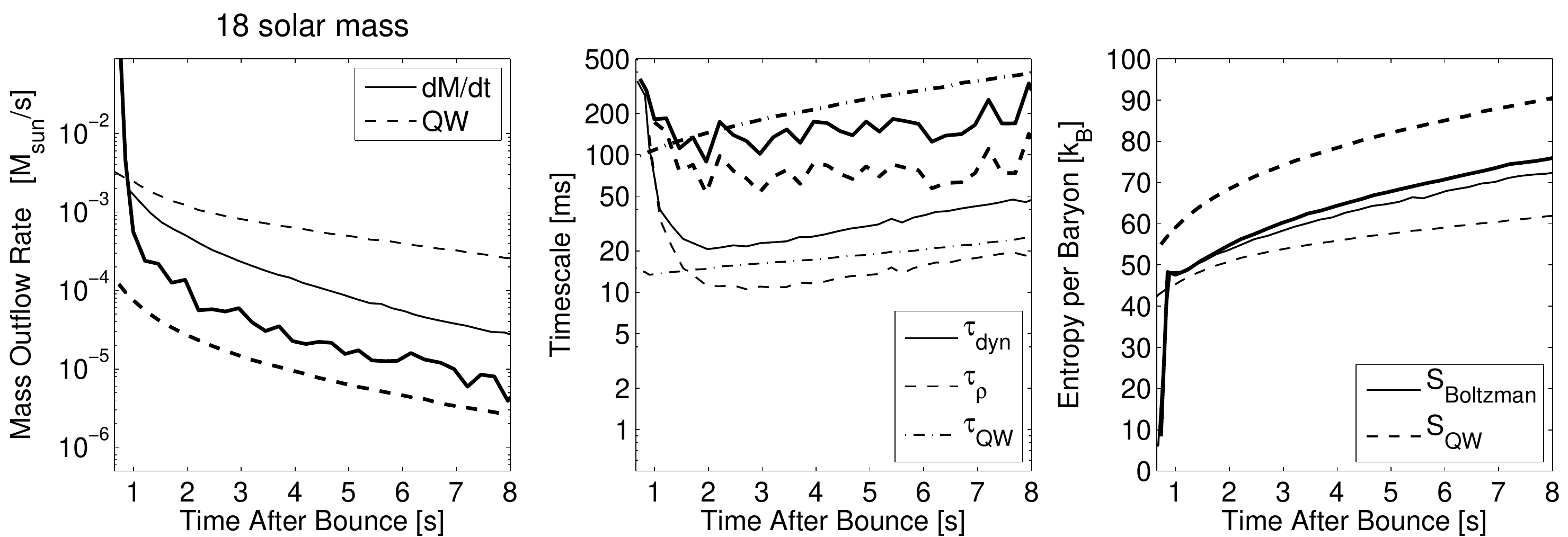}
\caption{Mass accretion rate, timescales and entropy in the wind
for the three progenitor models under investigation,
\(8.8\) M\(_\odot\) (top), 
\(10.8\) M\(_\odot\) (middle)
\(18\) M\(_\odot\) (bottom).
The thick lines show data from simulations using the standard rates from
\citet{Bruenn:1985}, where a relatively weak neutrino-driven wind was
obtained, and the thin lines show data from simulations using the
enhanced rates.}
\label{fig-accretion}
\end{figure*}

Previous wind models have long been investigated as a possible site for
the production of heavy elements via the \(r\)-process, motivated by the
expectation of the ejection of neutron-rich material, the high entropies per
baryon in the neutrino-driven wind and the short timescale of the neutrino-driven
wind expansion (see \citet{Hoffman:etal:2007}, \citet{PanovJanka:2009}
and references therein).
The relevant quantities are shown in Figs.~\ref{fig-shellplots-h10a}
and \ref{fig-shellplot-n08c} for the \(10.8\) M\(_\odot\) and
\(8.8\) M\(_\odot\) progenitor models respectively.
Illustrated are several selected mass shells that are part of the region
where the neutrino-driven wind develops in our radiation hydrodynamics
model based on spectral three-flavor Boltzmann neutrino transport.
The inclusion of neutrino transport in a dynamical model is essential in
order to obtain consistent neutrino spectra which determine
the evolution of the electron fraction and the PNS contraction due to
deleptonization and mass accretion.
In comparison to previous static steady-state and dynamic wind models
- where these ingredients were assumed - we confirm several properties
of the accelerated material in the neutrino-driven wind,
such as the fast expansion and the high matter outflow rate
shown in Fig.~\ref{fig-accretion}, the high velocities in the
Figs.~\ref{fig-shellplots-h10a} (f) and \ref{fig-shellplot-n08c} (f)) and the
rapidly decreasing density and temperature of the accelerated material in
Fig.~\ref{fig-shellplots-h10a} (b) and (c) respectively.
The expansion timescale in Fig.~\ref{fig-accretion} is given by the following
expression
\begin{equation*}
\tau_\text{dyn} = \left. \frac{r}{v}\right\vert_{T=0.5\,\text{MeV}},
\end{equation*}
evaluated at the surface of constant temperature of \(T=0.5\) ~MeV,
compared with an alternative definition of the dynamic timescale which has
been introduced in \citet{Thompson:etal:2001}
\begin{equation*}
\tau_\rho = \left\vert
\frac{1}{v}\frac{1}{\rho}\frac{\partial\rho}{\partial r}
\right\vert_{T=0.5\text{MeV}}^{-1},
\end{equation*}
as well as a timescale approximation which has been derived in
\citet{QianWoosley:1996} Eq.(61)
\begin{equation*}
\tau_\text{QW} \propto \frac{1}{L_{\bar{\nu}_e}}
\frac{1}{\epsilon_{\bar{\nu}_e}}
R_\text{PNS} M_\text{PNS},
\end{equation*}
where additionally approximations for the mass outflow rate
and the entropy per baryon are derived as follows
\begin{equation*}
\left(\frac{dM}{dt}\right)_\text{QW}
\propto
L_{\bar{\nu}_e}^{5/3} \epsilon_{\bar{\nu}_e}^{10/3}
R_\text{PNS}^{5/3} M_\text{PNS}^{-2},
\end{equation*}
\begin{equation*}
S_\text{QW} \propto 
\frac{1}{L_{\bar{\nu}_e}^{1/6}}\frac{1}{\epsilon_{\bar{\nu}_e}^{1/3}}
\frac{1}{R_\text{PNS}^{2/3}} M_\text{PNS},
\end{equation*}
where \(R_\text{PNS}\) and \(M_\text{PNS}\) are the PNS radius and mass
respectively, which we take to be given by the electron-antineutrinosphere.
\(\epsilon_{\bar{\nu}_e }\) is again the ratio of mean-square electron-antineutrino
energy to average electron-antineutrino energy and \(L_{\bar{\nu}_e}\) is the
electron-antineutrino luminosity, both taken at the neutrinosphere.
In comparison with previous wind studies (see for example Fig.~4 of
\citet{Arcones:etal:2007}), we find generally a longer timescale of
\(\tau_\text{dyn}=10-50\) ms shown in Fig.~\ref{fig-accretion} (middle column)
for the \(10.8\) (middle panel) and \(18\) M\(_\odot\) (bottom panel) Fe-core
progenitors using the enhanced opacities (thin lines) and
\(\tau_\text{dyn}=40-200\) ms
in Fig.~\ref{fig-accretion} (middle column, top panel)
for the \(8.8\) M\(_\odot\) O-Ne-Mg-core
using the standard rates based on \citet{Bruenn:1985}.
This corresponds to a mass outflow rate of
\(10^{-3}-10^{-4}\) M\(_\odot\) s\(^{-1}\)
shown in Fig.~\ref{fig-accretion} (left column)
for the \(10.8\) (middle panel) and \(18\) M\(_\odot\) (bottom panel)
Fe-core progenitors using the enhanced opacities (thin lines) and
\(10^{-3}-10^{-5}\) M\(_\odot\) s\(^{-1}\)
in Fig.~\ref{fig-accretion} (left column, top panel)
for the \(8.8\) M\(_\odot\) O-Ne-Mg-core using the standard rates.
Fig.~\ref{fig-accretion} also compares the mass outflow rate, timescale
and entropy per baryon with the approximations derived in \citet{QianWoosley:1996},
which are in qualitative agreement with our data obtained via
Boltzmann neutrino transport.
The differences for the mass outflow and the entropy at later times are
maximally on the order of \(2-5\) and relate most likely to the crucial
assumptions made during the derivation of the above expressions,
e.g. hydrostatic equilibrium, \(R_\text{PNS}=10\) km, 
\(L_{\nu_e}\simeq L_{\bar{\nu}_e}\), \(\epsilon_{\bar{\nu}_e }=20\) MeV,
which differ significantly from our findings.
Furthermore, since the \(10.8\) and \(18\) M\(_\odot\) wind models are
obtained using enhanced opacities, Fig.~\ref{fig-accretion} compares
additionally simulations using these tuned rates (thin lines) with data
obtained using the standard rates (thick lines) where again the
approximations from \citet{QianWoosley:1996} are in qualitative
agreement with our findings.
The differences between simulations based on enhanced rates (thin lines)
and standard rates (thick lines) in Fig.~\ref{fig-accretion},
i.e. higher mass outflow rates, shorter timescales and slightly lower
entropies per baryon, are due to the stronger dynamic effect of the more
pronounced wind in the models using the enhanced opacities.

However, the wind entropies of \(40 - 100\) k\(_\text{B}\) found
(initially driven due to neutrino heating and additionally
due to the deceleration in the reverse shock) are significantly
smaller than often assumed in the literature and the previously
accelerated matter does not become neutron-rich as the neutrino-driven
wind decelerates behind the explosion ejecta but stays slightly
proton-rich where \(Y_e = 0.51-0.54\) for more than \(10\) seconds.
This, in combination with the much slower PNS contraction
illustrated via the neutrinospheres in Fig.~\ref{fig-nuspheres-h10a-wind}
in comparison to static steady-state and dynamic wind models suggest that
the assumptions made in previous wind studies should be carefully
reconsidered.
With respect to \citet{Woosley:etal:1994} (e.g. Fig.~3), we find
generally smaller mean neutrino energies which decrease with respect
to time after bounce.
This results in a decreasing difference between the electron flavor neutrino
mean energies, while in \citet{Woosley:etal:1994} this difference increases.
This fact in combination with the different PNS properties found in
\citet{Woosley:etal:1994}, enabled a strong neutrino-driven wind
where high entropies up to \(400\) k\(_\text{B}\)/baryon
and a low electron fraction of \(Y_e\simeq0.35-0.45\) was obtained.
These properties of the neutrino-driven wind differ quantitatively from our
results, where lower entropies per baryon are obtained and matter
stays proton-rich for more than \(10\)~seconds.

\section{Summary and Outlook}

For the first time, spherically symmetric core collapse supernova simulations
based on general relativistic radiation hydrodynamics and three-flavor Boltzmann
neutrino transport are performed consistently for more than \(20\) seconds.
We follow the dynamical evolution of low- and intermediate-mass progenitors
through the collapse, bounce, post-bounce, explosion and neutrino-driven wind phases.
The explosions of Fe-core progenitors of \(10.8\) and \(18\) M\(_\odot\) are modeled
using artificially enhanced opacities, while the explosion of the \(8.8\) M\(_\odot\)
O-Ne-Mg-core is obtained using the standard opacities.
For all models under investigation, we confirm the formation and illustrate the
conditions for the appearance of the neutrino-driven wind during the dynamical
evolution after the explosions have been launched.
For the O-Ne-Mg-core and the \(10.8\) M\(_\odot\) Fe-core progenitor models,
the supersonic neutrino-driven wind collides with the slower expanding explosion
ejecta where due to the deceleration the neutrino-driven wind termination shock
appears.
We discuss the impact of the reverse shock for several properties of the ejecta
and find general agreement with \citet{Arcones:etal:2007}.

The comparison with approximate and static steady-state as well as parametrized
dynamic wind models leads to a discrepancy in the obtained physical properties
of the neutrino-driven wind.
Although the evolution of the hydrodynamic variables are in general agreement,
we find smaller neutrino luminosities and a different behavior of the mean
neutrino energies.
In particular, the differences between the neutrino and antineutrino
luminosities and mean neutrino energies are smaller.
These differences reduce with time as the PNSs contract, which results in
generally proton-rich neutrino-driven winds over more than \(10\) seconds
for all our models.
Hence, the suggestion that the physical conditions in the neutrino-driven wind
could be favorable for the nucleosynthesis of heavy elements via the \(r\)-process
could not be confirmed.
For the accurate determination of the yields of the neutrino-driven wind,
detailed nucleosynthesis analysis based on a large nuclear reaction network,
taking the \(r\)-, \(p\)- and \(\nu p\)-processes into account, is required.
In order to further support the robustness, improvements of the input
physics such as weak magnetism and nucleon-nucleon recoil
(following e.g. \citet{Horowitz:2002}), taking the presence of light
and heavy clusters of nuclei into account as well as different EoSs with
respect to different PNS contraction behaviors, should be considered.
These may have a strong influence on the properties of the neutrino spectra
at the neutrinospheres and may therefore modify some of the results found
in the present study of the neutrino-driven wind.

Our simulations are carried out until the neutrino-driven wind settles down
to a quasi-stationary state leading to the initial and neutrino dominated
PNS cooling phase.
There, the simulations have to be stopped because important neutrino cooling
processes like the direct and modified URCA processes are not taken into
account yet.
However, a smooth connection to isolated neutron or protoneutron
star cooling studies comes into reach for future work
(\citet{HendersonPage:2007}).

\section*{Acknowledgment}

The authors would like to thank A.~Arcones and G.~Mart{\'{\i}}nez-Pinedo
for discussions and helpful comments.
The project was funded by the Swiss National Science Foundation
grant numbers PP002-106627/1, PP00P2-124879 and 200020-122287.
The authors are additionally supported by CompStar,
a research networking program of the European Science Foundation,
and the Scopes project funded by the
Swiss National Science Foundation grant. no. IB7320-110996/1.
A.~Mezzacappa is supported at the Oak Ridge National Laboratory,
which is managed by UT-Battelle, LLC for the
U.S. Department of Energy under contract
DE-AC05-00OR22725.
%


\begin{thebibliography}{48}
\expandafter\ifx\csname natexlab\endcsname\relax\def\natexlab#1{#1}
\fi
  \expandafter\ifx\csname url\endcsname\relax
  \def\url#1{{\tt #1}}
\fi

\bibitem[{Arcones} et~al.(2007){Arcones}, {Janka}, and
  {Scheck}]{Arcones:etal:2007}
{Arcones}, A.; {Janka}, H.-Th.; and {Scheck} L. 2007,
\newblock {\em \aap}, 467, 1227

\bibitem[{Arcones} et~al.(2008){Arcones}, {Mart{\'{\i}}nez-Pinedo}, and
  {Connor}, and {Schwenk}, and {Janka}, and {Horowitz},
  and {Langanke}]{Arcones:etal:2008}
{Arcones}, A.; {Mart{\'{i}}nez-Pinedo}, G.; {O'Conner}, E.;
{Schwenk}, A.; {Janka}, H.-Th.; {Horowitz}, C.~J.; and
{Langanke}, K. 2008,
\newblock {\em \prc}, 78, 015806-+

\bibitem[{Bethe} and {Wilson}(1985)]{BetheWilson:1985}
{Bethe}, H.~A. and {Wilson}, J.~R. 1985,
\newblock {\em \apj}, 295, 14


\bibitem[{Bruenn}(1985)]{Bruenn:1985}
{Bruenn}, S.~W. 1985, \newblock {\em \apjs}, 58,771

\bibitem[{Bruenn} et~al.(2006){Bruenn}, {Dirk}, {Mezzacappa}, {Hayes},
  {Blondin}, {Hix}, and {Messer}]{Bruenn:etal:2006}
{Bruenn}, S.~W.; {Dirk}, C.~J.; {Mezzacappa}, A.; {Hayes}, J.~C.;
{Blondin}, J.~M.; {Hix}, W.~R. and {Messer}, O.~E.~B. 2006,
\newblock {\em Journal of Physics Conference Series}, 46, 393

\bibitem[{Burrows} et~al.(1995){Burrows}, {Hayes}, and
{Fryxell}]{Burrows:etal:1995}
{Burrows}, A.; {Hayes}, J. and {Fryxell}, B.~A. 1995,
\newblock {\em \apj}, 450, 830

\bibitem[{Duncan} et~al.(1986){Duncan}, {Shapiro}, and
  {Wasserman}]{Duncan:etal:1986}
{Duncan}, R.~C.; {Shapiro}, S.~L. and {Wasserman}, I. 1986,
\newblock {\em \apj}, 309, 141

\bibitem[{Fischer} et~al.(2009){Fischer}, {Whitehouse}, {Mezzacappa},
  {Thielemann}, and {Liebend{\"o}rfer}]{Fischer:etal:2009}
{Fischer}, T.; {Whitehouse}, S.~C.; {Mezzacappa}, A.;
{Thielemann}, F.-K. and {Liebend{\"o}rfer}, M. 2009,
\newblock {\em \aap}, 499, 1

\bibitem[{Fr{\"o}hlich} et~al.(2006{\natexlab{a}}){Fr{\"o}hlich}, {Hauser},
  {Liebend{\"o}rfer}, {Mart{\'{\i}}nez-Pinedo}, {Thielemann}, {Bravo},
  {Zinner}, {Hix}, {Langanke}, {Mezzacappa}, and
  {Nomoto}]{Froehlich:etal:2006a}
{Fr{\"o}hlich}, C.; {Hauser}, P.; {Liebend{\"o}rfer}, M.;
{Mart{\'{\i}}nez-Pinedo}, G.; {Thielemann}, F.-K.; {Bravo}, E.;
{Zinner}, N.~T.; {Hix}, W.~R.; {Langanke}, K.; {Mezzacappa}, A.
and {Nomoto}; K. 2006{\natexlab{a}}
\newblock {\em \apj}, 637, 415

\bibitem[{Fr{\"o}hlich} et~al.(2006{\natexlab{b}}){Fr{\"o}hlich}, {Hix},
  {Mart{\'{\i}}nez-Pinedo}, {Liebend{\"o}rfer}, {Thielemann}, {Bravo},
  {Langanke}, and {Zinner}]{Froehlich:etal:2006c}
{Fr{\"o}hlich}, C.; {Hix}, W.~R.; {Mart{\'{\i}}nez-Pinedo}, G.;
{Liebend{\"o}rfer}, M.; {Thielemann}, F.-K.; {Bravo}, E.; {Langanke}, K.
and {Zinner}, N.~T. 2006{\natexlab{b}},
\newblock {\em New Astronomy Review}, 50, 496

\bibitem[{Fr{\"o}hlich} et~al.(2006{\natexlab{c}}){Fr{\"o}hlich},
  {Mart{\'{\i}}nez-Pinedo}, {Liebend{\"o}rfer}, {Thielemann}, {Bravo}, {Hix},
  {Langanke}, and {Zinner}]{Froehlich:etal:2006b}
{Fr{\"o}hlich}, C.; {Mart{\'{\i}}nez-Pinedo}, G.; {Liebend{\"o}rfer}, M.;
{Thielemann}, F.-K.; {Bravo}, E.; {Hix}, W.-R.; {Langanke}, K.
and {Zinner}, N.~T. 2006{\natexlab{c}}
\newblock {\em Physical Review Letters}, 96 (14), 142502

\bibitem[{Henderson} and {Page}(2007)]{HendersonPage:2007}
{Henderson}, J~A. and {Page}, D. 2007,
\newblock {\em \apss}, 308, 513

\bibitem[{Herant} et~al. (1994)]{Herant:etal:1994}
{Herant}, M.; {Benz}, W.; {Hix}, W.~R.; {Fryer}, C.~L. and {Colgate}, S.~A. 1994,
\newblock {\em \apj}, 435, 339


\bibitem[{Hix} et~al. (2003)]{Hix:etal:2003}
{Hix}, W.~R.; {Messer}, O.~B.~E.; {Mezzacappa}, A.;
{Liebend\"orfer}, M.; {Sampaio}, J.; {Langanke}, K.; {Dean}, D.~J. and
{Mart{\'{\i}}nez-Pinedo}, G. 2003,
\newblock {\em Physical Review Letters}, 91, 201102

\bibitem[{Hoffman} et~al.(2007){Hoffman}, {Pruet}, {Fisker}, {Janka}, {Burras}, and {Woosley}]{Hoffman:etal:2007}
{Hoffman}, R.~D.; {Pruet}, J.; {Fisker}, J.~L.; {Janka}, H.-Th.;
{Burras}, R. and {Woosley}, S.~E. 2007,
\newblock {\em ArXiv e-prints} 0712.2847

\bibitem[{Hoffman} et~al.(1997){Hoffman}, {Woosley}, and
  {Qian}]{Hoffman:etal:1997a}
{Hoffman}, R.~D.; {Woosley}, S.~E. and {Qian}, Y.~Z. 1997,
\newblock {\em \apj}, 482, 951

\bibitem[{Horowitz} (2002){Horowitz}]{Horowitz:2002}
{Horowitz}, C.~J. 2002,
\newblock {\em \prd}, 65, 043001

\bibitem[{Janka} and {M\"uller}(1995)]{JankaMueller:1995}
{Janka}, H.-Th. and {M\"uller}, E. 1995,
\newblock {\em \apjl}, 448, L109

\bibitem[{Janka} and {M\"uller} (1996)]{JankaMueller:1996}
{Janka}, H.-Th. and M{\"u}ller, E.  1996,
\newblock {\em \aap}, 306, 167

\bibitem[{Janka}(2001)]{Janka:2001}
{Janka}, H.-Th. 2001,
\newblock {\em \aap}, 368, 527

\bibitem[{Janka} et~al.(2005){Janka}, {Buras}, {Kitaura Joyanes}, {Marek},
  {Rampp}, and {Scheck}]{Janka:etal:2005}
{Janka}, H.-Th.; {Buras}, R.; {Kitaura Joyanes}, F.~S.;
{Marek}, A.; {Rampp}, M. and {Scheck}, L. 2005,
\newblock {\em Nuclear Physics A}, 758, 19

\bibitem[{Janka} et~al.(2008){Janka}, {Marek}, {M{\"u}ller}, and
  {Scheck}]{Janka:etal:2008a}
{Janka}, H.-Th.; {Marek}, A.; {M{\"u}ller}, B. and {Scheck}, L. 2008,
\newblock In C.~{Bassa}, Z.~{Wang}, A.~{Cumming}, and V.~M. {Kaspi}, editors,
  {\em 40 Years of Pulsars: Millisecond Pulsars, Magnetars and More}, volume
  983 of {\em American Institute of Physics Conference Series},
  pages 369--378

\bibitem[{Kitaura} et~al.(2006){Kitaura}, {Janka}, and
  {Hillebrandt}]{Kitaura:etal:2006}
{Kitaura}, F.~S.; {Janka}, H.-Th. and {Hillebrandt}, W. 2006,
\newblock {\em \aap}, 450, 345

\bibitem[{Kotake} et~al.(2006){Kotake}, {Sato}, and
  {Takahashi}]{Kotake:etal:2006}
{Kotake}, K.; {Sato}, K. and {Takahashi}, K. 2006,
\newblock {\em Reports on Progress in Physics}, 69, 971

\bibitem[{Langanke} et~al. (2003)]{Langanke:etal:2003}
{Langanke}, K.; {Mart{\'{\i}}nez-Pinedo}, G.; {Sampaio}, J.~M.;
{Dean}, D.~J.; {Hix}, W.~R.; {Messer}, O.~B.~E.; {Mezzacappa}, A.;
{Liebend{\"o}rfer}, M.; {Janka}, H.-Th. and {Rampp}, M. 2003,
\newblock {\em Physical Review Letters}, 90, 241102

\bibitem[{Liebend{\"o}rfer} et~al.(2001{\natexlab{a}})
{Liebend{\"o}rfer}, {Mezzacappa}, {Thielemann}, {Messer}, {Hix}, {Bruenn}]
{Liebendoerfer:etal:2001a}
{Liebend{\"o}rfer}, M.; {Mezzacappa}, A.; {Thielemann}, F.-K.;
{Messer}, O.~E.~B.; {Hix}, W.~R. and {Bruenn}, S. 2001{\natexlab{a}},
\newblock {\em \prd}, 63, 103004

\bibitem[{Liebend{\"o}rfer} et~al.(2001{\natexlab{b}})
{Liebend{\"o}rfer}, {Mezzacappa}, {Thielemann}]
{Liebendoerfer:etal:2001b}
{Liebend{\"o}rfer}, M.; {Mezzacappa}, A. and {Thielemann}, F.-K. 
2001{\natexlab{b}},
\newblock {\em \prd}, 63, 104003

\bibitem[{Liebend{\"o}rfer} et~al.(2002){Liebend{\"o}rfer}, {Rosswog}, and
  {Thielemann}]{Liebendoerfer:etal:2002}
{Liebend{\"o}rfer}, M.; {Rosswog}, S. and {Thielemann}, F.-K. 2002,
\newblock {\em \apjs}, 141, 229

\bibitem[{Liebend{\"o}rfer} (2004){Liebend{\"o}rfer}]{Liebendoerfer:2004}
{Liebend{\"o}rfer}, M. 2004,
\newblock \emph{Proceedings of the 12th Workshop on Nuclear Astrophysics},
Rep. No. MPA/P14, 143, \emph{eprint}, astro-ph/0405029

\bibitem[{Liebend{\"o}rfer} et~al.(2004){Liebend{\"o}rfer}, {Messer},
  {Mezzacappa}, {Bruenn}, {Cardall}, and 
  {Thielemann}]{Liebendoerfer:etal:2004}
{Liebend{\"o}rfer}, M.; {Messer}, O.~E.~B.; {Mezzacappa}, A.;
{Bruenn}, S.~W.; {Cardall}, C.~Y. and {Thielemann}, F.-K. 2004,
\newblock {\em \apjs}, 150, 263

\bibitem[{Liebend{\"o}rfer} et~al.(2005){Liebend{\"o}rfer}, {Ramp},
{Janka} and {Mezzacappa}]{Liebendoerfer:etal:2005}
{Liebend{\"o}rfer}, M.; {Ramp}, M; {Janka}, H.-Th. and
{Mezzacappa}, A. 2005,
\newblock {\em \apj}, 620, 840

\bibitem[{Marek} and {Janka}(2009)]{MarekJanka:2009}
{Marek}, A. and {Janka}, H.-Th. 2009,
\newblock {\em \apj}, 694, 664

\bibitem[{Mayle} and {Wilson}(1988)]{MayleWilson:1988}
{Mayle}, R. and {Wilson}, J.~R. 1988,
\newblock {\em \apj}, 334, 909

\bibitem[{Messer} and {Bruenn}(2003)]{MesserBruenn:2003}
{Messer}, O.~E.~B. and {Bruenn}, S.~W. 2003,
\newblock {private communications}

\bibitem[{Mezzacappa} et~al.(2006){Mezzacappa}, {Blondin}, {Messer}, and
  {Bruenn}]{Mezzacappa:etal:2006}
{Mezzacappa}, A.; {Blondin}, J.~M.; {Messer}, O.~E.~B. and {Bruenn}, S.~W.
2006,
\newblock In {\em Origin of Matter and Evolution of Galaxies},
volume 847 of {\em American Institute of Physics Conference Series},
pages 179--189

\bibitem[{Mezzacappa} and {Bruenn}(1993{\natexlab{a}})]{MezzacappaBruenn:1993a}
{Mezzacappa}, A. and {Bruenn}, S.~W. 1993{\natexlab{a}},
\newblock {\em \apj}, 405, 637

\bibitem[{Mezzacappa} and {Bruenn}(1993{\natexlab{b}})]{MezzacappaBruenn:1993b}
{Mezzacappa}, A. and {Bruenn}, S.~W. 1993{\natexlab{b}},
\newblock {\em \apj}, 405, 669

\bibitem[{Mezzacappa} and {Bruenn}(1993{\natexlab{c}})]{MezzacappaBruenn:1993c}
{Mezzacappa}, A. and {Bruenn}, S.~W. 1993{\natexlab{c}},
\newblock {\em \apj}, 410, 740

\bibitem[{Mezzacappa} and {Messer}(1999)]{MezzacappaMesser:1999}
{Mezzacappa}, A. and {Messer}, O.~E.~B. 1999,
\newblock {\em Journal of Computational and Applied Mathematics}, 109, 281

\bibitem[{Miller} et~al. (1993)]{Miller:etal:1993}
{Miller}, D.~S.; {Wilson}, J.~R. and {Mayle}, R.~W. 1993,
\newblock {\em \apj}, 415, 278

\bibitem[{Misner} and {Sharp}(1964)]{MisnerSharp:1964}
{Misner}, C.W. and {Sharp}, D.~H. 1964,
\newblock {\em Physical Review}, 136, 571

\bibitem[{Nomoto} (1983)]{Nomoto:1983}
{Nomoto}, K. 1983,
\newblock {\em IAU Symposium}, 101, 139

\bibitem[{Nomoto} (1984)]{Nomoto:1984}
{Nomoto}, K. 1984,
\newblock {\em \apj}, 277, 791

\bibitem[{Nomoto} (1987)]{Nomoto:1987}
{Nomoto}, K. 1987,
\newblock {\em \apj}, 322, 206

\bibitem[{Otsuki} et~al.(2000){Otsuki}, {Tagoshi}, {Kajino}, and
  {Wanajo}]{Otsuki:etal:2000}
{Otsuki}, K.; {Tagoshi}, H.; {Kajino}, T. and {Wanajo}, S.~Y. 2000,
\newblock {\em \apj}, 533, 424

\bibitem[{Panov} and {Janka}(2009)]{PanovJanka:2009}
{Panov}, I.~V. and {Janka}, H.-Th. 2009,
\newblock {\em \aap}, 494, 829

\bibitem[{Qian} and {Woosley}(1996)]{QianWoosley:1996}
{Qian}, Y.~Z. and {Woosley}, S.~E. 1996,
\newblock {\em \apj}, 471, 331

\bibitem[{Sagert} et~al.(2009){Sagert}, {Fischer}, {Hempel}, {Pagliara},
  {Schaffner-Bielich}, {Mezzacappa}, {Thielemann}, and
  {Liebend{\"o}rfer}]{Sagert:etal:2009}
{Sagert}, I.; {Fischer}, T.; {Hempel}, M.; {Pagliara}, G.;
{Schaffner-Bielich}, J.; {Mezzacappa}, A.; {Thielemann}, F.-K.
and {Liebend{\"o}rfer}, M. 2009,
\newblock {\em Physical Review Letters}, 102 (8) 081101

\bibitem[{Scheck} et~al.(2006){Scheck}, {Kifondis}, {Janka},
and {M{\"u}ller}]{Scheck:etal:2006}
{Scheck}, L.; {Kifondis}, K.; {Janka}, H.-T. and
{M{\"u}ller}, E. 2006,
\newblock {\em \aap}, 457, 963 

\bibitem[{Schinder} and {Shapiro}(1982)]{SchinderShapiro:1982}
{Schinder}, P.~J. and {Shapiro}, S.~L. 1982,
\newblock {\em \apjs}, 50, 23

\bibitem[{Shen} et~al.(1998){Shen}, {Toki}, {Oyamatsu}, and
  {Sumiyoshi}]{Shen:etal:1998}
{Shen}, H.; {Toki}, H.; {Oyamatsu}, K. and {Sumiyoshi}, K. 1989,
\newblock {\em Progress of Theoretical Physics}, 100, 1013

\bibitem[{Takahashi} et~al.(1994){Takahashi}, {Witti}, and
  {Janka}]{Takahashi:etal:1994}
{Takahashi}, K.; {Witti}, J. and {Janka}, H.-Th. 1994,
\newblock {\em \aap}, 286, 857

\bibitem[{Thielemann} et~al.(2004){Thielemann}, {Brachwitz}, {H{\"o}flich},
  {Martinez-Pinedo}, and {Nomoto}]{Thielemann:etal:2004}
{Thielemann}, F.-K.; {Brachwitz}, F.; {H{\"o}flich}, P.;
{Martinez-Pinedo}, G. and {Nomoto}, K. 2004,
\newblock {\em New Astronomy Review}, 48, 605

\bibitem[{Thompson} and {Burrows}(2001)]{ThompsonBurrows:2001}
{Thompson}, T.~A. and {Burrows}, A. 2001,
\newblock {\em Nuclear Physics A}, 688, 377

\bibitem[{Thompson} et~al.(2001){Thompson}, {Burrows}, and
  {Meyer}]{Thompson:etal:2001}
{Thompson}, T.~A.; {Burrows}, A. and {Meyer}, B.~S. 2001,
\newblock {\em \apj}, 562, 887

\bibitem[{Timmes} and {Arnett}(1999)]{TimmesArnett:1999}
{Timmes}, F.~X. and {Arnett}, D. 1999,
\newblock {\em \apjs}, 125, 277

\bibitem[{Wanajo}(2006{\natexlab{a}})]{Wanajo:2006a}
{Wanajo}, S. 2006{\natexlab{a}},
\newblock {\em \apj}, 647, 1323

\bibitem[{Wanajo}(2006{\natexlab{b}})]{Wanajo:2006b}
{Wanajo}, S. 2006{\natexlab{b}},
\newblock {\em \apjl}, 650, L79

\bibitem[{Wilson} and {Mayle}(1993){Wilson}, and {Mayle}]
{WilsonMayle:1993}
{Wilson}, J.~R. and {Mayle},W.~R. 1993,
\newblock {\em Phys. Rep.}, 227, 97

\bibitem[{Witti} et~al.(1994){Witti}, {Janka}, and
  {Takahashi}]{Witti:etal:1994}
{Witti}, J.; {Janka}, H.-Th. and {Takahashi}, K. 1994,
\newblock {\em \aap}, 286, 841

\bibitem[{Woosley} and {Baron}(1992)]{WoosleyBaron:1992}
{Woosley}, S.~E. and {Baron}, E. 1992,
\newblock {\em \apj}, 391, 228

\bibitem[{Woosley} et~al.(1994)]{Woosley:etal:1994}
{Woosley}, S.~E.; {Wilson}, J.~R.; {Mathews}, G.~J.; {Hoffman}, R.~D.
and {Meyer}, B.~S. 1994,
\newblock {\em \apj}, 433, 229

\bibitem[{Woosley} et~al.(2002){Woosley}, {Heger}, and
  {Weaver}]{Woosley:etal:2002}
{Woosley}, S.~E.; {Heger}, A. and {Weaver}, T.~A. 2002,
\newblock {\em Reviews of Modern Physics}, 74, 1015

\bibitem[{Yueh} and {Buchler}(1976){Yueh} and {Buchler}]
{YuehBuchler:1976}
{Yueh}, W.~R. and {Buchler}, J.~R. 1976,
\newblock {\em Ap\&SS}, 41, 221

\end{thebibliography}

\end{document}